\documentclass[12pt, onecolumn, oneside, draftclsnofoot]{IEEEtran}
\usepackage{ amssymb, graphicx, dsfont, setspace, url}

\usepackage[cmex10]{amsmath}
\usepackage{amsmath, amssymb, graphicx, setspace,url, amsfonts, amsthm, multirow,stmaryrd,mathabx,MnSymbol}
\usepackage{graphicx,caption,subcaption, dsfont}

\DeclareMathOperator{\e}{e}

\DeclareMathOperator{\supr}{sup}

\DeclareMathOperator{\Pro}{Pr}

\newtheoremstyle{Amin}% name
  {3pt}%      Space above
  {3pt}%      Space below
  {}%         Body font
  {}%         Indent amount (empty = no indent, \parindent = para indent)
  {\bfseries}% Thm head font
  {:}%        Punctuation after thm head
  {.5em}%     Space after thm head: " " = normal interword space;
  {}%         Thm head spec (can be left empty, meaning `normal')

\theoremstyle{Amin}
\newtheorem{lemma}{Lemma}
\newtheorem{remark}{Remark}
\newtheorem{defin}{Definition}
\newtheorem{prop}{Proposition}

\newtheorem{theorem}{Theorem}

\usepackage[T1]{fontenc}
\usepackage[latin9]{inputenc}

\usepackage{setspace}

\begin{document}
\setcounter{page}{1}

\title{Semi-Quantitative Group Testing:\\ A Unifying Framework for Group Testing\\with Applications in Genotyping}
\author{Amin~Emad and Olgica Milenkovic\\
\today
\thanks{This work was presented in part at the IEEE 2012 International Symposium on Information Theory (ISIT'12)~\cite{AM11}.}
\thanks{The authors are with the Department of Electrical and Computer Engineering, University of Illinois at Urbana-Champaign, Urbana, IL. (e-mail: emad2@illinois.edu; milenkov@illinois.edu).}
\thanks{This work was supported in part by a Natural Sciences and Engineering Research Council of Canada (NSERC) scholarship and NSF Grants CIF 1218764, CIF 1117980, and STC Class 
2010, CCF 0939370.}% <-this % stops a space
\thanks{The authors would like to thank Alexander Barg, Arkadii D'yachkov, and Yaniv Erlich for useful discussions.}}

\maketitle
\thispagestyle{empty}

\begin{abstract} We propose a novel group testing method, termed \emph{semi-quantitative group testing}, motivated by a class of problems arising in 
genome screening experiments. 
Semi-quantitative group testing (SQGT) is a (possibly) non-binary pooling scheme that may be viewed as a concatenation of an adder channel and an integer-valued quantizer. In its full generality, SQGT may be viewed as a unifying framework for group testing, in the sense that most group testing models are special instances of SQGT. For the new testing scheme, we define the notion of SQ-disjunct and SQ-separable codes, representing generalizations of classical disjunct and separable codes. We describe several combinatorial and probabilistic constructions for such codes. While for most of these constructions we assume that the number of defectives is much smaller than total number of test subjects, we also consider the case in which there is no restriction on the number of defectives and they may be as large as the total number of subjects. For the codes constructed in this paper, we describe a number of efficient decoding algorithms. In addition, we describe a belief propagation decoder for sparse SQGT codes for which no other efficient decoder is currently known. Finally, we define the notion of capacity of SQGT and evaluate it for some special choices of parameters using information theoretic methods. 

\end{abstract}

%%%%%%%%%%%%%%%%%%%%%%%%%%%%%%%%%%%%%%%%%%%%%%%%%%%%%%%%%%%%%%%%%%%%%%%%%%%%%%%%%%%%%%%%%%%%%%%%%%%%%%%%%%%%%%%%%%%%%%%%%%%%%%%%
\newpage
\section {Introduction}
Group testing (GT) is a general term for a family of test schemes designed to identify a number of subjects with some particular characteristic -- called \emph{defectives} (or \emph{positives}) -- among a large pool of subjects. 
The idea behind GT is that if the number of defectives is much smaller than the number of subjects, one can reduce the number of experiments required for identifying the defectives by testing properly 
chosen subgroups of subjects rather than testing each subject individually. In its full generality, GT may be viewed as the problem of inferring the state of a system from the superposition of the state 
vectors of a subset of the system's elements. As such, GT has found many applications in communication theory~\cite{TLP83}-\cite{FDH95}, signal processing~\cite{CRT06}-\cite{DH00}, computer science~\cite{AT09}-\cite{DDTW09}, and mathematics~\cite{CKRS10}. Some examples of these applications include error-correcting coding~\cite{A94},~\cite{AK92},~\cite{MS91}, identifying users accessing a multiple access channel (MAC)~\cite{DR81}, ~\cite{BPR93}, reconstructing sparse signals from low-dimensional projections~\cite{CRT06},~\cite{DO06}, and many others.

The group testing literature examines two partially overlapping categories of problems, based on the way the number of defectives is modeled: probabilistic GT and combinatorial GT. In the former case, 
a probability distribution is considered for the number of defectives, and the goal is to minimize the \emph{expected} number of tests (see for example~\cite{D43}-\cite{H75})\footnote{In some papers, "probabilistic group testing" refers to a probabilistic construction of tests in a combinatorial GT model. In this paper, we refer to such constructions as ``probabilistic constructions'' as opposed to ``explicit constructions''. }. 
In the latter case, the number of defectives (or at least an upper bound on the number of defectives) is known in advance~\cite{DH00}. 

Another way to distinguish between different GT schemes is through the way the tests are performed. In nonadaptive group testing \emph{all} the tests are designed in advance\footnote{The design of a single test reduces to selecting the subjects that are present in that test.}. In other words, the tests are designed in one pass, and the outcome of one test does not affect the design of another test. On the other hand, in sequential (adaptive) 
group testing, the result of one test may be used to govern the design of other tests, leading to more efficient pooling schemes (see~\cite{DH00} and references therein). 
Although, in general, sequential GT requires fewer tests, in most practical applications nonadaptive GT is preferred since it allows one to perform all tests simultaneously. This reduces the overall time required for testing. In what follows, we focus on combinatorial, nonadaptive GT.

Many different models have been considered for combinatorial GT; in the original setting described by Dorfman~\cite{D43} (henceforth, conventional GT or CGT) the result of a test indicates 
if there exist \emph{at least one} defective in the test. Hence, the test output equals $0$ if there are no defectives in the test, and $1$ otherwise. Another important model is the 
additive model~\cite{DH00}, also known as quantitative GT (QGT). In this model, the result of a test equals the exact number of defectives in that test. In the threshold group testing (TGT) model~\cite{D06}, if the number of defectives in a test is smaller than a fixed lower threshold, the test outcome is negative (or equal to $0$); if the number of defectives is larger than a fixed upper threshold, the test outcome is positive (or equal to $1$); and if the number of defectives is between the lower and upper threshold, the test result is arbitrary (either equal to $0$ or $1$). The difference between the upper and lower thresholds is called the gap. In yet another model introduced in~\cite{DR84}, a threshold is fixed beforehand and the test output corresponds to an additive model output whenever the number of defectives does not exceed the threshold. If the number
of defectives exceeds the threshold, the output of the test is some value outside the range of the sub-thresholded additive model output. 

In all these models, each subject is assigned a unique binary vector (codeword) of length equal to the total number of tests. 
Each coordinate of a subject's codeword corresponds to a test and equals $1$ if the subject is present in the test, and equals $0$ otherwise. 
Since in nonadaptive GT all the tests are designed in parallel, it is convenient to group all the codewords into a matrix (code) termed the \emph{test matrix} (test code). 
The test matrix is a binary matrix of size $m\times n$, where $m$ is the number of tests and $n$ is the number of subjects. 
The design of efficient test matrices has been a topic of interest for many years: for a comprehensive survey of such codes, see~\cite{DH00}, \cite{DH06}, and \cite{D04}. The two main families of test codes were 
originally designed for CGT by Kautz and Singleton~\cite{KS64}. The first family is known as \emph{disjunct codes} (or zero-false-drop codes), while the second family is usually referred to as \emph{separable codes} 
(or uniquely decipherable codes). Disjunct codes satisfy an \emph{inclusion} constraint: a $d$-disjunct code has the property that no codeword is included in -- or is covered by -- the component-wise Boolean ORs of any other $\leq d$ number of codewords. This property enables disjunct codes to uniquely identify up to $d$ defectives and also endows them with an efficient decoding algorithm. 
Separability is a weaker notion than disjunctness as it only requires the component-wise Boolean ORs of any two distinct sets of $\leq d$ codewords to be different. 

Despite the significant interest the subject has garnered in computer science, coding and combinatorial theory, and despite the analysis of 
many diverse extensions of the underlying problem, group testing has still not seen widespread use in medical sciences and biology. 
Two notable exceptions were the early use of group testing for DNA sequence analysis~\cite{DH06} and the very recent work on group testing for genotyping and 
biosensing~\cite{EGBHM10}-\cite{DSMB09}. The reason behind this practical failure of group testing in life sciences is that
most analytical models do not capture the full complexity of bioengineering systems. Model simplifications are necessarily introduced in order to derive closed-form expressions 
on the smallest number of tests required to perform the experiments or to guarantee code constructions with provable performance guarantees, 
thereby neglecting the fact that in practical applications such simplifications may not be appropriate. 
For example, one would be inclined to accept a number of tests higher than those predicted to be theoretically optimal for a coarse model if there is evidence that the scheme is suitable for practical implementation. 

This work represents the first step in developing a novel framework for group testing that caters to the unique needs of the emerging field of genotyping through high-throughput sequencing\footnote{Although this work was motivated by applications in genotyping, the model, results, and code constructions are applicable to a wide variety of applications in biology, communication theory, signal processing, etc.}, as motivated below.

\subsection{Challenges in Genotyping, and Semi-quantitative Group Testing}
Genotyping is an emerging field in systems biology concerned with determining genetic variations in the traits of individuals.  At the core of every genotyping method is DNA sequencing -- determining the genetic blueprint of an individual -- and a comparative analysis of the sequences obtained from different individuals. Comparative studies of the DNA makeup play an indispensable role in medical genetics, the goals of which are to efficiently determine ``outliers'' in genetic codes that may lead to devastating disorders or illnesses~\cite{EGBHM10}. 

One of the most important applications of genotyping is detecting the \emph{carriers} of a particular genetic disorder. Since the human genome consists of pairs of chromosomes, and paired chromosomes contain genes with matching functionalities, a human who has inherited a mutated gene may not display the symptoms of the genetic disease. In this situation, the individual has a normal (unmutated) copy of a gene, which prohibits the disease from being expressed. Although the carrier does not display disease symptoms, the offspring of two carriers may have the disease. While affected individuals can be diagnosed based on their symptoms, a carrier can only be identified via DNA screening. 

In the screening process of genotyping, one targets genomic regions known to harbor genetic mutations.  Until recently, only serial sequencing of the genome of one individual was possible; however, the introduction of the new class of genome sequencing methods dubbed \emph{the next-generation sequencing technologies}~\cite{C08} enabled parallel sequencing of the genome. These platforms break the genomic region of interest into short fragments and perform millions of sequence reads in a single run (for the description of one such platform, see Illumina~\cite{Ill}). Due to the high cost of sample preparation for sequencing, and, in order to fully utilize the potential of the sequencing platforms, multiplexing a large number of specimens in a single batch is essential. As a result, group testing presents itself as a natural paradigm to address these challenges, and the first steps in this direction were taken in~\cite{STR03,SMB07,EGBHM10,SAZ10}. Despite the promising results of applying the existing group testing models to genotyping, many practical problems still stand in the way of the wide-scale use of this method. 

One such problem arises from the fact that genotyping methods allow for more precise readings at the output than classical GT detectors, but still do not provide full information about the abundance of a target gene in the test. As a result, codes constructed for CGT or TGT underutilize the potential of these sequencers, while codes constructed for QGT are prone to errors due to ``overestimating'' the sequencers' precision. Specifically, since the precision of a sequencer often depends on the number of defectives and the amount of genetic material in the test, the error is signal/design dependent and cannot be modeled easily. In order to overcome this problem, in what follows we propose a new framework called semi-quantitative group testing (SQGT). 

In SQGT, the result of a test is a non-binary value that depends on the number of defectives through a given set of thresholds. The thresholds depend on the sequencer and represent its precision. The SQGT paradigm may be viewed as a combination of the adder model (QGT) and a decimator (quantizer). Although QGT has been widely studied in literature, the addition of a system-dependent decimator makes test construction and analysis quite challenging. It is worth emphasizing that the application of SQGT model is not limited to genotyping, and in general any scheme in which tests are obtained using a test device with \emph{limited} precision may be modeled as an instance of SQGT. In particular, CGT, TGT (with zero gap), and QGT are all special cases of SQGT. 

We also allow for the possibility of having different amounts of sample material for different test subjects, which results in non-binary test matrices. Although binary testing is required for some applications -- such as the classic coin weighing problem -- in other applications, such as conflict resolution in multiple access channel (MAC) and genotyping, non-binary tests may be used to further reduce the number of tests. While in binary test matrices a value $0$ or $1$ corresponds to the absence or presence of a subject in a test, respectively, in non-binary SQGT the value of an entry of the test matrix reflects the ``strength'' or ``concentration'' of a subject in a test. For example in conflict resolution in MAC, different non-binary values in a test correspond to different power levels of the users, while in genotyping they correspond to different amounts of genetic material of different subjects. For example, if the value corresponding to the $j^{\text{th}}$ subject in a genotyping test equals $2$, while the value corresponding to the $k^{\text{th}}$ subject is equal to $1$, this indicates that the amount of DNA of subject $j$ in this test is twice the amount of the DNA of subject $k$. 

The reason for focusing on integer-valued test matrices, as opposed to real-valued matrices, is that the sample preparation in genotyping is performed by robotic arms that are usually programed to sample the same amount of DNA. One could program the robotic arm to dispense different amounts of DNA into test wells, but such a process would be extremely complicated and imprecise. A better alternative is to program the robotic arm to dispense the same amount of DNA into a test well multiple times. Since all test wells contain integer multiples of the same volume of DNA, one can model the test parameters using bounded integers. 

Note that non-binary integer-valued group testing can be also used in applications where:
\begin{itemize}
\item The subjects to be tested come as a whole and cannot be divided into \emph{real-valued} parts. For example, in the coin-weighing problem, if one has $n$ bags of coins, where each bag contains $q-1$ identical coins, and some of the bags have counterfeit coins, one can use tests with an alphabet of size $q$ to find the counterfeit bag with fewer experiments than when using binary tests.
\item A real-valued alphabet may not be practical due to ``limited precision''. With unlimited precision, one could design \emph{one} single experiment to find any number of defectives among any number of subjects. 
\item Some robustness to errors and noise is needed in the testing schemes; integers, unlike reals, are spaced discretely, which ensures a form of error protection (see for example~\cite{DM09}). 
\end{itemize}
While there exist information theoretic approaches applicable to the study of non-binary test matrices~\cite[Ch. 6]{D04}, to the best of the authors' knowledge, the results on non-binary code construction relevant to group testing are limited to a handful of papers, including~\cite{J95} and~\cite{CW99}, where constructions are considered for an \emph{adder} MAC channel (i.e. QGT).

For the new model of SQGT with $Q$-ary test results and $q$-ary test sample sizes, $Q,q \geq 2$, we define a new generalization of \emph{disjunct} and \emph{separable} codes, called ``SQ-disjunct'' and ``SQ-separable'' codes, respectively. Probabilistic constructions as well as explicit constructions are provided for these two families of codes when the number of defectives is much smaller than the total number of subjects. In addition, the important special case of SQGT with equidistant thresholds is discussed in detail, and test constructions are provided for this model as well\footnote{SQGT with equidistant thresholds may be viewed as a special instance of \emph{quantized integer compressive sensing}, introduced in~\cite{DM09}, where the entries of the sensing matrices as well as the sparse vectors are allowed to be bounded integers. Another topic in the compressive sensing literature related to this SQGT model is quantized compressive sensing, one instance of which was discussed in~\cite{DPM09}.}. Furthermore, a generalization of the Lindstr{\"o}m construction for QGT~\cite{L75} is described, capable of identifying any number of defectives, even as large as the total number of subjects. Our derivations also have an information theoretic underpinning and are centered around the notion of \emph{capacity} of SQGT, which we study in relation to the minimal number of tests required to identify defectives with an average probability of error converging to zero.

Other problems arising in the context of genotyping -- such as \emph{copy number variation}~\cite{CAN05}-\cite{FPFRMAAJTHC}, probabilistic modeling of family trees within the GT framework, as well as multiple gene mutation disorder screening, and the resulting notion of two-dimensional group testing, will be discussed elsewhere.

The paper is organized as follows. Section~\ref{sec:model} describes the SQGT model. Section~\ref{sec:disjunct} introduces SQ-disjunct and SQ-separable codes and their properties. 
In Section~\ref{sec:construction}, we describe a number of combinatorial and probabilistic constructions for SQGT codes. The characteristics and parameters of these codes are summarized at the end of this section. Belief propagation decoders for probabilistic construction of SQGT codes are described in Section~\ref{sec:bp}, while Section~\ref{sec:informationtheory} includes information theoretic bounds and the capacity of SQGT. Section~\ref{sec:conclusion} concludes the paper.

%%%%%%%%%%%%%%%%%%%%%%%%%%%%%%%%%%%%%%%%%%%%%%%%%%%%%%%%
%%%%%%%%%%%%%%%%%%%%%%%%%%%%%%%%%%%%%%%%%%%%%%%%%%%%%%%%%%%%%%%%%%%%%%%%%%%%%%
\section{Semi-quantitative Group Testing: The Model}\label{sec:model}

Throughout the paper, we adopt the following notation. Bold-face upper-case and bold-face lower-case letters denote matrices and vectors, respectively. Calligraphic letters are used to denote sets. Asymptotic symbols such as $\sim$, $o(\cdot)$, and $O(\cdot)$ are used in a standard manner. For a positive integer $k$, we define $[k]:=\{0,1,\dots,k-1\}$, and $\llbracket{k}\rrbracket :=\{1,2,\dots,k\}$. For simplicity, 
we sometimes use $\mathcal{X}=\{\mathbf{x}_i\}_1^s$ to denote a set of $s$ codewords, $\mathcal{X}=\{\mathbf{x}_1,\mathbf{x}_2,\dots,\mathbf{x}_s\}$. 

Let $n$, $m$, and $d$ denote the number of test subjects, the number of tests, and the number of defectives, respectively. Let $S_i$ denote the $i^{\textnormal{th}}$ subject, $i\in\llbracket n\rrbracket$, and let $D_j$ be the $j^{\textnormal{th}}$ defective, $j\in\llbracket d\rrbracket $. Furthermore, let $\mathcal{D}$ denote the set of defectives, so that $|\mathcal{D}|=d$. Let $\mathbf{w}\in{[2]}^n$ be a binary vector with its $i^\textnormal{th}$ coordinate equal to $1$ if the $i^\text{th}$ subject is defective, and $0$ otherwise.

We assign to each subject a unique $q$-ary vector of length $m$, termed the codeword of the subject. Each coordinate of the codeword corresponds to a test. 
If $\mathbf{x}_i\in[q]^m$ denotes the codeword of the $i^{\textnormal{th}}$ subject, then the $k^{\textnormal{th}}$ coordinate of $\mathbf{x}_i$, denoted by $\mathbf{x}_i(k)$, may be viewed as the ``amount'' of $S_i$ used in the $k^{\textnormal{th}}$ test\footnote{Note that $q$ is actually the \emph{available} alphabet size and not necessarily the \emph{effective} alphabet size. In many constructions in this paper, we use an effective alphabet size smaller than $q$, but if the maximum available entry of the alphabet is $q-1$, we still call the alphabet size $q$.}. Note that the symbol $0$ indicates that $S_i$ is not present in the test. We denote the test matrix, or equivalently, the code, by $\mathbf{C}\in[q]^{m\times n}$. The goal is to construct a code such that the defectives can be uniquely identified in an SQGT model.  

\begin{table}[t!]\centering
\caption{Table of symbols and their definitions}
\begin{tabular}{|c|c|}
			\hline 
			Symbol &  Definition \\ 
			\hline\hline
		
			$n$ & Total number of subjects \\	\hline
			$m$ & Number of tests  \\				\hline
			$d$ & Number of defectives  \\	\hline				
			$Q$ & Size of the output alphabet\\	\hline
			$q$	& Size of the test matrix alphabet \\ \hline	
			$\eta_l$ & The $l^{\textnormal{th}}$ threshold where $l\in\llbracket Q\rrbracket$  \\	\hline
			$\mathcal{D}$ & Set of defectives  \\				\hline		
			$\mathbf{w}\in[2]^n$ & Indicator vector of defectives  \\	\hline								
			$\mathbf{y}\in[Q]^m$ & Vector of test results  \\	\hline										
			$\mathbf{x}_i\in [q]^m$ & Codeword assigned to the $i^{\textnormal{th}}$ subject   \\	\hline																	
			$\mathbf{C}\in[q]^{m\times n}$ & Code (test matrix)  \\\hline
			$e$ & Number of errors in $\mathbf{y}$ that $\mathbf{C}$ can correct  \\				\hline
			
			\end{tabular}
			\label{table:syms}

%			\vspace*{-15pt}	
\end{table}

The result of each test in SQGT is an integer from the set $[Q]$. Each test outcome depends on the number of defectives and their sample amount in the test through $Q$ thresholds, $\eta_l$ ($l\in\{1,2,\dots,Q\}$). Table~\ref{table:syms} summarizes the previously described notation. 

\begin{figure}
\includegraphics[width=0.8\textwidth]{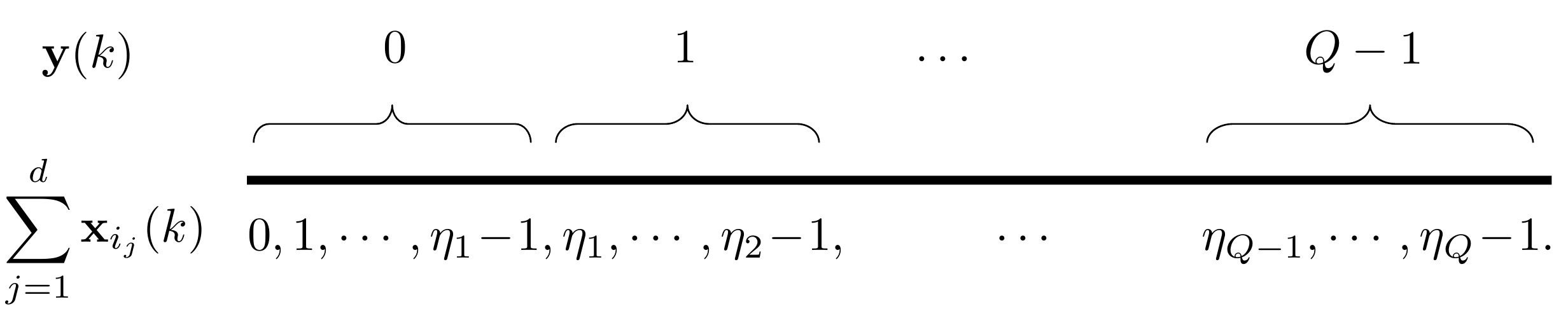}
\centering
\caption{The outcome of the $k^{\text{th}}$ test and its relationship with $\sum_{j=1}^d \mathbf{x}_{i_j}(k)$ through the thresholds in a SQGT model with (possibly) non-binary test design.}
\label{fig:test-result}
\end{figure}

\begin{figure}
\includegraphics[width=0.6\textwidth]{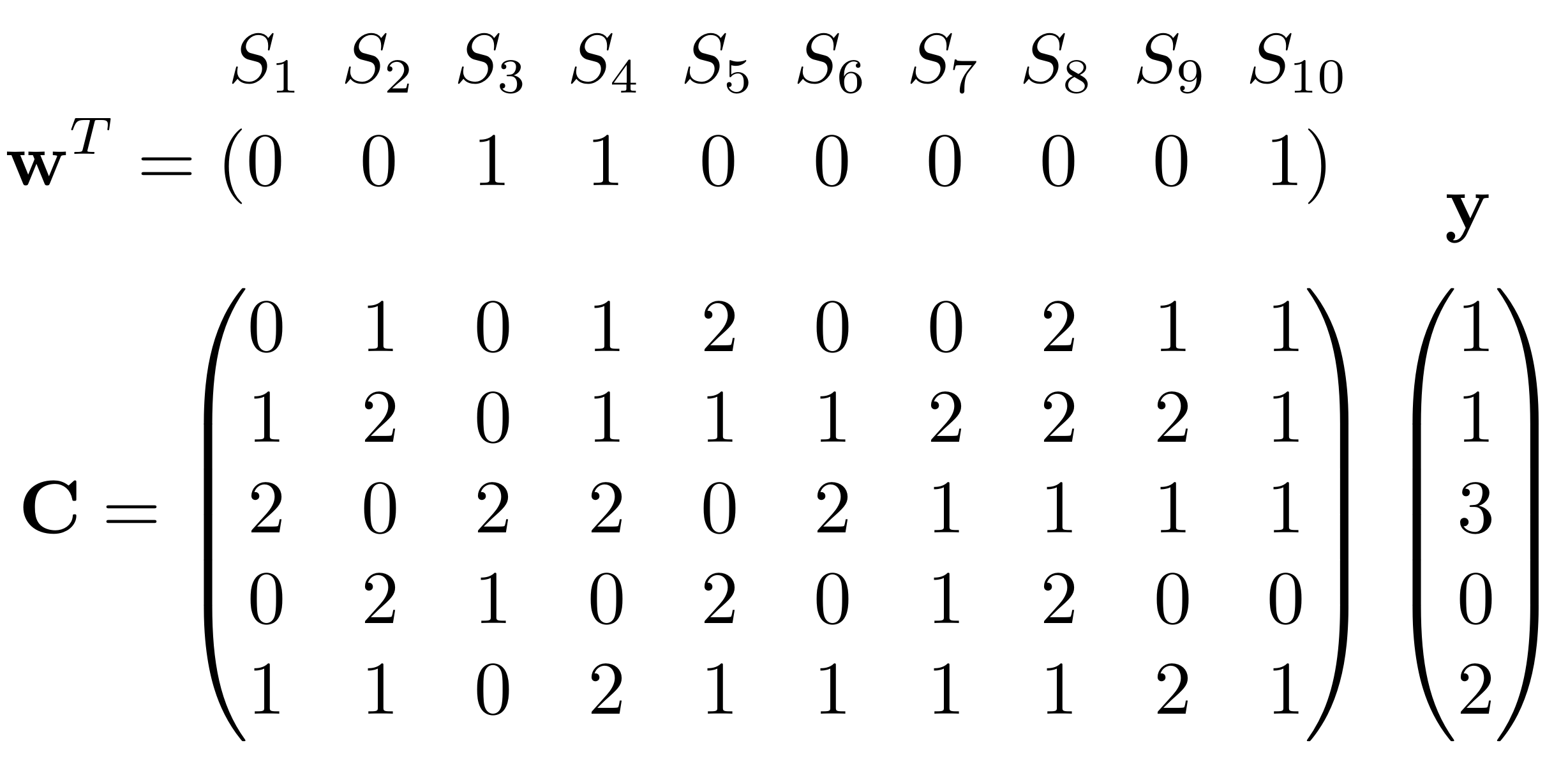}
\centering
\caption{A test matrix $\mathbf{C}$, indicator vector of defectives $\mathbf{w}$, and the corresponding vector of test results $\mathbf{y}$, for an SQGT scheme with $d=3$, $m=5$, $n=10$, $q=3$, $Q=4$, and $\boldsymbol{\eta}=[0,2,3,5,7]^T$.}
\label{fig:model}
\end{figure}

In order to simplify the relationship between the test results and the codewords assigned to the defectives, we use the following definition.
\begin{defin}\label{def1}
The ``SQ-sum'' of a set of $s\geq 1$ codewords, $\mathcal{X}=\{\mathbf{x}_1,\mathbf{x}_2,\dots,\mathbf{x}_s\}=\{\mathbf{x}_j\}_{1}^s$,  in a SQGT model with thresholds $\boldsymbol{\eta}=[\eta_0=0,\eta_1,\eta_2,\dots,\eta_Q]^T$, is represented by $\mathbf{y}_{\!_\mathcal{X}}=\bigoasterisk_{j=1}^{s}\mathbf{x}_j=\mathbf{x}_1\oasterisk\mathbf{x}_2\oasterisk\dots\oasterisk\mathbf{x}_s$, and describes a vector of length $m$ with its $k^{\textnormal{th}}$ coordinate equal to 
\begin{equation}\nonumber
{\mathbf{y}_{\!_\mathcal{X}}}(k)=r\ \ \ \ \ \textnormal{if}\ \ \ \eta_r\leq\sum_{j=1}^s \mathbf{x}_{j}(k)< \eta_{r+1},\  \  \  \  0 \leq r < Q,
\end{equation}
where $\mathbf{x}_{j}(k)$ is the $k^{\textnormal{th}}$ coordinate of $\mathbf{x}_{j}$, and ``$+$'' stands for real-valued addition. We refer to $\mathbf{y}_{\!_\mathcal{X}} \in [Q]^m$ as the \emph{syndrome} of $\mathcal{X}$, and the underlying $\oasterisk$ operation as the SQ-sum.
\end{defin}
Using this definition, the vector of test results for a SQGT model takes the form
\begin{equation}\nonumber
\mathbf{y}=\bigoasterisk_{j=1}^{d}\mathbf{x}_{i_j},
\end{equation}
where $\mathbf{x}_{i_j}$ is the codeword of the $j^{\textnormal{th}}$ defective. This equation implies that the result of the $k^{\textnormal{th}}$ test depends on the sum of the $k^{\textnormal{th}}$ coordinate of the defectives' codewords, $\sum_{j=1}^d \mathbf{x}_{i_j}(k)$, as shown in Fig.~\ref{fig:test-result}. Fig.~\ref{fig:model} provides an example of a SQGT code, an incidence vector of the defectives, and vector of test results, with $d=3$, $m=5$, $n=10$, $q=3$, $Q=4$, and $\boldsymbol{\eta}=[0,2,3,5,7]^T$.

\begin{figure}
        \centering
        \begin{subfigure}[b]{0.4\textwidth}
                \centering
                \includegraphics[width=\textwidth]{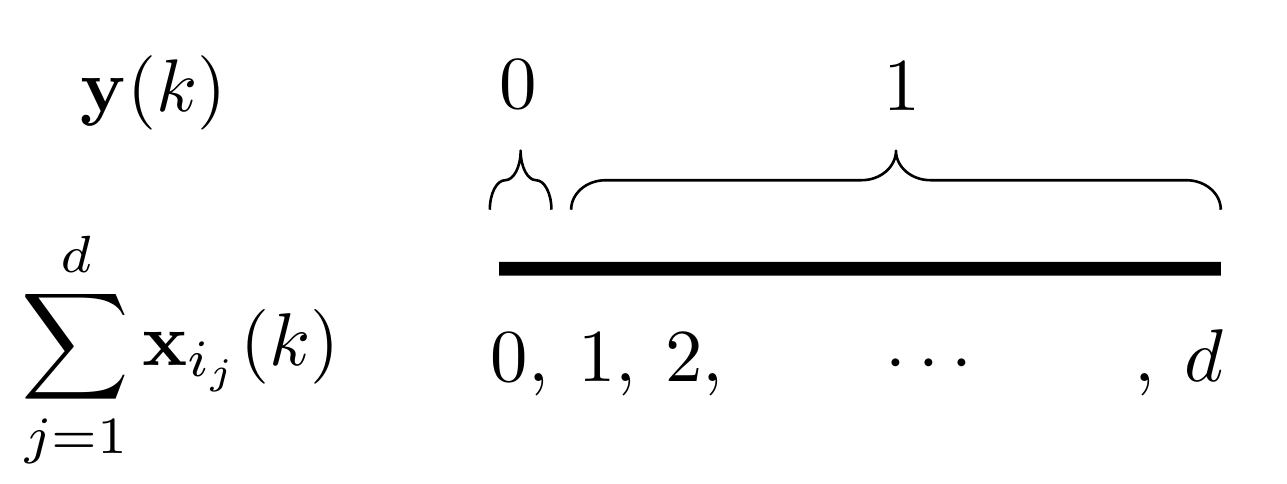}
                \caption{CGT}
                \label{fig:CGT}
        \end{subfigure}%
        \qquad\qquad
        %add desired spacing between images, e. g. ~, \quad, \qquad etc. 
        \begin{subfigure}[b]{0.4\textwidth}
                \centering
                \includegraphics[width=\textwidth]{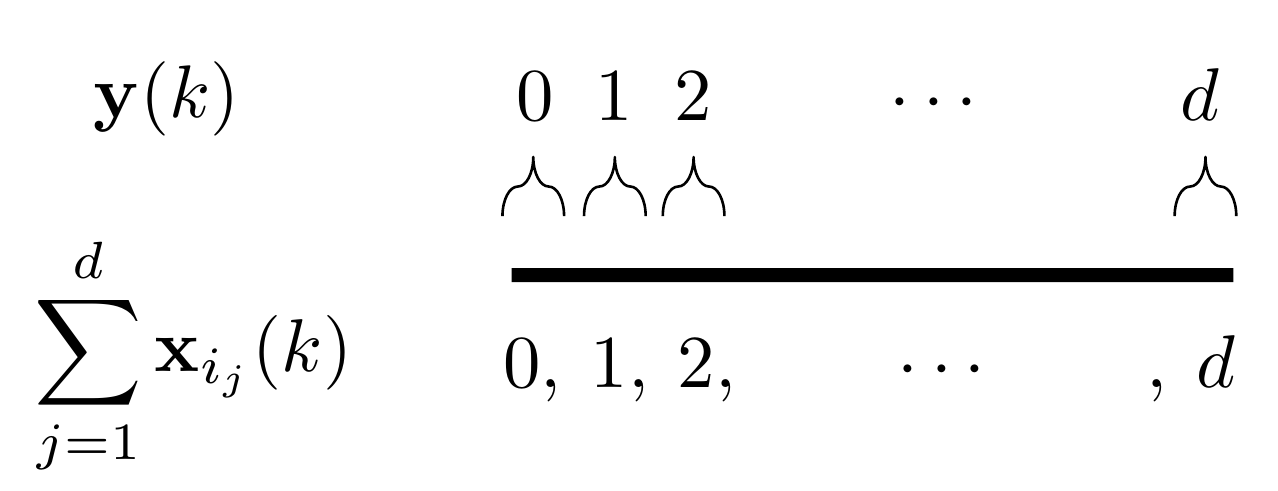}
                \caption{QGT}
                \label{fig:QGT}
        \end{subfigure}
        ~ %add desired spacing between images, e. g. ~, \quad, \qquad etc. 
          \vspace{20pt}
        
        \begin{subfigure}[b]{0.4\textwidth}
                \centering
                \includegraphics[width=\textwidth]{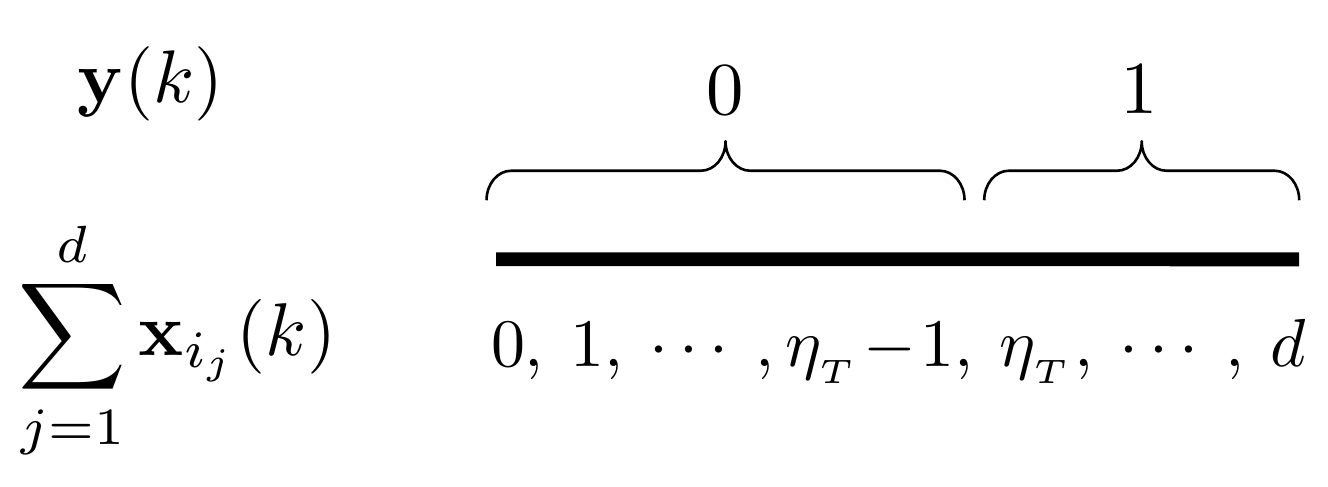}
                \caption{TGT with zero gap}
                \label{fig:TGT}
        \end{subfigure}
                \qquad\qquad%add desired spacing between images, e. g. ~, \quad, \qquad etc. 
          %(or a blank line to force the subfigure onto a new line
        \begin{subfigure}[b]{0.48\textwidth}
                \centering
                \includegraphics[width=\textwidth]{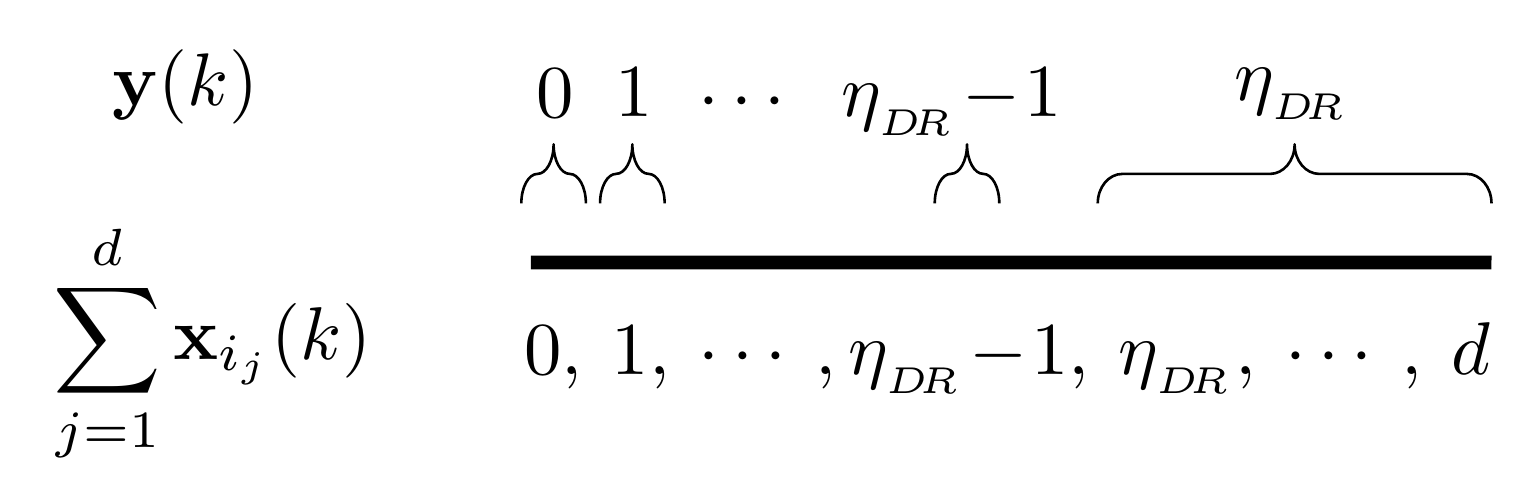}
                \caption{The model in~\cite{DR84}}
                \label{fig:DR84}
        \end{subfigure}
        \caption{Different group testing models for the case $q=2$. In the figures, $\eta_{_{T}}$ denotes the threshold in TGT and $\eta_{_{DR}}$ denotes the threshold in the model described in~\cite{DR84}.}
        \label{fig:oth_models}
\end{figure}

Based on the definition, it is clear that SQGT may be viewed as a concatenation of an adder channel and a decimator (quantizer). 
Also, if $q=Q=2$ and $\eta_1=1$, the SQGT model reduces to CGT. Furthermore, if $Q-1=d(q-1)$ and $\forall r\in[Q]$, $\eta_r=r$, then SQGT reduces to the adder model (QGT), with a possibly non-binary test matrix. Similarly, TGT with zero gap and the model in~\cite{DR84} also represent special instances of SQGT. Fig.~\ref{fig:oth_models} describes all these models for $q=2$.

 Note that in the SQGT model, we assume that $\eta_Q>(q-1)d$. Of special interest is a SQGT model with a uniform quantizer - i.e. SQGT with equidistant thresholds. In this case, $\eta_r=r\eta$, where $r\in[Q+1]$, 
and the SQ-sum of $s$ codewords, $\mathbf{y} _{\!_\mathcal{X}}=\bigoasterisk_{j=1}^{s}\mathbf{x}_j$, simplifies to $\mathbf{y} _{\!_\mathcal{X}}(k)=\left\lfloor\frac{\mathbf{x}_{1}(k)+\mathbf{x}_{2}(k)+\dots+\mathbf{x}_{s}(k)}{\eta}\right\rfloor$, where $\lfloor \cdot \rfloor$ denotes the floor function. We discuss code constructions for the uniform model in more detail in the next sections. 

%%%%%%%%%%%%%%%%%%%%%%%%%%%%%%%%%%%%%%%%%%%%%%%%%%%%%%%%%%%%%%%%%%%%%%%%%%%%%%%%%%%%%%%%
\section{Generalized Disjunct and Separable Codes for SQGT}\label{sec:disjunct}
In what follows, we introduce two new families of codes suitable for SQGT, termed \emph{SQ-disjunct} and \emph{SQ-separable}. These codes are generalizations of binary disjunct and binary separable codes introduced in~\cite{KS64} for efficient zero-error identification of defectives in CGT. SQ-disjunct codes, similar to their CGT counterparts, benefit from a simple decoding algorithm with complexity of $O(mn)$. For both of these codes, we use a set of parameters as explained below. 

A $[q;Q;\boldsymbol{\eta};(l\!:\!u);e]$-SQ-disjunct/separable code is a $q$-ary code for a SQGT model with thresholds $\boldsymbol{\eta}=[0,\eta_1,\eta_2,\dots,\eta_Q]^T$. Such a code is capable of uniquely identifying any number of defectives between $l$ and $u$, $l\leq d\leq u$, from a $Q$-ary vector of test results containing up to $e$ erroneous test results. For simplicity, when the code can only identify exactly $d$ defectives (i.e. $l=u=d$), we use $d$ instead of $(l:u)$. Also, in the case of equidistant SQGT, we use $\eta$ instead of $\boldsymbol{\eta}$.

\subsection{SQ-disjunct codes}
In what follows, we define a new family of disjunct codes for SQGT that shares many of the properties of binary disjunct codes. We start by providing the following definitions.

\begin{defin}
A set of codewords $\mathcal{X}=\{\mathbf{x}_j\}_{1}^s$ with syndrome $\mathbf{y}_{\!_\mathcal{X}}$ is said to be \emph{included} in another set of codewords $\mathcal{Z}=\{\mathbf{z}_j\}_{1}^t$ with syndrome $\mathbf{y}_{\!_\mathcal{Z}}$, if $\forall i\in\llbracket m\rrbracket $, ${\mathbf{y}_{\!_\mathcal{X}}}(i)\leq {\mathbf{y}_{\!_\mathcal{Z}}}(i)$. We denote this inclusion property by $\mathcal{X}\lhd\mathcal{Z}$, or equivalently, $\mathbf{y}_{\!_\mathcal{X}}\lhd\mathbf{y}_{\!_\mathcal{Z}}$.  
\end{defin}
\begin{remark}
Using this definition, it can be easily verified that if $\mathcal{X}\subseteq\mathcal{Z}$, then $\mathcal{X}\lhd \mathcal{Z}$.
\end{remark}

Note that for $q=Q=2$ and $\eta_1=1$, Definition 2 is equivalent to the definition of inclusion for disjunct codes in CGT~\cite{KS64}. 
Based on the notion of inclusion, we may define SQ-disjunct codes for the error-free scenario, $e=0$.

\begin{defin}\label{def:SQ-disj}
A code is called a $[q;Q;\boldsymbol{\eta};(1\!:\!d);0]$-SQ-disjunct code of length $m$ and size $n$ if $\forall s,t\leq d$  and for any sets of $q$-ary codewords $\mathcal{X}=\{\mathbf{x}_j\}_1^s$ and $\mathcal{Z}=\{\mathbf{z}_j\}_1^t$, $\mathcal{X}\lhd \mathcal{Z}$ implies $\mathcal{X}\subseteq \mathcal{Z}$. 
\end{defin}

The next two theorems describe some properties of SQ-disjunct codes.

\begin{theorem}
A $[q;Q;\boldsymbol{\eta};(1\!:\!d);0]$-SQ-disjunct code is capable of identifying any number of defectives less than or equal to $d$ in the absence of test errors. In other words, given  an error-free vector of test results $\mathbf{y}\in[Q]^m$, any codeword with a syndrome included in $\mathbf{y}$ corresponds to a defective, and any codeword with a syndrome not included in $\mathbf{y}$ corresponds to a non-defective. 
\end{theorem}
\begin{IEEEproof}
Let $\mathbf{x}_i$, $i\in\llbracket n\rrbracket$, be a codeword of a $[q;Q;\boldsymbol{\eta};(1\!:\!d);0]$-SQ-disjunct code. Since $\mathbf{y}=\bigoasterisk_{j=1}^{|\mathcal{D}|}\mathbf{x}_{i_j}$, for $i_j\in\mathcal{D}$, if $i$ corresponds to a defective, i.e. $i\in\mathcal{D}$, we have $\mathbf{y}_{\!_{\{\mathbf{x}_i\}}}\lhd\mathbf{y}$. Conversely, by Definition~\ref{def:SQ-disj}, it can be easily verified that if $i\notin\mathcal{D}$ and $|\mathcal{D}|\leq d$, then $\mathbf{y}_{\!_{\{\mathbf{x}_i\}}}\ntriangleleft\mathbf{y}$.
\end{IEEEproof}

We also prove the following result used in subsequent derivations.

\begin{theorem}\label{SQDisj}
 A code is $[q;Q;\boldsymbol{\eta};(1\!:\!d);0]$-SQ-disjunct if and only if no codeword is included in a set of $d$ other codewords. 
 \end{theorem}
\begin{IEEEproof}
It is easy to verify that if a code is $[q;Q;\boldsymbol{\eta};(1\!:\!d);0]$-SQ-disjunct, then no codeword is included in the set of $d$ other codewords. 

Conversely, let $\mathcal{X}=\{\mathbf{x}_j\}_1^s$ and $\mathcal{Z}=\{\mathbf{z}_j\}_1^t$ be two sets of codewords where $s,t\leq d$. From the assumption that no codeword is included in a set of $d$ other codewords, one can conclude that no codeword is included in a set of $t$ other codewords whenever $t\leq d$. If $\mathcal{X}\lhd\mathcal{Z}$ but $\mathcal{X}\nsubseteq\mathcal{Z}$, then there exists a codeword $\mathbf{x}_j\in\mathcal{X}$, $j\in\llbracket s\rrbracket $, such that $\{\mathbf{x}_j\}\nsubseteq\mathcal{Z}$. But since $\{\mathbf{x}_j\}\lhd \mathcal{X}\lhd\mathcal{Z}$, then $\{\mathbf{x}_j\}\lhd\mathcal{Z}$, which contradicts the assumption that no codeword is included in $t$ other codewords.
\end{IEEEproof}
 
\begin{remark}\label{remark:1}
From Theorem~\ref{SQDisj}, one can conclude that a code is $[q;Q;\eta;(1\!:\!d);0]$-SQ-disjunct if and only if for any set of $d+1$ codewords, $\mathcal{X}=\{\mathbf{x}_j\}_1^{d+1}$, and for any codeword $\mathbf{x}_i\in\mathcal{X}$, there exists at least one \emph{unique coordinate} ${k_i}$ for which 
\begin{equation}\label{condition}
\mathbf{y}_{\!_{\{\mathbf{x}_i\}}}(k_i)>\mathbf{y}_{\!_{\mathcal{X}\backslash\{\mathbf{x}_i\}}}(k_i),
\end{equation}
where $\mathbf{y}_{\!_{\{\mathbf{x}_i\}}}$ is the syndrome of ${\{\mathbf{x}_i\}}$, and $\mathbf{y}_{\!_{\mathcal{X}\backslash\{\mathbf{x}_i\}}}$ is the syndrome of the other $d$ codewords in $\mathcal{X}$.
%By unique coordinate, we mean that for any $\mathbf{x}_i,\mathbf{x}_j\in\mathcal{X}$ such that $i\neq j$, one has $k_i\neq k_j$. 
Note that for equidistant SQGT,~\eqref{condition} implies
\begin{align}\nonumber
\left\lfloor\frac{\mathbf{x}_{i}(k_i)}{\eta}\right\rfloor>\left\lfloor\frac{\sum_{j=1,j\neq i}^{d+1}\mathbf{x}_{j}(k_i)}{\eta}\right\rfloor.
\end{align}
\end{remark}

The uniqueness property in Remark~\ref{remark:1} can be proved as follows. Fix a set $\mathcal{X}$ and $\mathbf{x}_i,\mathbf{x}_j\in\mathcal{X}$ such that $i\neq j$ and $k_i=k_j$. Using Definition~\ref{def1}, it can be easily verified that for any coordinate $k$, 
\begin{equation}\label{eq:uniqueness}
\mathbf{y}_{\!_{\mathcal{X}\backslash\{\mathbf{x}_i\}}}(k)=\mathbf{y}_{\!_{(\mathcal{X}\backslash\{\mathbf{x}_i,\mathbf{x}_j\})\cup\{\mathbf{x}_j\}}}(k)\geq\mathbf{y}_{\!_{\{\mathbf{x}_j\}}}(k).
\end{equation}
Using~\eqref{condition} and~\eqref{eq:uniqueness}, one has
\begin{equation}\label{eq:cont1}
\mathbf{y}_{\!_{\{\mathbf{x}_i\}}}(k_i)>\mathbf{y}_{\!_{\mathcal{X}\backslash\{\mathbf{x}_i\}}}(k_i)\geq \mathbf{y}_{\!_{\{\mathbf{x}_j\}}}(k_i).
\end{equation}
Applying condition~\eqref{condition} to $\mathbf{x}_j$ and using~\eqref{eq:uniqueness}, one similarly obtains
\begin{equation}\label{eq:cont2}
\mathbf{y}_{\!_{\{\mathbf{x}_j\}}}(k_j)>\mathbf{y}_{\!_{\mathcal{X}\backslash\{\mathbf{x}_j\}}}(k_j)\geq \mathbf{y}_{\!_{\{\mathbf{x}_i\}}}(k_j).
\end{equation}
Since $k_i=k_j$, \eqref{eq:cont1} and \eqref{eq:cont2} contradict each other, which completes the proof.

Using the notion of unique coordinate, we can generalize Definition~\eqref{def:SQ-disj} to SQ-disjunct codes that are capable of correcting up to $e>0$ errors.
\begin{defin}[\textbf{SQ-disjunct codes}]\label{def:SQ-disj2}
A code is called a $[q;Q;\boldsymbol{\eta};(1\!:\!d);e]$-SQ-disjunct code of length $m$ and size $n$ if for any set of $d+1$ codewords, $\mathcal{X}=\{\mathbf{x}_j\}_1^{d+1}$, and for any codeword $\mathbf{x}_i\in\mathcal{X}$, there exists a set of coordinates, $\mathcal{R}_i$, of size at least $2e+1$ such that $\forall k_i\in\mathcal{R}_i$,
\begin{equation}\label{condition3} 
\mathbf{y}_{\!_{\{\mathbf{x}_i\}}}(k_i)>\mathbf{y}_{\!_{\mathcal{X}\backslash\{\mathbf{x}_i\}}}(k_i),
\end{equation}
and $\mathcal{R}_i$ is disjoint of any $\mathcal{R}_l$ for which $\mathbf{x}_l\in\mathcal{X}$ and $l\neq i$; in this equation $\mathbf{y}_{\!_{\{\mathbf{x}_i\}}}$ is the syndrome of ${\{\mathbf{x}_i\}}$, and $\mathbf{y}_{\!_{\mathcal{X}\backslash\{\mathbf{x}_i\}}}$ is the syndrome of the remaining $d$ codewords in $\mathcal{X}$.
\end{defin}
Such a code is capable of uniquely identifying up to $d$ defectives, in the presence of up to $e$ errors in the vector of test results. If a codeword $\mathbf{x}_i$ does not correspond to a defective, its syndrome contains \emph{at least} $e+1$ coordinates satisfying $\mathbf{y}_{\!_{\{\mathbf{x}_i\}}}(k)>\mathbf{y}(k)$. On the other hand, if $\mathbf{x}_i$ corresponds to a defective, its syndrome contains \emph{at most} $e$ coordinates satisfying $\mathbf{y}_{\!_{\{\mathbf{x}_i\}}}(k)>\mathbf{y}(k)$.

\begin{remark}
It can be easily seen from~\eqref{condition} and~\eqref{condition3} that a necessary condition for the existence of a $[q;Q;\boldsymbol{\eta};(1\!:\!d);e]$-SQ-disjunct code is that $q-1\geq\eta_1$. 
As a result, there exist no \emph{binary} $[2;Q;\boldsymbol{\eta};(1\!:\!d);e]$-SQ-disjunct codes when $\eta_1>1$.
\end{remark}

\begin{remark}[\textbf{Decoding Algorithm:}]\label{SQ-disjunct_decoding}
Definition~\ref{def:SQ-disj2} suggests an efficient decoding algorithm for SQ-disjunct codes with complexity $O(mn)$, which resembles the decoding algorithm for binary disjunct codes for CGT. The decoding algorithm for a $[q;Q;\boldsymbol{\eta};(1\!:\!d);e]$-SQ-disjunct code of length $m$ and size $n$ works as follows. For each codeword $\mathbf{x}_i$, $i\in\llbracket n\rrbracket$, count the number of coordinates of $\mathbf{y}_{\!_{\{\mathbf{x}_i\}}}$ for which $\mathbf{y}_{\!_{\{\mathbf{x}_i\}}}(k)>\mathbf{y}(k)$. If the number of such coordinates is at least $e+1$, $\mathbf{x}_i$ does not correspond to a defective. On the other hand, if the number of such coordinates is at most $e$, the codeword corresponds to a defective.
\end{remark}
%%%%%%%%%%%%%%%%%%%%%%%%%%%%%%%%%%%%%%

%%%%%%%%%%%%%%%%%%%%%%%%%%%%%%%%%%%%%%%%%%%%%%%%%%%%%%%%

\subsection{SQ-separable Codes}
Although SQ-disjunct codes can be used to find defectives in a SQGT design via a simple decoding procedure, the requirements imposed on such codes may appear too restrictive for certain applications. As a result, relaxing these structural constraints may lead to a reduction in the number of tests for fixed values of $n$. Since SQ-disjunct codes cannot be used for the case when $q\leq\eta_1$, one may be interested in designing codes with smaller alphabet size. SQ-separable codes are a family of $q$-ary codes that are capable of overcoming these issues.

\begin{defin}[\textbf{SQ-separable codes}]\label{SQsep}
A code is called a $[q;Q;\boldsymbol{\eta};(l\!:\!u);e]$-SQ-separable code of length $m$ and size $n$ if for any two distinct sets of codewords $\mathcal{X}$ and $\mathcal{Z}$ that satisfy $l\leq|\mathcal{X}|,|\mathcal{Z}|\leq u$, there exists a set of coordinates $\mathcal{R}$, with size $|\mathcal{R}|\geq 2e+1$, such that $\forall k\in\mathcal{R}$
\begin{align}\nonumber
\mathbf{y}_{\!_{\mathcal{X}}}(k)\neq \mathbf{y}_{\!_{\mathcal{Z}}}(k).
\end{align}
\end{defin}
Such codes are capable of identifying defectives when the vector of test results contains at most $e$ errors, given that the number of defectives is at least $l$ and at most $u$. Note that as next proposition demonstrates, SQ-disjunct codes are special cases of SQ-separable codes.
\begin{prop} 
Any $[q;Q;\boldsymbol{\eta};(1\!:\!d);e]$-SQ-disjunct code is a $[q;Q;\boldsymbol{\eta};(1\!:\!d);e]$-SQ-separable code.
\end{prop}
\begin{IEEEproof}
Consider any $[q;Q;\boldsymbol{\eta};(1\!:\!d);e]$-SQ-disjunct code, and any two distinct sets of codewords $\mathcal{X}$ and $\mathcal{Z}$ that satisfy $1\leq|\mathcal{X}|,|\mathcal{Z}|\leq d$. Without loss of generality, assume that $|\mathcal{X}|\leq|\mathcal{Z}|$. Since these two sets are distinct, $\mathcal{Z}\backslash\mathcal{X}\neq\varnothing$; let $\mathbf{z}$ be a codeword such that $\mathbf{z}\in\mathcal{Z}\backslash\mathcal{X}$. Since $|\mathcal{X}\cup\{\mathbf{z}\}|\leq d+1$, using the definition of SQ-disjunct codes, one can conclude that there exists a set of coordinates, $\mathcal{R}$, of size at least $2e+1$, such that $\forall k\in\mathcal{R}$,
\begin{equation}\nonumber
\mathbf{y}_{\!_{\{\mathbf{z}\}}}(k)>\mathbf{y}_{\!_{\mathcal{X}}}(k).
\end{equation}
On the other hand since $\mathbf{z}\in\mathcal{Z}$, Definition~\ref{def1} implies that $\forall k\in\mathcal{R}$, $\mathbf{y}_{\!_{\mathcal{Z}}}(k)\geq\mathbf{y}_{\!_{\{\mathbf{z}\}}}(k)>\mathbf{y}_{\!_{\mathcal{X}}}(k)$, which completes the proof.
\end{IEEEproof}

\begin{remark}
From Definition~\ref{SQsep}, one can see that a necessary condition for the existence of a $[q;Q;\boldsymbol{\eta};(l\!:\!u);e]$-SQ-separable code is that $l(q-1)\geq \eta_1$. If $l=1$, this condition simplifies to $q-1\geq \eta_1$, 
which is the same as the necessary condition for the existence of a $[q;Q;\boldsymbol{\eta};(1:d);e]$-SQ-disjunct code. This is expected, since any SQ-disjunct code is also a SQ-separable code, while the converse is not true. On the other hand, if $q=2$, the condition simplifies to $l\geq \eta_1$. This implies that if the number of defectives is smaller than $\eta_1$, one cannot identify the defectives using a \emph{binary} code.  
\end{remark}

%%%%%%%%%%%%%%%%%%%%%%%%%%%%%%%%%%%%%%%%%%%%%%%%%%
\section{Code Construction for SQGT}\label{sec:construction}

Next, we discuss both probabilistic and explicit combinatorial constructions of SQ-disjunct and SQ-separable codes. For each of these code families, we first describe constructions with \emph{arbitrary} thresholds, $\boldsymbol{\eta}$. While such constructions are applicable to any set of thresholds, one may be able to construct codes with smaller test numbers designed specifically for a certain choice of thresholds. For example, QGT is a special case of SQGT; while there are many interesting code constructions for QGT, these constructions do not apply to CGT, another special case of SQGT. Therefore, after introducing some general constructions, we focus on one of the most important special cases of SQGT: SQGT with equidistant thresholds.

The section is organized as follows. In Subsections~\ref{subsec:q_disj} and~\ref{subsec:q_sep}, we describe constructions of $q$-ary SQ-disjunct and $q$-ary SQ-separable codes, respectively. The construction of \emph{binary} SQ-separable codes are described in~\ref{subsec:2_sep}. In~\ref{subsec:arb_d}, construction of SQ-separable codes for arbitrary number of defectives are described. Finally, the parameters of the codes constructed in this section are summarized and compared to each other in~\ref{sec:comparison}.

In some of the constructions described in this section, we take advantage of the properties of binary disjunct and separable codes designed for CGT and QGT. These codes are defined in what follows.

\begin{defin}[\textbf{Binary $\boldsymbol{d}$-disjunct codes for CGT}]\label{CGTdisjunct}
A binary $d$-disjunct code designed for CGT, capable of correcting up to $e$ errors, is a code of length $m$ and size $n$ such that for any set of $d+1$ codewords, $\mathcal{X}=\{\mathbf{x}_j\}_1^{d+1}$, and for any codeword $\mathbf{x}_i\in\mathcal{X}$, there exists a set of coordinates $\mathcal{R}_i$ of size at least $2e+1$, such that $\forall k\in\mathcal{R}_i$, $\mathbf{x}_i(k)=1$ and $\mathbf{x}_j(k)=0$, for $\mathbf{x}_j\in\mathcal{X}$ and $j\neq i$.
\end{defin}

\begin{defin}[\textbf{Binary $\boldsymbol{d}$-separable codes for CGT}]\label{CGTsep}
A binary $d$-separable code designed for CGT, capable of correcting up to $e$ errors, is a code of length $m$ and size $n$ such that for any two distinct sets of codewords $\mathcal{X}$ and $\mathcal{Z}$, $1\leq|\mathcal{X}|,|\mathcal{Z}|\leq d$, the \emph{Boolean sum} of the codewords in $\mathcal{X}$ differs from the \emph{Boolean sum} of the codewords in $\mathcal{Z}$ in at least $2e+1$ coordinates.
\end{defin}

\begin{defin}[\textbf{Binary $\boldsymbol{d}$-separable codes for QGT}]\label{QGTsep}
A binary $d$-separable code designed for QGT, capable of correcting up to $e$ errors, is a code of length $m$ and size $n$ such that for any two distinct sets of codewords $\mathcal{X}$ and $\mathcal{Z}$, $1\leq|\mathcal{X}|,|\mathcal{Z}|\leq d$, the \emph{arithmetic sum} of the codewords in $\mathcal{X}$ differs from the \emph{arithmetic sum} of the codewords in $\mathcal{Z}$ in at least $2e+1$ coordinates.
\end{defin}

%%%%%%%%%%%%%%%%%%%%%%%%%
\subsection{Construction of $q$-ary SQ-disjunct codes}\label{subsec:q_disj}
SQ-disjunct codes represent generalizations of conventional binary disjunct codes. As a result, it is expected that one can construct SQ-disjunct codes using conventional disjunct codes. The following proposition describes 
one such construction. 

\begin{prop}[\textbf{Construction 1}]\label{prop:disj}
Any code generated by multiplying a conventional binary $d$-disjunct code capable of correcting $e$ errors\footnote{For constructions of binary $d$-disjunct codes with error correcting capabilities, see~\cite{DH00},~\cite{M97},~\cite{DHMVW05} and references therein.} by $q-1$, where $q-1\geq\eta_1$, is a $[q;Q;\boldsymbol{\eta};(1\!:\!d);e]$-SQ-disjunct code. 
\end{prop}
\begin{IEEEproof}
A conventional binary $d$-disjunct code, capable of correcting $e$ errors, satisfies the condition that for any set of $d+1$ codewords, $\mathcal{Z}=\{\mathbf{z}_j\}_1^{d+1}$, and for any codeword $\mathbf{z}_i\in\mathcal{Z}$, there exists a set of coordinates $\mathcal{R}_i$ of size at least $2e+1$, such that $\forall k\in\mathcal{R}_i$,
\begin{align}\nonumber
&\mathbf{z}_i(k)=1,\\\nonumber
&\mathbf{z}_j(k)=0,\  \  \  \  \text{for $\mathbf{z}_j\in\mathcal{Z}$ and $j\neq i$}.
\end{align}
Multiplying such a code with $q-1$, where $q-1\geq\eta_1$, produces a $q$-ary code such that for any set of $d+1$ codewords, $\mathcal{X}=\{\mathbf{x}_j\}_1^{d+1}$, and for any codeword $\mathbf{x}_i\in\mathcal{X}$, there exists a unique set of coordinates, $\mathcal{R}_i$, of size at least $2e+1$, such that $\forall k\in\mathcal{R}_i$,
\begin{align}\nonumber
&\mathbf{y}_{\!_{\{\mathbf{x}_i\}}}(k)>0,\\\nonumber
&\mathbf{x}_j(k)=0,\  \  \  \  \text{for $\mathbf{x}_j\in\mathcal{X}$ and $j\neq i$}.
\end{align}
As a result, $\forall k\in\mathcal{R}_i$, 
\begin{equation}\nonumber
\mathbf{y}_{\!_{\{\mathbf{x}_i\}}}(k)>\mathbf{y}_{\!_{\mathcal{X}\backslash\{\mathbf{x}_i\}}}(k)=0.
\end{equation}
\end{IEEEproof}

Next, we focus on SQGT with equidistant thresholds, i.e., codes for which $\eta_r=r\eta$, where $ r\in[Q+1]$. The following lemma will be used for constructing SQ-disjunct codes with equidistant thresholds. 
\begin{lemma}\label{alphabet}
Given a $[q;Q;\eta;(1\!:\!d);e]$-SQ-disjunct code $\mathbf{C}\in [q]^{m\times n}$ exists, one can construct a $[q;Q;\eta;(1\!:\!d);e]$-SQ-disjunct code $\mathbf{C}'$ that effectively uses only an $(I+1)-$ary alphabet, $\{0,\eta,2\eta,\dots,I\eta\}$, where $I=\lfloor\frac{q-1}{\eta}\rfloor$. 
\end{lemma}
\begin{IEEEproof}
Form $\mathbf{C}'$ by the following substitution: $\forall i\in \llbracket m\rrbracket$ and $\forall j\in \llbracket n\rrbracket$, let $\mathbf{C}'({i,j})=\lfloor\frac{\mathbf{C}(i,j)}{\eta}\rfloor\eta \in \{0,{\eta,2\eta,\ldots,I\eta\}}$. Consider a set of $d+1$ column-indices $\mathcal{S}$ and fix a column-index $l\in\mathcal{S}$. If $\mathbf{C}(i,l)$, $i\in\llbracket m\rrbracket$, is a unique coordinate of the $l^\text{th}$ column of $\mathbf{C}$ for which~\eqref{condition3} is satisfied for the given set $\mathcal{S}$, the same condition will still be satisfied in $\mathbf{C}'$ for $l$ and $\mathcal{S}$. The reason is that after the substitution, the $i^{\text{th}}$ coordinate of the syndrome of the $l^{\text{th}}$ column remains unchanged, while the $i^{\text{th}}$ coordinate of the syndrome of the other $d$ codewords indexed by $\mathcal{S}\backslash \{l\}$ will have a smaller value. Since this is true for any $\mathcal{S}\subseteq\llbracket n\rrbracket$ with $|\mathcal{S}|=d+1$ and for any $l\in\mathcal{S}$, $\mathbf{C}'$ is a $[q;Q;\eta;(1\!:\!d);e]$-SQ-disjunct code. On the other hand, if for $i\in\llbracket m\rrbracket$, none of the columns of $\mathbf{C}$ indexed by $\mathcal{S}$ has a unique coordinate in the $i^{\text{th}}$ row, then this substitution may generate a unique coordinate in a column and therefore improve the error correcting capability of the code.
\end{IEEEproof}

\begin{remark}
Lemma~\ref{alphabet} implies that given an available alphabet $[q]$, in order to design a $[q;Q;\eta;(1\!:\!d);e]$-SQ-disjunct code with minimum length $m$ for a fixed size $n$, one only needs to use a $(I+1)$-ary alphabet, $\{0,\eta,2\eta,\dots,I\eta\}$, where $I=\lfloor\frac{q-1}{\eta}\rfloor$.
\end{remark}

We use this lemma and remark to describe a probabilistic construction for SQ-disjunct codes with equidistant thresholds. 

\begin{theorem}[\textbf{Construction 2}]
Form a matrix $\mathbf{C}\in\{0,\eta,2\eta,\dots,I\eta\}^{m\times n}$ by choosing each entry independently according to the following probability distribution,
\begin{align}\nonumber
P_X(x)= 
\left\{
     \begin{array}{ll}
         P_0,   & \textnormal{if}\  \  \  x=0\\
              P_1,   & \textnormal{if} \  \  \  x\in\{\eta,2\eta,\dots,I\eta\}
     \end{array},
   \right.
\end{align}
where $I=\lfloor\frac{q-1}{\eta}\rfloor$, $P_0=\frac{d}{d+1}$, and $P_1=\frac{1}{I(d+1)}$. Then $\mathbf{C}$ is a $[q;Q;{\eta};(1\!:\!d);e]$-SQ-disjunct code of length $m_I$ and size $n_I$ with probability at least $1-o(1)$; asymptotically, $m_I$ equals
\begin{align}\nonumber
m_I\sim \frac{m_1}{\left(1+\frac{1}{I^{d+1}d^{d}}\sum_{k=0}^{d-1}{d\choose k}{I\choose d-k+1}(Id)^k\right)},
\end{align}
where $m_1$ is the length of a $[q;Q;\eta;(1\!:\!d);e]$-SQ-disjunct code of size $n_1=n_I$, obtained by multiplying the best probabilistically constructed\footnote{By ``best'', we mean a code designed probabilistically in a way to have the minimum $m$ for a fixed $n$.} binary $d$-disjunct code, capable of correcting up to $e$ errors, by $\eta$. 
\end{theorem}

\begin{IEEEproof}
Fix a choice of $d+1$ column indices, $\mathcal{S}\subseteq\llbracket n\rrbracket$, and among them choose one index, $l\in \mathcal{S}$. There are ${n\choose d+1}(d+1)$ ways to choose $\mathcal{S}$ and $l$. 
Let $\pi_I$ be the probability of ``success'' of a row, i.e., the probability that for a row of $\mathbf{C}$ denoted by $\mathbf{r}$, one has $\lfloor\frac{\mathbf{r}(l)}{\eta}\rfloor>\lfloor\frac{\sum_{i\in\mathcal{S}\,\backslash\{l\}}\mathbf{r}(i)}{\eta}\rfloor$. Due to the fact that the alphabet consists of integer multiples of $\eta$, the aforementioned conditioned is equivalent to
\begin{align}\label{eq:prob}
\mathbf{r}(l)>{\sum_{i\in\mathcal{S}\,\backslash\{l\}}\mathbf{r}(i)}.
\end{align}
Let $\mathcal{E}_\beta$ be the event that~\eqref{eq:prob} is satisfied and that $\mathbf{r}(l)=\beta\eta$. From this definition, and the law of total probability, it follows that
\begin{align}\label{PI}
\pi_I=\Pro\left(\bigcup_{\beta=1}^I\mathcal{E}_\beta\right)=\sum_{\beta=1}^I\Pro(\mathcal{E}_\beta).
\end{align}
On the other hand, one has
\begin{align}\nonumber
\Pro(\mathcal{E}_\beta)=P_1\left(P_0^d+P_1^d\sum_{k=0}^{d-1}{d\choose k}\left(\frac{P_0}{P_1}\right)^k    \left(\sum_{i=d-k}^{\beta-1}{i-1\choose d-k-1}\right)\right),
\end{align}
where ${i-1\choose d-k-1}$ counts the number of compositions of $i$ with $d-k$ parts, or equivalently the number of positive integer solutions to $\sum_{j=1}^{d-k}x_j=i$. 
Since 
\begin{align}\nonumber
\sum_{i=d-k}^{\beta-1}{i-1\choose d-k-1}={\beta-1\choose d-k},
\end{align}
equation~\eqref{PI} simplifies to
\begin{align}\nonumber
\pi_I&=\sum_{\beta=1}^I P_1\left(P_0^d+P_1^d\sum_{k=0}^{d-1}{d\choose k}\left(\frac{P_0}{P_1}\right)^k   {\beta-1\choose d-k}\right)\\\nonumber
&=IP_1P_0^d+P_1^{d+1}\sum_{k=0}^{d-1}{d\choose k}   \left(\frac{P_0}{P_1}\right)^k  \sum_{\beta=2}^{I}{\beta-1\choose d-k}\\\nonumber
&=IP_1P_0^d+P_1^{d+1}\sum_{k=0}^{d-1}{d\choose k}   \left(\frac{P_0}{P_1}\right)^k {I\choose d-k+1}\\\label{PiI}
&=(1-P_0)P_0^d+(1-P_0)^{d+1}I^{-(d+1)}\sum_{k=0}^{d-1}{d\choose k}   \left(\frac{P_0I}{1-P_0}\right)^k {I\choose d-k+1}.
\end{align}

Consequently, using the union bound, we can derive an upper bound on the probability that $\mathbf{C}$ is not a $[q;Q;\eta;(1\!:\!d);0]$-SQ-disjunct code,
\begin{align}\nonumber
P_F&={n\choose d+1}(d+1)(1-\pi_I)^m\leq{n\choose d+1}(d+1)\exp(-m\pi_I)\\\nonumber
&\leq\exp\left((d+1)\log n-d\log(d+1)+d+1-m\pi_I\right).
\end{align}
As a result, for any $\delta>0$, one has $P_F=o(1)$ if 
\begin{align}\nonumber
m=\left(\frac{d+1}{\pi_I}+\delta\right)\log \frac{n}{d}.
\end{align}

This result can be generalized for $[q;Q;\eta;(1\!:\!d);e]$-SQ-disjunct codes, where $e$ is allowed to grow with $n$. For a fixed $\mathcal{S}$ and $l$, $\forall j\in\llbracket m\rrbracket$, let $N_j$ be a Bernoulli random variable with value $1$ if the $j^{\text{th}}$ row of $\mathbf{C}$ satisfies~\eqref{eq:prob}, and $0$ otherwise. By definition, the random variables $N_j$ are independent identically distributed (i.i.d.) and $\Pro(N_j=1)=\pi_I$, for $j\in\llbracket m\rrbracket$. Based on the Chernoff bound, for $0<\delta<1$, one obtains
\begin{align}\nonumber
\Pro\left(\sum_{j=1}^mN_j\leq(1-\delta)m\pi_I\right)\leq\exp\left(-\frac{\delta^2m\pi_I}{2}\right).
\end{align}
By setting $\delta=1-\frac{2e}{m\pi_I}$, it follows that
\begin{align}\nonumber
\Pro\left(\sum_{j=1}^mN_j\leq2e\right)\leq\exp\left({-\frac{m\pi_I}{2}{\left(1-\frac{2e}{m\pi_I}\right)^2}}\right),
\end{align}
which provides an upper bound on the probability that for a fixed $\mathcal{S}$ and $l$, at most $2e$ rows of $\mathbf{C}$ satisfy~\eqref{eq:prob}. As a result, the probability that $\mathbf{C}$ is not a $[q;Q;\eta;(1\!:\!d);e]$-SQ-disjunct code is upper bounded by 
\begin{align}\nonumber
P_F&\leq{n\choose d+1}(d+1)\exp\left({-\frac{m\pi_I}{2}{\left(1-\frac{2e}{m\pi_I}\right)^2}}\right)\\\nonumber
&\leq\exp\left((d+1)\log n+d+1-d\log(d+1)-\frac{m\pi_I}{2}-\frac{2e^2}{m\pi_I}+2e\right).
\end{align}
It can be easily seen that for any $\delta>0$, $P_F=o(1)$ if 
\begin{align}\nonumber
m=\left(\frac{2(d+1)}{\pi_I}+\delta\right)\log \frac{n}{d}+\frac{4e}{\pi_I}.
\end{align}

We can compare the number of tests $m_I$ for a code constructed using this method with the number of tests $m_1$ in a code constructed by multiplying a conventional binary $d$-disjunct code with $\eta$ (Construction 1), provided that they have the same number of codewords $n$. It can be easily verified -- see for example~\cite{DH00} -- that for a fixed $n$, the distribution $P_X(x)$ that minimizes the number of tests of a conventional binary $d$-disjunct code is the one that assigns $P_0=\frac{d}{d+1}$ to $x=0$ and $P_1=\frac{1}{d+1}$ to $x=1$. Consequently, $\pi_1=\frac{d^d}{(d+1)^{d+1}}$ maximizes the probability of ``success'' of a row\footnote{Note that even though $\pi_1$ is the optimal probability of success of a row when $q-1<2\eta$, the same statement does not necessarily hold for $\pi_I$ found in this construction.}. Since Construction 1 does not change the size and length of the underlying binary $d$-disjunct code, asymptotically it holds that
\begin{align}\label{rate1}
\frac{m_I}{m_1}\sim\frac{\pi_1}{\pi_I}.
\end{align} 
On the other hand, 
\begin{align}\nonumber
\pi_I=\pi_1+\gamma_I,
\end{align}
where $\gamma_I=\frac{1}{I^{d+1}(d+1)^{d+1}}\sum_{k=0}^{d-1}{d\choose k}{I\choose d-k+1}(Id)^k$. Consequently, 
\begin{align}\nonumber 
\lim_{n\rightarrow\infty}\frac{m_1}{m_I}=1+\frac{1}{I^{d+1}d^{d}}\sum_{k=0}^{d-1}{d\choose k}{I\choose d-k+1}(Id)^k.
\end{align}
\end{IEEEproof}

\begin{figure}
%\vspace{-130pt}
\includegraphics[width=0.9\textwidth]{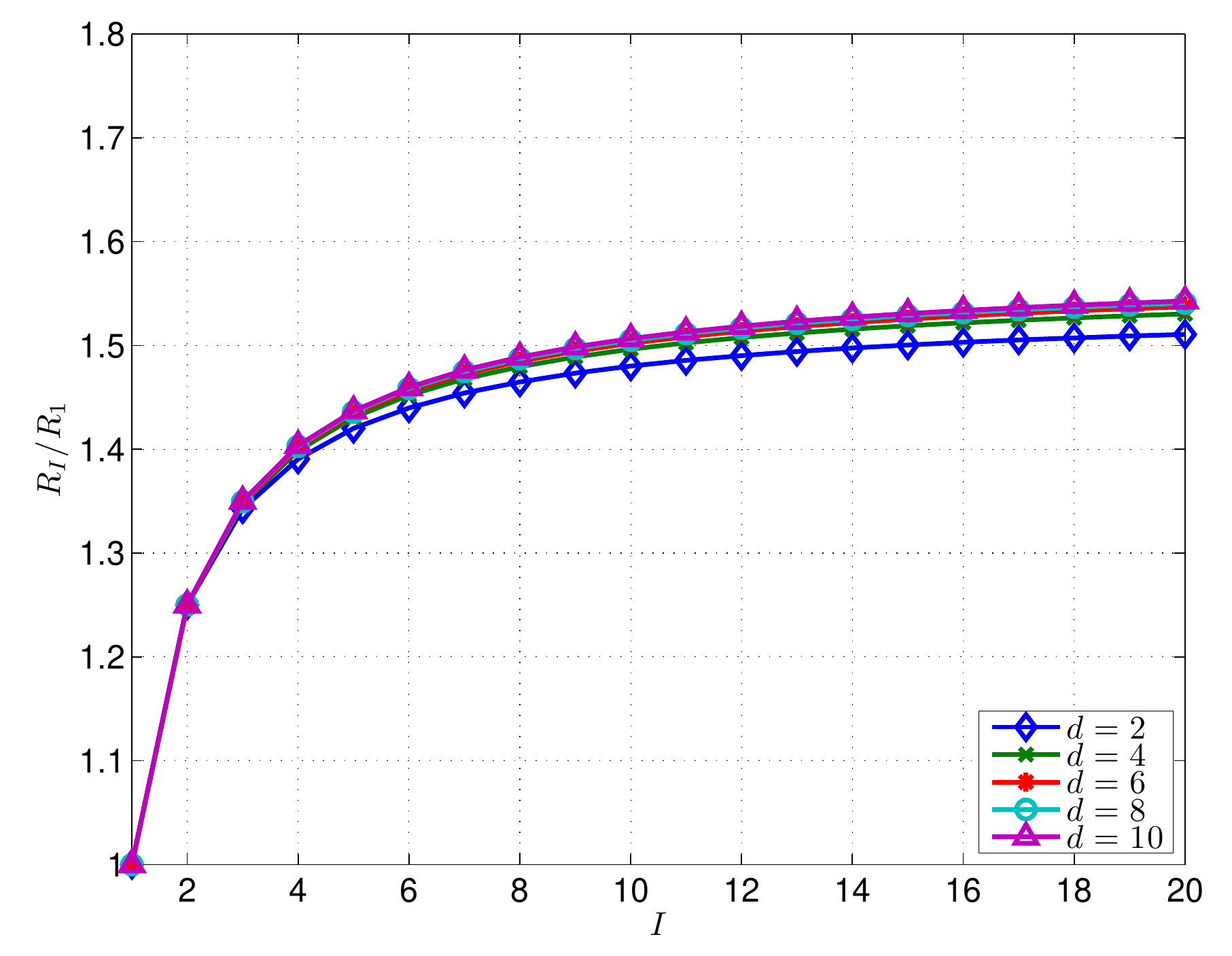}
\centering
%\vspace{-130pt}
\caption{Reduction in the number of tests of a SQ-disjunct code based on Construction 2 for a simple choice of the probability $P_0$.}
\label{Improvement_simple}
\end{figure}

Fig.~\ref{Improvement_simple} shows the asymptotic reduction in the number of tests, $\frac{m_1}{m_I}$, as a function of $I$ for different values of $d$. Note that in this theorem, we assumed that $I$ and $d$ do not grow with $n$. However, we can also consider the case in which $d\rightarrow\infty$ (for a fixed value of $I$) to obtain
\begin{align}\nonumber 
\lim_{d\rightarrow\infty}\lim_{n\rightarrow\infty}\frac{m_1}{m_I}&=\lim_{d\rightarrow\infty}\left(1+\frac{1}{I^{d+1}d^{d}}\sum_{k=0}^{d-1}{d\choose k}{I\choose d-k+1}(Id)^k\right)\\\nonumber
&=\lim_{d\rightarrow\infty}\left(1+\frac{1}{I^{d+1}d^{d}}\sum_{k=d-I+1}^{d-1}{d\choose k}{I\choose d-k+1}(Id)^k\right)\\\nonumber
&=\lim_{d\rightarrow\infty}\left(1+\sum_{k=0}^{I-2}{I\choose k}\frac{1}{I^{I-k}}\  \frac{{d\choose I-k-1}}{d^{I-k-1}}\right)\\\nonumber
&=1+\sum_{k=0}^{I-2}{I\choose k}\frac{1}{I^{I-k}}\  \lim_{d\rightarrow\infty}\frac{{d\choose I-k-1}}{d^{I-k-1}}\\\nonumber
&=1+\sum_{k=0}^{I-2}{I\choose k}\frac{1}{I^{I-k}}\frac{1}{(I-k-1)!},
\end{align}
where we changed the order of the limit and the summation operations, since the sum was over a finite number of terms.

\begin{figure}
%\vspace{-130pt}
\includegraphics[width=0.9\textwidth]{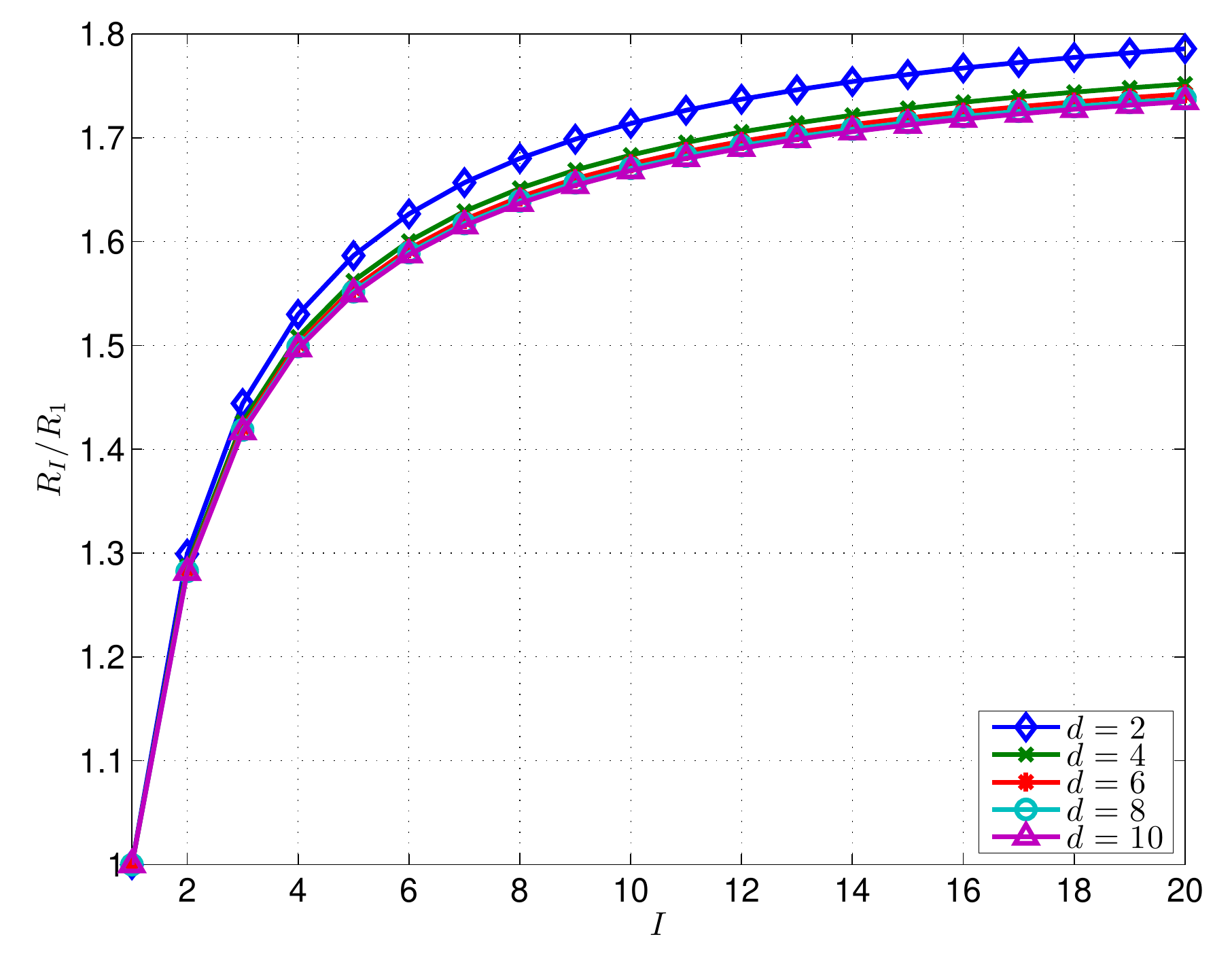}
\centering
%\vspace{-130pt}
\caption{Reduction in the number of tests of a SQ-disjunct code constructed based on Construction 2 for the optimum choice of $P_0$. The parameter $u$, as before, denotes a known upper bound on the number of
defectives.}
\label{Improvement_optim}
\end{figure}

\begin{remark}
It is worth mentioning that instead of setting $P_0=\frac{d}{d+1}$, one can consider $P_0$ to be a parameter that may be optimized so as to minimize the number of tests in the code. 
Making this change does not affect the validity of~\eqref{PiI} and \eqref{rate1}, but it may increase the ratio $\frac{m_1}{m_I}$.  Although finding a simple closed-form expression for the maximum 
$\pi_I$ over $P_0$ does not seem to be straightforward, we evaluated~\eqref{PiI} numerically to find the maximum probability of ``success'' of a row. 
The resulting ratio $\frac{m_1}{m_I}$ is shown in Fig.~\ref{Improvement_optim} as a function of $I$, for different values of $d$.
\end{remark}

As discussed earlier (see Remark~\ref{SQ-disjunct_decoding}), SQ-disjunct codes are endowed with a simple decoding algorithm of complexity $O(mn)$. The next theorem describes an explicit construction for a code that is based on SQ-disjunct codes as building blocks; even though this code is not SQ-disjunct, but only SQ-separable, it iteratively employs a decoder for SQ-disjunct codes and hence maintains a decoding complexity of $O(mn)$. 

%~\footnote{In order to clarify this method of concatenation, as an example, assume that we want to form $\mathbf{C}=[\mathbf{C}_1,\mathbf{C}_2]$ by concatenating $\mathbf{C}_1=\begin{pmatrix}1 & 1\\1 & 1\end{pmatrix}$ and $\mathbf{C}_2=\begin{pmatrix}2 & 2\\2 & 2\end{pmatrix}$. The result will be $\mathbf{C}=\begin{pmatrix}1 & 1 & 2 & 2\\1 & 1 & 2 & 2\end{pmatrix}$.}

\begin{theorem}[\textbf{Construction 3}]\label{const3}
Fix a binary $d$-disjunct code matrix $\mathbf{C}_b$ of dimensions $m_b\times n_b$, capable of correcting up to $e$ errors. Let $K=\left\lfloor\log_d\left(\left(\frac{q-1}{\eta}\right)(d-1)+1\right)\right\rfloor$. Form a matrix $\mathbf{C}$ of length $m=m_b$ and size $n=Kn_b$ by concatenating $K$ matrices horizontally, such that for $j\in\llbracket K\rrbracket$ and $l\in\llbracket m\rrbracket$, the $((j-1)n_b+l)^{\text{th}}$ column of $\mathbf{C}$ is equal to the $l^{\text{th}}$ column of $\mathbf{C}_j$, where $\mathbf{C}_j=\left(\sum_{i=0}^{j-1}d^i\eta\right)\mathbf{C}_b$\footnote{Henceforth, we use the notation $\mathbf{C}=[\mathbf{C}_1,\mathbf{C}_2,\dots, \mathbf{C}_K]$,  $1\leq j\leq K$, to refer to this form of concatenation}. The constructed code is a $[q;Q;\eta;(1\!:\!d);e]$-SQ-separable code with decoding complexity $O(mn)$.
\end{theorem} 
\begin{IEEEproof}
First, we show that the value of the largest entry of $\mathbf{C}$ is at most $q-1$. In order to prove this claim, it suffices to focus on $\mathbf{C}_K$. The largest entry of this matrix is equal to $\sum_{i=0}^{K-1}d^i\eta=\eta\frac{d^K-1}{d-1}$. Since $K=\left\lfloor\log_d\left(\left(\frac{q-1}{\eta}\right)(d-1)+1\right)\right\rfloor$, the largest entry of $\mathbf{C}_K$ (and therefore the largest entry of $\mathbf{C}$) is at most equal to $\eta\frac{\left(\frac{q-1}{\eta}\right)(d-1)+1-1}{d-1}=\eta\frac{\left(\frac{q-1}{\eta}\right)(d-1)}{d-1}=\eta\left(\frac{q-1}{\eta}\right)=q-1$.
The remainder of the proof is based on describing the decoding procedure and proving that the procedure allows for distinguishing between any two different sets of not more than $d$ defectives. 

Let $\mathbf{y}$ be the $Q$-ary vector of test outcomes, or equivalently, the syndrome of the defectives. For a rational vector $\mathbf{z}$, 
let $\left\lfloor\mathbf{z}\right\rfloor$ and $\left\langle\mathbf{z}\right\rangle$ denote the vector of integer parts of $\mathbf{z}$ and fractional parts of $\mathbf{z}$, respectively. 
If $d=1$, decoding reduces to finding the column of $\mathbf{C}$ equal to $\eta \mathbf{y}$. 
 If $d>1$, decoding proceeds as follows.

\textbf{Step 1:} Set $\mathbf{y}'_{K}=\mathbf{y}$ and form vectors $\mathbf{y}_j$, $1\leq j\leq K$, using the rules:
\begin{equation}\nonumber
\mathbf{y}_j=\left(\frac{d^j-1}{d-1}\right)\left\lfloor\left(\frac{d-1}{d^j-1}\right)\mathbf{y}'_{j}\right\rfloor,
\end{equation}
and
\begin{equation}\nonumber
\mathbf{y}'_{j-1}=\left(\frac{d^j-1}{d-1}\right)\left\langle\left(\frac{d-1}{d^j-1}\right)\mathbf{y}'_{j}\right\rangle.
\end{equation}

\textbf{Step 2:} Use the decoding algorithm in Remark~\ref{SQ-disjunct_decoding} for $\mathbf{C}_j$ and $\mathbf{y}_j$ to find the defectives among the subjects corresponding to the columns of $\mathbf{C}_j$.

The result is obviously true for $d=1$. Therefore, we focus on the case $d>1$. If there are no errors, using induction one can prove that each $\mathbf{y}_j$, $1\leq j\leq K$, is the syndrome of a subset of columns of $\mathbf{C}_j$ corresponding to defectives. Let $\mathbf{C}'_j=[\mathbf{C}_1,\mathbf{C}_2,\dots, \mathbf{C}_j]$, where $1\leq j\leq K$. Since the non-zero entries of $\mathbf{C}$ are multiples of $\eta$, $\eta\mathbf{y}$ is the sum of columns of $\mathbf{C}$ corresponding to a subset of defectives. Also, the maximum value of the entries of $\mathbf{C}'_{K-1}$ equals $\eta\frac{d^{K-1}-1}{d-1}$. Since there are at most $d$ defectives, the maximum value of their sum does not exceed $\eta\frac{d^K-d}{d-1}$. This bound is strictly smaller than $\eta\frac{d^K-1}{d-1}$, the minimum non-zero entry of $\mathbf{C}_K$. As a result, $\mathbf{y}_K$ is the syndrome of the defectives with codewords in $\mathbf{C}_K$, and $\mathbf{y}'_{K-1}$ is the syndrome of defectives with codewords in $\mathbf{C}'_{K-1}$. Similarly, it can be shown that $\forall j, 1\leq j\leq K-1$, $\mathbf{y}_j$ is the syndrome of the defectives with codewords in $\mathbf{C}_j$, and $\mathbf{y}'_{j-1}$ is the syndrome of the defectives with codewords in $\mathbf{C}'_{j-1}$.

On the other hand if there are $e>0$ errors in $\mathbf{y}$, for each $\mathbf{y}_j$, $1\leq j\leq K$, there are at most $e$ erroneous coordinates. Since from Theorem~\ref{prop:disj} we know that each $\mathbf{C}_j$ is a $[q;Q;\eta;(1\!:\!d);e]$-SQ-disjunct code, using Step 2 one can uniquely identify the defectives with codewords from $\mathbf{C}_j$.
\end{IEEEproof}
In order to gain a better understanding of this construction, consider the binary $2$-disjunct code from~\cite[Ch. 3]{DH06} shown below
\begin{align}\nonumber
\mathbf{C}_b=\begin{pmatrix}
1 \  \  0 \  \  0 \  \  0 \  \  1 \  \  0 \  \  0 \  \  1 \  \  1 \  \  0 \  \  0 \  \  0\\
1 \  \  0 \  \  0 \  \  0 \  \  0 \  \  1 \  \  1 \  \  0 \  \  0 \  \  1 \  \  0 \  \  0\\
0 \  \  1 \  \  0 \  \  0 \  \  1 \  \  0 \  \  1 \  \  0 \  \  0 \  \  0 \  \  1 \  \  0\\
0 \  \  1 \  \  0 \  \  0 \  \  0 \  \  1 \  \  0 \  \  1 \  \  0 \  \  0 \  \  0 \  \  1\\
0 \  \  0 \  \  1 \  \  0 \  \  1 \  \  0 \  \  0 \  \  0 \  \  0 \  \  1 \  \  0 \  \  1\\
0 \  \  0 \  \  1 \  \  0 \  \  0 \  \  1 \  \  0 \  \  0 \  \  1 \  \  0 \  \  1 \  \  0\\
0 \  \  0 \  \  0 \  \  1 \  \  0 \  \  0 \  \  1 \  \  0 \  \  1 \  \  0 \  \  0 \  \  1\\
0 \  \  0 \  \  0 \  \  1 \  \  0 \  \  0 \  \  0 \  \  1 \  \  0 \  \  1 \  \  1 \  \  0\\
1 \  \  1 \  \  1 \  \  1 \  \  0 \  \  0 \  \  0 \  \  0 \  \  0 \  \  0 \  \  0 \  \  0
\end{pmatrix},
\end{align}
capable of correcting $e=0$ error with $m_b=9$ and $n_b=12$. Assume that $q=7$ and consider an equidistant SQGT model with $\eta=2$. Consequently, $K=\left\lfloor\log_2\left(\left(\frac{7-1}{2}\right)(2-1)+1\right)\right\rfloor=2$, and therefore $\mathbf{C}_1=2\mathbf{C}_b$ and $\mathbf{C}_2=6\mathbf{C}_b$. Concatenating these matrices according to the rule $\mathbf{C}=[\mathbf{C}_1,\mathbf{C}_2]$ yields
\begin{align}\nonumber
\mathbf{C}=\begin{pmatrix}
2 \  \  0 \  \  0 \  \  0 \  \  2 \  \  0 \  \  0 \  \  2 \  \  2 \  \  0 \  \  0 \  \  0\  \  6 \  \  0 \  \  0 \  \  0 \  \  6 \  \  0 \  \  0 \  \  6 \  \  6 \  \  0 \  \  0 \  \  0\\
2 \  \  0 \  \  0 \  \  0 \  \  0 \  \  2 \  \  2 \  \  0 \  \  0 \  \  2 \  \  0 \  \  0\  \  6 \  \  0 \  \  0 \  \  0 \  \  0 \  \  6 \  \  6 \  \  0 \  \  0 \  \  6 \  \  0 \  \  0\\
0 \  \  2 \  \  0 \  \  0 \  \  2 \  \  0 \  \  2 \  \  0 \  \  0 \  \  0 \  \  2 \  \  0\  \  0 \  \  6 \  \  0 \  \  0 \  \  6 \  \  0 \  \  6 \  \  0 \  \  0 \  \  0 \  \  6 \  \  0\\
0 \  \  2 \  \  0 \  \  0 \  \  0 \  \  2 \  \  0 \  \  2 \  \  0 \  \  0 \  \  0 \  \  2\  \  0 \  \  6 \  \  0 \  \  0 \  \  0 \  \  6 \  \  0 \  \  6 \  \  0 \  \  0 \  \  0 \  \  6\\
0 \  \  0 \  \  2 \  \  0 \  \  2 \  \  0 \  \  0 \  \  0 \  \  0 \  \  2 \  \  0 \  \  2\  \  0 \  \  0 \  \  6 \  \  0 \  \  6 \  \  0 \  \  0 \  \  0 \  \  0 \  \  6 \  \  0 \  \  6\\
0 \  \  0 \  \  2 \  \  0 \  \  0 \  \  2 \  \  0 \  \  0 \  \  2 \  \  0 \  \  2 \  \  0\  \  0 \  \  0 \  \  6 \  \  0 \  \  0 \  \  6 \  \  0 \  \  0 \  \  6 \  \  0 \  \  6 \  \  0\\
0 \  \  0 \  \  0 \  \  2 \  \  0 \  \  0 \  \  2 \  \  0 \  \  2 \  \  0 \  \  0 \  \  2\  \  0 \  \  0 \  \  0 \  \  6 \  \  0 \  \  0 \  \  6 \  \  0 \  \  6 \  \  0 \  \  0 \  \  6\\
0 \  \  0 \  \  0 \  \  2 \  \  0 \  \  0 \  \  0 \  \  2 \  \  0 \  \  2 \  \  2 \  \  0\  \  0 \  \  0 \  \  0 \  \  6 \  \  0 \  \  0 \  \  0 \  \  6 \  \  0 \  \  6 \  \  6 \  \  0\\
2 \  \  2 \  \  2 \  \  2 \  \  0 \  \  0 \  \  0 \  \  0 \  \  0 \  \  0 \  \  0 \  \  0\  \  6 \  \  6 \  \  6 \  \  6 \  \  0 \  \  0 \  \  0 \  \  0 \  \  0 \  \  0 \  \  0 \  \  0
\end{pmatrix},
\end{align}
which is a $[7;Q;2;(1\!:\!2);0]$-SQ-separable code, for any $Q>6$, with $m=9$ and $n=24$. 

Now assume that there are $2$ defectives, $S_2$ and $S_{20}$. In this case, the syndrome in the absence of any errors is equal to
\begin{align}\nonumber
\mathbf{y}=\begin{pmatrix}
3 \  \  0 \  \  1 \  \  4 \  \  0 \  \  0 \  \  0 \  \  3 \  \  1 
\end{pmatrix}^T.
\end{align}
Step 1 of the decoding procedure begins by setting $\mathbf{y}_2'=\mathbf{y}$. Then, we form the vectors 
\begin{align}\nonumber\mathbf{y}_2&=\begin{pmatrix}
3 \  \  0 \  \  3 \  \  3 \  \  0 \  \  0 \  \  0 \  \  3 \  \  0 
\end{pmatrix}^T,\\\nonumber
\mathbf{y}_1'&=\begin{pmatrix}
0 \  \  0 \  \  1 \  \  1 \  \  0 \  \  0 \  \  0 \  \  0 \  \  1 
\end{pmatrix}^T,\\\nonumber
\mathbf{y}_1&=\begin{pmatrix}
0 \  \  0 \  \  1 \  \  1 \  \  0 \  \  0 \  \  0 \  \  0 \  \  1 
\end{pmatrix}^T.
\end{align}
Since the syndrome of $\mathbf{x}_{20}$, $\mathbf{y}_{\{\mathbf{x}_{20}\}}=\begin{pmatrix}
3 \  \  0 \  \  0 \  \  3 \  \  0 \  \  0 \  \  0 \  \  3 \  \  0 
\end{pmatrix}^T$, is included in $\mathbf{y}_2$ and the syndrome of no other codeword in $\mathbf{C}_2$ is included in $\mathbf{y}_2$, we conclude that $S_{20}$ is a defective and no other defectives exist among the set $\{S_{13},S_{14},\dots,S_{24}\}$. Also, since the syndrome of $\mathbf{x}_{2}$, $\mathbf{y}_{\{\mathbf{x}_2\}}=\begin{pmatrix}
0 \  \  0 \  \  1 \  \  1 \  \  0 \  \  0 \  \  0 \  \  0 \  \  1 
\end{pmatrix}^T$, is included in $\mathbf{y}_1$ and the syndrome of no other codeword in $\mathbf{C}_1$ is included in $\mathbf{y}_1$, we conclude that the only defectives among the subjects are $S_{2}$ and $S_{20}$.

%%%%%%%%%%%%%%%%%%%%%%%%%
\subsection{Construction of $q$-ary SQ-separable codes}\label{subsec:q_sep}

Similar to the case of SQ-disjunct codes, SQ-separable codes may also be constructed from classical binary separable codes. 
\begin{prop}[\textbf{Construction 4}]\label{const4}
Any code generated by multiplying a conventional binary $d$-separable code capable of correcting up to $e$ errors by $q-1$, where $q-1\geq\eta_1$, represents a $[q;Q;\boldsymbol{\eta};(1\!:\!d);e]$-SQ-separable code. 
\end{prop}
\begin{IEEEproof}
The proof follows easily from the definition of SQ-separable codes and separable codes and is consequently omitted.
\end{IEEEproof}

While the proposition describes the construction of $q$-ary SQ-separable codes for an arbitrary set of thresholds, it is also of interest 
to consider $q$-ary SQ-separable codes for the equidistant SQGT model. In this case, SQ-separable codes are closely related to separable codes for the additive model (QGT). Similar to Construction 4, one can use $\mathbf{C}_b$, a binary $d$-separable code for QGT capable of correcting up to $e$ errors, in order to form $\mathbf{C}=(q-1)\mathbf{C}_b$, where $q-1\in\{\eta,2\eta,\dots\}$. Then $\mathbf{C}$ represents a $[q;Q;{\eta};(1\!:\!d);e]$-SQ-separable code. 

An interesting code design for the additive model is the construction by Lindstr{\"o}m, described in~\cite[Theorem 8]{L75}. In his approach, Lindstr{\"o}m used a theorem by Bose and Chowla in additive number theory~\cite{BC62} to construct binary codes for an adder channel. Multiplying this code with $\eta$ results in a $[q;Q;{\eta};d;0]$-SQ-separable code of size $n$ and length $m=\lceil d\log_2 L\rceil$, where $L$ is a power of a prime such that $n\leq L$. A similar idea can be used to further improve the rate of SQ-separable codes for equidistant SQGT. The idea is based on a result, proved in~\cite{BC62}, that shows that if $L$ is power of a prime, there exist $L$ nonzero integers smaller than $L^d$ such that the sums of any $d$ such integers, i.e., their $d$-sums, are all distinct modulo $L^d-1$. 

\begin{theorem}[\textbf{Construction 5}]
Let $L$ be a power of a prime such that $n\leq L$; also, let $q'=\lfloor\frac{q-1}{\eta}\rfloor+1$. Using the construction in~\cite{BC62},  find $L$ non-zero integers with distinct $d$-sums. Let the
 $q'$-ary representation of these integers serve as columns of a code $\mathbf{C}_{q'}$. Form the code $\mathbf{C}=\eta\  \mathbf{C}_{q'}$ of length $m=\lceil d\log_{q'}L\rceil$ and size $L$. 
 A code obtained by choosing any $n$ columns of $\mathbf{C}$ is a $[q;Q;{\eta};d;0]$-SQ-separable code of length $m$ and size $n$.
\end{theorem}
\begin{IEEEproof}
We only need to show that $\mathbf{C}_{q'}$ is capable of identifying $d$ defectives in an adder model. Assume that there exists two sets of $d$ codewords $\mathcal{X}=\{\mathbf{x}_i\}_{i=1}^{d}$ and $\mathcal{Z}=\{\mathbf{z}_j\}_{j=1}^d$ such that $|\mathcal{X}\cap\mathcal{Z}|<d$, and $\sum_{i=1}^d\mathbf{x}_i=\sum_{j=1}^d\mathbf{z}_j$. Consequently, $\forall k\in\llbracket m\rrbracket$, $\sum_{i=1}^d\mathbf{x}_{i}(k)=\sum_{j=1}^d\mathbf{z}_{j}(k)$. Then, 
\begin{align}\nonumber
\sum_{k=1}^m \left( \sum_{i=1}^d\mathbf{x}_{i}(k) \right) q'^{k-1}=\sum_{k=1}^m \left( \sum_{j=1}^d\mathbf{z}_{j}(k)  \right) q'^{k-1},
\end{align}
which implies that there exists two sets of $d$ integers with the same sum. This contradicts the assumptions behind the construction of $\mathbf{C}_{q'}$, and completes the proof.
\end{IEEEproof}

\begin{remark}[\textbf{Construction 6}]
A corollary of Construction 4 is that the same concatenation method used in Theorem~\ref{const3} along with binary $d$-disjunct codes may be combined with binary $d$-separable codes for CGT and QGT in order to construct $q$-ary SQ-separable codes for equidistant thresholds with high rates. This claim can be easily verified using the same steps performed in the proof of Theorem~\ref{const3}. Note that the decoding complexity of these codes, unlike that of the codes in Construction 3, may not be $O(mn)$ as it depends on the decoding complexity of the underlying $d$-separable codes.
\end{remark}
To illustrate the aforementioned construction, consider the binary $2$-separable code from~\cite{KS64}
\begin{align}\nonumber
\mathbf{C}_b=\begin{pmatrix}
1 \  \  1 \  \  0 \  \  0 \  \  0 \  \  0 \  \  0 \  \  0 \\
1 \  \  0 \  \  1 \  \  0 \  \  0 \  \  0 \  \  0 \  \  0 \\
0 \  \  1 \  \  0 \  \  1 \  \  0 \  \  1 \  \  0 \  \  0\\
0 \  \  0 \  \  0 \  \  1 \  \  1 \  \  0 \  \  0 \  \  0 \\
0 \  \  0 \  \  1 \  \  0 \  \  1 \  \  0 \  \  1 \  \  0\\
0 \  \  0 \  \  0 \  \  0 \  \  0 \  \  1 \  \  0 \  \  1\\
0 \  \  0 \  \  0 \  \  0 \  \  0 \  \  0 \  \  1 \  \  1
\end{pmatrix},
\end{align}
capable of correcting $e=0$ error with $m_b=7$ and $n_b=8$. Assume that $q=7$ and consider an equidistant SQGT model with $\eta=2$. Consequently, $K=\left\lfloor\log_2\left(\left(\frac{7-1}{2}\right)(2-1)+1\right)\right\rfloor=2$, and therefore $\mathbf{C}_1=2\mathbf{C}_b$ and $\mathbf{C}_2=6\mathbf{C}_b$. Concatenating these matrices according to $\mathbf{C}=[\mathbf{C}_1,\mathbf{C}_2]$ yields
\begin{align}\nonumber
\mathbf{C}_b=\begin{pmatrix}
2 \  \  2 \  \  0 \  \  0 \  \  0 \  \  0 \  \  0 \  \  0 \  \  6 \  \  6 \  \  0 \  \  0 \  \  0 \  \  0 \  \  0 \  \  0 \\
2 \  \  0 \  \  2 \  \  0 \  \  0 \  \  0 \  \  0 \  \  0 \  \  6 \  \  0 \  \  6 \  \  0 \  \  0 \  \  0 \  \  0 \  \  0 \\
0 \  \  2 \  \  0 \  \  2 \  \  0 \  \  2 \  \  0 \  \  0 \  \  0 \  \  6 \  \  0 \  \  6 \  \  0 \  \  6 \  \  0 \  \  0\\
0 \  \  0 \  \  0 \  \  2 \  \  2 \  \  0 \  \  0 \  \  0 \  \  0 \  \  0 \  \  0 \  \  6 \  \  6 \  \  0 \  \  0 \  \  0 \\
0 \  \  0 \  \  2 \  \  0 \  \  2 \  \  0 \  \  2 \  \  0 \  \  0 \  \  0 \  \  6 \  \  0 \  \  6 \  \  0 \  \  6 \  \  0\\
0 \  \  0 \  \  0 \  \  0 \  \  0 \  \  2 \  \  0 \  \  2 \  \  0 \  \  0 \  \  0 \  \  0 \  \  0 \  \  6 \  \  0 \  \  6\\
0 \  \  0 \  \  0 \  \  0 \  \  0 \  \  0 \  \  2 \  \  2 \  \  0 \  \  0 \  \  0 \  \  0 \  \  0 \  \  0 \  \  6 \  \  6
\end{pmatrix},
\end{align}
which is a $[7;Q;2;(1\!:\!2);0]$-SQ-separable code, for any $Q>6$, with $m=7$ and $n=16$. 

As a parting note, $d$-separable codes for QGT can be used in conjunction with the same concatenation method to form SQ-separable codes.

%%%%%%%%%%%%%%%%%%%%%%%%%
\subsection{Construction of binary SQ-separable codes}\label{subsec:2_sep}
The constructions considered up to this point used an alphabet size of $q\geq\eta_1+1$. On the other hand, it is important to address the issue of constructing SQGT codes with alphabet size $q\leq\eta_1$, and in particular $q=2$. This problem may be solved by noticing that SQGT can be viewed as a generalization of TGT with a zero gap. While in TGT with zero gap there exist only one threshold, in SQGT one may have more than one threshold if $Q$-ary test results are allowed. This implies that any code constructed for TGT is also a SQ-separable code. In~\cite{CF09}, Chen and Fu observed that a variation of binary disjunct codes, also studied under the name of cover-free families (see~\cite{DVMT02}-\cite{CDH07}), can be used for TGT. In~\cite{C10} Cheraghchi showed that a weaker notion of disjunct codes, so called \emph{threshold disjunct} codes, are also applicable to the TGT problem and provided constructions with high rates. In the following theorem, we describe a generalization of these codes that are particularly useful for the SQGT model. This generalization provides binary and non-binary codes for arbitrary thresholds, $\boldsymbol{\eta}$.

\begin{theorem}\label{SQsep_bin}
Let $\eta_{\alpha}$ be the $\alpha^{\text{th}}$ threshold in a SQGT model. Consider a matrix $\mathbf{C}\in [q]^{m\times n}$ such that for any subset of column-indices $\mathcal{S}\subseteq\llbracket n\rrbracket $ with $\frac{\eta_{\alpha}}{q-1}\leq|\mathcal{S}|\leq d$, and for any index $l\in\mathcal{S}$, any set $\mathcal{N}\in\llbracket n\rrbracket $, where $|\mathcal{N}|\leq|\mathcal{S}|$, and $\mathcal{S}\cap\mathcal{N}=\varnothing$, there exists a set of row-indices $\mathcal{R}$ with size at least $2e+1$, such that  $\forall j\in\mathcal{R}$ it holds that
\begin{align}\label{sep_bin1}
& \sum_{k\in\mathcal{S}}\mathbf{C}(j,k)\  \in\{\eta_1,\eta_2,\dots,\eta_{\alpha}\},\\
& \sum_{k\in\mathcal{N}}\mathbf{C}(j,k)=0, \\\label{sep_bin3}
 &\mathbf{C}(j,l)\neq 0.
 \end{align}
 Then,  $\mathbf{C}$ is a $[q;Q;\boldsymbol{\eta};(\lceil\frac{\eta_{\alpha}}{q-1}\rceil\!:\!d);e]$-SQ-separable code.
\end{theorem}

\begin{IEEEproof}
Consider two distinct sets of codewords (i.e. columns of $\mathbf{C}$), denoted by $\mathcal{X}$ and $\mathcal{Z}$, such that $\lceil\frac{\eta_{\alpha}}{q-1}\rceil\leq|\mathcal{X}|,|\mathcal{Z}|\leq d$. Without loss of generality, assume that $|\mathcal{X}|\geq|\mathcal{Z}|$. Let $\mathcal{S}$ be the set of column-indices corresponding to $\mathcal{X}$. Also, let $\mathcal{N}$ be the set of column-indices corresponding to $\mathcal{Z}\backslash\mathcal{X}$. Consequently, $\frac{\eta_{\alpha}}{q-1}\leq|\mathcal{S}|\leq d$, $|\mathcal{N}|\leq|\mathcal{S}|$, and  $\mathcal{S}\cap\mathcal{N}=\varnothing$. Let $l$ be the index of the codeword $\mathbf{x}_l\in\mathcal{X}\backslash\mathcal{Z}$. Such a codeword always exists due to the manner in which $\mathcal{X}$ and $\mathcal{Z}$ are chosen.

From the definition of $\mathbf{C}$, there exists a set of row-indices with size $|\mathcal{R}|\geq2e+1$ such that $\forall k\in\mathcal{R}$, conditions~\eqref{sep_bin1}-\eqref{sep_bin3} are satisfied. 
This implies that $\forall k\in\mathcal{R}$,
\begin{align}\nonumber
\mathbf{y}_{\!_{\mathcal{X}}}(k)>\mathbf{y}_{\!_{\mathcal{Z}}}(k). 
\end{align}
As a result, $\mathbf{C}$ is a $[q;Q;\boldsymbol{\eta};(\lceil\frac{\eta_{\alpha}}{q-1}\rceil\!:\!d);e]$-SQ-separable code.
 \end{IEEEproof}

The next theorem describes a probabilistic  construction for this type of SQ-separable codes with $q=2$. This construction can be generalized for $q>2$ in a similar manner. 

\begin{theorem}[\textbf{Construction 7}]\label{const5}
Let $r=\lfloor \log_2\frac{d}{\eta_{\alpha}}\rfloor +1$, $\mu=\frac{1}{2^3}\left(1-\frac{1}{\eta_\alpha}\right)$, and $\rho= \frac{1}{2}\sum_{\beta=1}^{\alpha}\left(\frac{\mu}{\eta_\beta-1}\right)^{\eta_\beta}\frac{\eta_\beta-1}{d-1}$. 
Assume that $d=o(n)$. For any $i\in\llbracket r\rrbracket$, form a binary matrix $\mathbf{C}_i\in{[2]}^{({m}/{r})\times n}$ by choosing each entry independently according to a Bernoulli distribution such that the probability of choosing $1$ equals $P_i=\frac{1}{2^{i+2}\eta_\alpha}$. Now, form a matrix $\mathbf{C}=[\mathbf{C}_1^T,\mathbf{C}_2^T,\dots,\mathbf{C}_r^T]^T$, where $T$ denotes the matrix transpose operator. Then $\mathbf{C}$ is a $[2;Q;\boldsymbol{\eta};(\eta_{\alpha}\!:\!d);0]$-SQ-separable code with probability at least $1-o(1)$, provided that $m=r\left(\frac{2d}{\rho}+\delta\right)\log \frac{n}{d}$, $\forall \delta>0$. Similarly, $\mathbf{C}$ is a $[2;Q;\boldsymbol{\eta};(\eta_{\alpha}\!:\!d);e]$-SQ-separable code with probability at least $1-o(1)$, if $m=r\left[\left(\frac{4d}{\rho}+\delta\right)\log \frac{n}{d}+\frac{4e}{\rho}\right]$, $\forall\delta>0$.
\end{theorem}

\begin{IEEEproof}
The idea behind this construction is that each sub-matrix $\mathbf{C}_i$, $i\in\llbracket r\rrbracket$, satisfies  conditions~\eqref{sep_bin1}-\eqref{sep_bin3} for different sizes of $\mathcal{S}$. 

From Theorem~\ref{SQsep_bin}, we know that for $q=2$ it is only required to consider $\mathcal{S}$ with size $\eta_\alpha\leq|\mathcal{S}|\leq d$; therefore, for any such choice of $\mathcal{S}$ we can find $i\in\llbracket r\rrbracket$ such that $\eta_\alpha 2^{i-1}\leq |\mathcal{S}|< 2^i\eta_\alpha$. Fix a choice of $\mathcal{S}$, a choice of $l\in\mathcal{S}$, and a choice of $\mathcal{N}$ such that $|\mathcal{N}|\leq|\mathcal{S}|$. Let $A_i$ denote the total number of such choices. Form $\mathbf{C}_i$ by choosing each entry independently according to a Bernoulli distribution such that the probability of choosing $1$ equals $P_i=\frac{1}{2^{i+2}\eta_\alpha}$. Let $\pi_i$ denote the probability that a fixed row of $\mathbf{C}_i$ denoted by $\mathbf{r}$ satisfies conditions~\eqref{sep_bin1}-\eqref{sep_bin3}. Note that since the entries of $\mathbf{C}_i$ are chosen according to an i.i.d. probability distribution, the choice of $\mathbf{r}$ does not affect $\pi_i$. Let $\mathcal{E}_\beta$, $\beta\in\llbracket \alpha \rrbracket$, be the event that $\sum_{k\in\mathcal{S}}\mathbf{r}(k)\ =\eta_\beta$, and $\sum_{k\in\mathcal{N}}\mathbf{r}(k)=0$, and $\mathbf{r}(l)=1$. Consequently,
\begin{align}\nonumber
\pi_i=\Pro\left(\bigcup_{\beta=1}^\alpha\mathcal{E}_\beta\right)=\sum_{\beta=1}^{\alpha}\Pro\left(\mathcal{E}_\beta\right),
\end{align}
where the second equality follows from the disjointness of these events. A lower bound on the probability of the event $\mathcal{E}_\beta$ can be found using
\begin{align} \nonumber
\Pro\left(\mathcal{E}_\beta\right) &= \sum_{\substack{\mathcal{T}\,\subseteq\,\mathcal{S}\,\backslash\, \{l\},\\ |\mathcal{T}|=\eta_\beta-1}}\Pro\left(\mathbf{r}(k)=1,\   \forall k\in\mathcal{T}\right)\cdot\Pro(\mathbf{r}(l)=1)\cdot\Pro\left(\mathbf{r}(k)=0,\  \forall k\in(\mathcal{S}\cup\mathcal{N})\backslash(\mathcal{T}\cup\{l\})\right)\\\nonumber
&=\sum_{\mathcal{T}}P_i^{\eta_\beta-1}\cdot P_i\cdot(1-P_i)^{|\mathcal{S}|+|\mathcal{N}|-\eta_\beta}\geq\sum_{\mathcal{T}}P_i^{\eta_\beta}\left(1-(|\mathcal{S}|+|\mathcal{N}|-\eta_\beta)P_i\right).
\end{align}
On the other hand,
\begin{align}\nonumber
P_i(|\mathcal{S}|+|\mathcal{N}|-\eta_\beta)&\leq P_i(2|\mathcal{S}|-\eta_\beta)= 2P_i(|\mathcal{S}|-\frac{\eta_\beta}{2})\\\nonumber
&=\frac{1}{2^{i+1}\eta_{\alpha}}(|\mathcal{S}|-\frac{\eta_\beta}{2})\leq\frac{|\mathcal{S}|}{\eta_\alpha}\frac{1}{2^{i+1}}\leq\frac{1}{2}.
\end{align}
As a result,
\begin{align}\nonumber
\Pro\left(\mathcal{E}_\beta\right)&\geq\frac{1}{2}\sum_{\mathcal{T}}P_i^{\eta_\beta}=\frac{1}{2}{|\mathcal{S}|-1\choose \eta_\beta-1}P_i^{\eta_\beta}\geq\frac{1}{2}\left(\frac{|\mathcal{S}|-1}{\eta_\beta-1}\right)^{\eta_\beta-1}P_i^{\eta_\beta}\\\nonumber
&=\frac{1}{2}\frac{(P_i(|\mathcal{S}|-1))^{\eta_\beta}}{(\eta_\beta-1)^{\eta_\beta}}\frac{(\eta_\beta-1)}{|\mathcal{S}|-1}\geq\frac{1}{2}\frac{(2^{-3}-2^{-i-2}/\eta_\alpha)^{\eta_\beta}}{(\eta_\beta-1)^{\eta_\beta}}\frac{\eta_\beta-1}{|\mathcal{S}|-1}\\\nonumber
&\geq\frac{1}{2}\left(\frac{\mu}{\eta_\beta-1}\right)^{\eta_\beta}\frac{\eta_\beta-1}{|\mathcal{S}|-1}\geq\frac{1}{2}\left(\frac{\mu}{\eta_\beta-1}\right)^{\eta_\beta}\frac{\eta_\beta-1}{d-1}
\end{align}
where $\mu=\frac{1}{2^3}\left(1-\frac{1}{\eta_\alpha}\right)$.
Consequently, a lower bound on $\pi_i$ reads as
\begin{align}\label{eq:rho}
\pi_i=\sum_{\beta=1}^{\alpha}\Pro\left(\mathcal{E}_\beta\right)\geq \frac{1}{2}\sum_{\beta=1}^{\alpha}\left(\frac{\mu}{\eta_\beta-1}\right)^{\eta_\beta}\frac{\eta_\beta-1}{d-1}\coloneq\rho,
\end{align}
which is independent of $i$.

Using a union bound and~\eqref{eq:rho}, we arrive at an upper bound on the probability that $\mathbf{C}$ does not satisfy the conditions in Theorem~\ref{SQsep_bin}, i.e.
\begin{align}\label{Pf}
P_{F}\leq \sum_{i=1}^r A_iP_{F_i}(\pi_i).
\end{align}
Here, $P_{F_i}(\pi_i)$ is the probability that $\mathbf{C}_i$ does not satisfy the conditions in Definition~\ref{SQsep} for a choice of $\mathcal{S}$ that satisfies $\eta_\alpha 2^{i-1}\leq |\mathcal{S}|< 2^i\eta_\alpha$. 

Next, let $m'$ denote the number of rows of $\mathbf{C}_i$, for all $i\in\llbracket r\rrbracket$. If $e=0$, then
\begin{align}\label{pf1}
P_{F_i}(\pi_i)=(1-\pi_i)^{m'}\leq(1-\rho)^{m'}\leq\exp(-m'\rho)\triangleq{p_F(\rho)};
\end{align} 
otherwise, for $e>0$ we can use the Chernoff bound to find
\begin{align}\label{pfe}
P_{F_i}(\pi_i)\leq\exp\left({-\frac{m'\rho}{2}{\left(1-\frac{2e}{m'\rho}\right)^2}}\right)\triangleq{p_F(\rho)}.
\end{align}
Since these upper bounds are independent of $i$,~\eqref{Pf} simplifies to
\begin{align}\label{PF2}
P_F\leq A_\alpha \  p_F(\rho),
\end{align}
where $A_\alpha=\sum_{i=1}^r A_i$ and $p_F(\rho)$ are defined in~\eqref{pf1} and~\eqref{pfe} for $e=0$ and $e>0$, respectively.

Since $A_\alpha$ is equal to the total number of choices for $\mathcal{S}$, $l$, and $\mathcal{N}$, one has
\begin{align}\nonumber
A_\alpha=\sum_{s=\eta_\alpha}^d{n \choose s}s\sum_{z=0}^{\min(s,n-s)}{n-s\choose z},
\end{align}
where $s$ denotes the size of $\mathcal{S}$ and $z$ denotes the size of $\mathcal{N}$. Since ${n-s\choose z}\leq {n\choose s}$ for any $z\in \{0,1,\dots,\min(s,n-s)\}$, by assuming that $d\leq\frac{n}{2}$ for simplicity, we may write
\begin{align}\nonumber
A_\alpha&\leq\sum_{s=\eta_\alpha}^d{n\choose s}^2(s+1)s<\sum_{s=\eta_\alpha}^d\left(\frac{n\e}{s}\right)^{2s}(s+1)s\\\label{eq:Au}
&<(d-\eta_\alpha)(d+1)d\left(\frac{n\e}{d}\right)^{2d}<d^3\left(\frac{n\e}{d}\right)^{2d},
\end{align}
where $\e=\exp(1)$ denotes the base of the natural logarithm and is not to be confused with the number of errors $e$ that the code can correct.
Note that the third inequality follows from the fact that the largest term in $\sum_{s=\eta_\alpha}^d\left(\frac{n\e}{s}\right)^{2s}(s+1)s$ is indexed by $s=d$. This can be easily shown by noting that
\begin{align}\nonumber
\frac{\left(\frac{n\e}{s}\right)^{2s}(s+1)s}{\left(\frac{n\e}{s+1}\right)^{2s+2}(s+1)(s+2)}=\left(1+\frac{1}{s}\right)^{2s}\frac{(s+1)^2s}{s+2}\frac{\e^{-2}}{n^2}\leq\frac{1}{n^2}s(s+1)<1.
\end{align}
Using~\eqref{eq:rho},~\eqref{pf1},~\eqref{PF2}, and~\eqref{eq:Au}, the probability that $\mathbf{C}$ is not a $[2;Q;\boldsymbol{\eta},\alpha;d,0]$-SQ-separable code of size $n$ and length $m=rm'$ is upper bounded by
\begin{align}\nonumber
P_F\leq d^3\left(\frac{n\e}{d}\right)^{2d}\exp(-m'\rho)=\exp\left(2d\log n+3\log d+2d-2d\log d-m'\rho\right).
\end{align}
As a result, if $d=o(n)$, for any $\delta>0$, one has $P_F=o(1)$ if 
\begin{align}\nonumber
m=rm'=r\left(\frac{2d}{\rho}+\delta\right)\log \frac{n}{d}.
\end{align}
Similarly, the probability that $\mathbf{C}$ is not a $[2;Q;\boldsymbol{\eta},\alpha;d,e]$-SQ-separable code of size $n$ and length $m=rm'$ is upper bounded by
\begin{align}\nonumber
P_F&\leq d^3\left(\frac{n\e}{d}\right)^{2d}\exp\left({-\frac{m'\rho}{2}{\left(1-\frac{2e}{m'\rho}\right)^2}}\right)\\\nonumber
&=\exp\left(2d\log n+3\log d+2d-2d\log d-{\frac{m'\rho}{2}{\left(1-\frac{2e}{m'\rho}\right)^2}}\right).
\end{align}
Then, if $d=o(n)$, for any $\delta>0$, one has $P_F=o(1)$ if
\begin{align}\nonumber
m=rm'=r\left[\left(\frac{4d}{\rho}+\delta\right)\log \frac{n}{d}+\frac{4e}{\rho}\right].
\end{align}
\end{IEEEproof}

\begin{remark}
As discussed earlier, any code designed for TGT without a gap, such that $\eta_{_{T}}\in\{\eta_1,\eta_2,\dots,\eta_Q\}$, can be used for the purpose of SQGT. Hence, the threshold disjunct codes in~\cite{C10}, constructed probabilistically, provide an alternative to the codes in Construction 7 for the SQGT model. However, as the next lemma indicates, the rate of this family of threshold disjunct codes, $R_{TD}$, is a decreasing function of $\eta_{_{T}}$ and the highest rate is achieved if $\eta_{_{T}}=\eta_1$. Consequently, the codes described in Construction 7 provide an improvement in the rate, quantified as follows. For any $\eta_{_{T}}\in\{\eta_1,\eta_2,\dots,\eta_Q\}$,
it holds that
\begin{align}\nonumber
\frac{R_{SQ7}}{R_{TD}}\geq\min_{\eta_{_{T}}\in\{\eta_1,\eta_2,\dots,\eta_Q\}}\frac{R_{SQ7}}{R_{TD}(\eta_{_{T}})}= \frac{\lfloor \log_2\frac{d}{\eta_{1}}\rfloor +1}{\lfloor \log_2\frac{d}{\eta_{\alpha}}\rfloor +1}\  \  \frac{\sum_{\beta=1}^{\alpha}\left(\frac{\mu}{\eta_\beta-1}\right)^{\eta_\beta}\frac{\eta_\beta-1}{d-1}}{\left(\frac{\mu_{\eta_1}}{\eta_1-1}\right)^{\eta_1}\frac{\eta_1-1}{d-1}}>\frac{\lfloor \log_2\frac{d}{\eta_{1}}\rfloor +1}{\lfloor \log_2\frac{d}{\eta_{\alpha}}\rfloor +1},
\end{align}
where $R_{SQ7}=\frac{\log_2 n}{m}$ is the rate of the code in Construction 7, and $\mu_{\eta_1}=\frac{1}{2^3}\left(1-\frac{1}{\eta_1}\right)$. As an example, if $d=\eta_\alpha=4\eta_1$, then $\frac{R_{SQ7}}{R_{TD}}>3$.
\end{remark}

\begin{lemma}\label{lemma:TD}
The rate of the family of threshold disjunct codes constructed probabilistically in~\cite{C10}, denoted by $R_{TD}=\frac{\log_2 n}{m}$, is a decreasing function of $\eta_{_{T}}$ (for a fixed $d$) and the highest rate is achieved if $\eta_{_{T}}=\eta_1$.\end{lemma}
\begin{IEEEproof}
In order to show that for a fixed $d$, the rate $R_{TD}$ is a decreasing function of $\eta_{_{T}}=\eta$, $2\leq \eta\leq d$, we express the rate as $R_{TD}=\frac{C_d}{f(d,\eta)}$, where $C_d$ is a coefficient that depends on $d$, 
\begin{align}\nonumber
f(d,\eta)=\frac{\lfloor \log_2\frac{d}{\eta}\rfloor +1}{\left(\frac{\mu_\eta}{\eta-1}\right)^{\eta}\frac{\eta-1}{d-1}},
\end{align} 
and $\mu_\eta=\frac{1}{8}\left(\frac{\eta-1}{\eta}\right)$. Consequently, $f(d,\eta)=\left(\lfloor \log_2\frac{d}{\eta}\rfloor +1\right)\frac{d-1}{\eta-1}\left(8\eta\right)^{\eta}$.
Now, to prove that $R_{TD}$ is a decreasing in $\eta$, it suffices to show that $f(d,\eta)$ is an increasing function of $\eta$, $2\leq \eta\leq d$. Let $d$ be fixed, where $d\geq 3$. In what follows, we prove that $\forall \eta\in\{2,3,\cdots,d-1\}$,
\begin{align}\nonumber
\frac{f(d,\eta+1)}{f(d,\eta)}\geq1.
\end{align}
One has
\begin{align}\nonumber
\frac{f(d,\eta+1)}{f(d,\eta)}=\frac{\lfloor \log_2\frac{d}{\eta+1}\rfloor +1}{\lfloor \log_2\frac{d}{\eta}\rfloor +1}\ \left(\frac{\eta+1}{\eta}\right)^{\eta+1}8(\eta-1)\geq 27\  \frac{\lfloor \log_2\frac{d}{\eta+1}\rfloor +1}{\lfloor \log_2\frac{d}{\eta}\rfloor +1},
\end{align}
where the inequality follows since $\eta\geq 2$. Let $K=\lfloor\log_2\frac{d}{2}\rfloor$. Since $1\leq \frac{d}{\eta+1}<\frac{d}{\eta}\leq\frac{d}{2}$, we partition the closed interval $[1,d/2]$ into a union of disjoint intervals as
\begin{align}\nonumber
[1,d/2]=[1,2)\cup[2,4)\cup\dots\cup [2^{K-1},2^K)\cup[2^K,d/2]=\bigcup_{k=1}^{K+1}\mathcal{I}_k,
\end{align}
where for $1\leq k\leq K$, $\mathcal{I}_k=[2^{k-1},2^k)$, and $\mathcal{I}_{K+1}=[2^K,d/2]$.
If $\frac{d}{\eta+1}$ and $\frac{d}{\eta}$ are in the same interval $\mathcal{I}_k$, i.e. $2^{k-1}\leq\frac{d}{\eta+1}<\frac{d}{\eta}<2^k$, then 
\begin{align}\nonumber
\frac{f(d,\eta+1)}{f(d,\eta)}\geq 27\  \frac{k-1+1}{k-1+1}=27>1.
\end{align}
If for some $1\leq k\leq K$, one has $\frac{d}{\eta+1}\in\mathcal{I}_k$ and $\frac{d}{\eta}\geq 2^k$, then 
\begin{align}\nonumber
\frac{f(d,\eta+1)}{f(d,\eta)}\geq 27\  \frac{\lfloor \log_2\frac{d}{\eta+1}\rfloor +1}{\lfloor \log_2\frac{d}{\eta}\rfloor +1}=\frac{27\  k}{\left\lfloor \log_2\left(2^k(1+\Delta)\right)\right\rfloor +1}=\frac{27\  k}{\left\lfloor \log_2(1+\Delta)\right\rfloor +1+k}\  ,
\end{align}
where $\Delta=\frac{d}{\eta  2^k}-1$. Since $\Delta\geq 0$, one has $\log_2(1+\Delta)\leq \frac{\Delta}{\ln 2}$. Since $\frac{d}{\eta+1}< 2^k$, it follows that $\Delta<1/\eta$. Consequently, $\log_2(1+\Delta)< \frac{1}{\eta\ln 2}<1$ and therefore $\left\lfloor \log_2(1+\Delta)\right\rfloor=0$. As a result,
\begin{align}\nonumber
\frac{f(d,\eta+1)}{f(d,\eta)}\geq \frac{27\  k}{\left\lfloor \log_2(1+\Delta)\right\rfloor +1+k}= \frac{27\  k}{1+k}\geq \frac{27}{2}>1.
\end{align}
This proves the claim that $R_{TD}$ is a decreasing function in $\eta_{_{T}}$. \end{IEEEproof}

Next, we describe an explicit construction of the family of codes described in Theorem~\ref{SQsep_bin}. 
In~\cite{C10}, an explicit construction based on lossless condensers~\cite{TUZ07} for TGT codes was described. In what follows, we explain how to use the building blocks of~\cite[Construction 3]{C10} for TGT and leverage the fact that in SQGT we have $Q$ thresholds at our disposal. 

The key ingredient of our method are building block matrices for threshold disjunct codes (henceforth, BBTD matrices)~\cite[Construction 3]{C10}. BBTDs are obtained from a strong lossless $(\tilde{k},\tilde{\epsilon})$-condenser\footnote{For the definition and a detailed explanation of strong lossless condensers, see~\cite[Definition 1]{C10} and~\cite{TUZ07}.} $f:{\{0,1\}}^{\tilde{n}}\times {\{0,1\}}^{\tilde{t}}\rightarrow {\{0,1\}}^{\tilde{l}}$; if the parameters of the BBTD matrix are $m'\times n'$, then $n'=2^{\tilde{n}}$ and $m'=2^{\tilde{t}+\tilde{k}}{8\eta_{_T}2^{\tilde{l}-\tilde{k}}\choose \eta_{_T}}=2^{\tilde{t}+\tilde{k}}\:O_{\eta_{_T}}\!\!\left(2^{\eta_{_T}(\tilde{l}-\tilde{k})}\right)$, where $\eta_{_T}$ is the threshold in the TGT model, and $\tilde{k}$ and $\tilde{\epsilon}$ denote the entropy and the error in the definition of a lossless condenser, respectively. Also, $\tilde{\epsilon}<(1-p)/16$ for some real parameter $0\leq p<1$. Let $\tilde{\gamma}:=\max\{1,2^{\tilde{k}-\tilde{l}}2^{\tilde{k}}/(10\eta_{_T})\}$. The following lemma was proved in~\cite{C10}.
\begin{lemma}\label{BBTD}
In a BBTD matrix $\mathbf{B}$ with parameters described above, and for any subset of column-indices $\mathcal{S}\subseteq\llbracket n\rrbracket$ with $2^{\tilde{k}-2}\leq|\mathcal{S}|\leq2^{\tilde{k}-1}$, and for any $\mathcal{N}\in\llbracket n\rrbracket$, where $|\mathcal{N}|\leq|\mathcal{S}|$, and $\mathcal{S}\cap\mathcal{N}=\varnothing$, there exists a set of row-indices $\mathcal{R}$ with size at least $p\tilde{\gamma}2^{\tilde{t}}$, such that $\forall j\in\mathcal{R}$ 
\begin{align}\label{eq:BBTD1}
&\sum_{k\in\mathcal{S}}\mathbf{B}(j,k)=\eta_{_T}\\\label{eq:BBTD2}
& \sum_{k\in\mathcal{N}}\mathbf{B}(j,k)=0.
\end{align}
\end{lemma}

The BBTD matrices described above are used in~\cite{C10} to obtain the so-called ``regular'' matrices, which are then converted into threshold disjunct codes. 

In the next theorem, we use BBTD matrices to construct SQ-separable code with rates exceeding their threshold disjunct code counterparts with $\eta_{_T}\in\{\eta_1,\eta_2,\dots,\eta_Q\}$. 
\begin{theorem}[\textbf{Construction 8}]
Assume that $d\geq\eta_{\alpha}\geq\eta_1>1$. Let $\eta_{\alpha}'=2^{\lceil \log_2(\eta_\alpha-1)\rceil}$ be the smallest power of $2$ that is at least as large as $(\eta_\alpha-1)$, let $r=\lceil \log_2\left((d-1)/\eta_\alpha'\right)\rceil$, and let $p\in [0,1)$. Let $\mathcal{B}={\{\mathbf{B}_i\}}_0^r$ be a set of binary BBTD matrices constructed for parameter $\eta_{_T}=\eta_{_1}-1$ using a family of strong lossless $(\tilde{k}_i,\tilde{\epsilon})$-condensers $\mathcal{F}={\{f_i\}}_0^r$, where $\tilde{k}_i={\lceil \log_2(\eta_1-1)\rceil}+i+1$ and $\epsilon<(1-p)/16$. For each $ i\in[r+1]$, $f_i:\{0,1\}^{\tilde{n}}\times \{0,1\}^{\tilde{t}}\rightarrow\{0,1\}^{\tilde{l}_i}$, and for the corresponding BBTD matrix, one has $\mathbf{B}_i\in {[2]}^{m_i\times n}$ where $m_i=2^{\tilde{t}+\tilde{k}_i}\:O_{\eta_{_1}}\!\!\left(2^{(\eta_{_1}-1)\!(\tilde{l}_i-\tilde{k}_i)}\right)$ and $n=2^{\tilde{n}}$. In step 1, $\forall i\in[r+1]$ construct $\mathbf{B}_i'\in {[2]}^{2^{r-i}m_i\times n}$ by repeating $\mathbf{B}_i$, $2^{r-i}$ times according to the rule $\mathbf{B}_i'={[{\mathbf{B}_i}^T,{\mathbf{B}_i}^T,\dots,{\mathbf{B}_i}^T]}^T$. In step 2, form matrix $\mathbf{C}'={[{\mathbf{B}_0'}^T,{\mathbf{B}_1'}^T,\dots,{\mathbf{B}_r'}^T]}^T$. In step 3, fix a $d$-disjunct binary matrix $\mathbf{D}\in {[2]}^{m_d\times n}$ capable of correcting $e_1$ errors in the CGT model. Form the binary matrix $\mathbf{C}$ such that its $k^{\text{th}}$ row is equal to the bit-wise OR of the $i^{\text{th}}$ row of $\mathbf{C}'$ and the $j^{\text{th}}$ row of $\mathbf{D}$, where $i={\lceil\frac{k}{m_d}\rceil}$ and $j=k-(i-1)m_d$. Then $\mathbf{C}$ is a $[2;Q;\boldsymbol{\eta};(\eta_\alpha:d);e]$-SQ-separable code of size $m\times n$, where $m=2^{\tilde{t}}m_d(d-1)\frac{\eta_1-1}{\eta_\alpha-1}\left(\sum_{i=0}^rO_{\eta_{_1}}\!\!\left(2^{(\eta_{_1}-1)\!(\tilde{l}_i-\tilde{k}_i)}\right)\right)$, $e=\lfloor\frac{(2e_1+1)p2^{\tilde{t}}\tilde{\gamma}'-1}{2}\rfloor$, and $\tilde{\gamma}'=\max\left\{1,\frac{d-1}{5(\eta_{_1}-1)}\min_{i\in[r+1]}\{2^{\tilde{k}_i-\tilde{l}_i}\}\right\}$.

\end{theorem}
\begin{IEEEproof}
First, we provide the sketch of the proof in order to build some intuition. The idea behind the proof is to first show that the interval $[\eta_{\alpha}-1,d-1]$ is a subset of the interval $[2^{-1}\eta'_{\alpha}\:,\:2^r\eta_{\alpha}']$. Then, using the definition of $\tilde{k}_i$,  $i\in [r+1]$, we show that $[2^{-1}\eta'_{\alpha}\:,\:2^r\eta_{\alpha}']=\bigcup_{i=0}^{r}[2^{\tilde{k}_i-2}\:,\:2^{\tilde{k}_i-1}]$. Then by construction of $\mathbf{B}_i$, $i\in [r+1]$, we have that $\mathbf{B}_i$ has at least $p\tilde{\gamma}_i2^{\tilde{t}}$ rows that satisfy~\eqref{eq:BBTD1} and~\eqref{eq:BBTD2} for $\eta_{_T}=\eta_{_1}-1$ and $2^{\tilde{k}_i-2}\leq|\mathcal{S}|\leq2^{\tilde{k}_i-1}$, where $\tilde{\gamma}_i=\max\{1,2^{\tilde{k}_i-\tilde{l}_i}2^{\tilde{k}_i}/\left(10(\eta_{_1}-1)\right)\}$. Since each $\mathbf{B}'_i$ is formed by concatenating $\mathbf{B}_i$ vertically $2^{r-i}$ times, $i\in [r+1]$, then $\mathbf{B}'_i$ has at least $p\tilde{\gamma}_i2^{\tilde{t}+r-i}$ rows that satisfy~\eqref{eq:BBTD1} and~\eqref{eq:BBTD2} for $\eta_{_T}=\eta_{_1}-1$ and $2^{\tilde{k}_i-2}\leq|\mathcal{S}|\leq2^{\tilde{k}_i-1}$. 

Similarly, since $\mathbf{C}'$ is formed by concatenating the $\mathbf{B}'_i$ matrices vertically, $i\in[r+1]$, it follows that $\mathbf{C}'$ has at least $p\tilde{\gamma}_i2^{\tilde{t}}$ rows that satisfy~\eqref{eq:BBTD1} and~\eqref{eq:BBTD2} for $\eta_{_T}=\eta_{_1}-1$ and $|\mathcal{S}|\in[\eta_{\alpha}-1,d-1]\subseteq\bigcup_{i=0}^{r}[2^{\tilde{k}_i-2}\:,\:2^{\tilde{k}_i-1}]$. Upon proving these results, one can reduce the rest of the proof to showing that $\mathbf{C}$ formed by performing bit-wise OR on the rows of $\mathbf{C}'$ and $\mathbf{D}$ according to the description in the statement of the theorem gives a $[2;Q;\boldsymbol{\eta};(\eta_\alpha:d);e]$-SQ-separable code. 

Consider a set of column-indices $\mathcal{S}$ such that $\eta_\alpha-1\leq |\mathcal{S}|\leq d-1$. Since $\eta_{\alpha}'=2^{\lceil \log_2(\eta_\alpha-1)\rceil}$, one has
\begin{align}\label{thm7:ineq1}
 \eta_\alpha'/2=2^{\lceil \log_2(\eta_\alpha-1)\rceil-1}\leq\eta_\alpha-1.
\end{align}
In addition, since $r=\lceil \log_2\left((d-1)/\eta_\alpha'\right)\rceil$, one also has 
\begin{align}\label{thm7:ineq2}
2^r\eta_{\alpha}'=2^{\lceil \log_2\left((d-1)/\eta_\alpha'\right)\rceil}\  \eta_{\alpha}'\geq 2^{\log_2\left((d-1)/\eta_\alpha'\right)}\  \eta_{\alpha}'=\frac{d-1}{\eta_{\alpha}'}\   \eta_{\alpha}'=d-1.
\end{align}
Using inequalities~\eqref{thm7:ineq1} and~\eqref{thm7:ineq2}, one obtains $ \eta_\alpha'/2\leq\eta_\alpha-1\leq|\mathcal{S}|\leq  d-1\leq 2^r\eta_{\alpha}'$. Since $\forall i\in [r+1]$, $\tilde{k}_i$ is chosen as $\tilde{k}_i={\lceil \log_2(\eta_1-1)\rceil}+i+1$, one has
\begin{align}\nonumber
2^{\tilde{k}_0-2}=2^{{\lceil \log_2(\eta_1-1)\rceil}-1}=\eta_\alpha'/2\leq|\mathcal{S}|\leq 2^r\eta_{\alpha}'=2^{{\lceil \log_2(\eta_1-1)\rceil}+r}=2^{\tilde{k}_r-1}.
\end{align} 
This implies that for any set of column indices $\mathcal{S}$, where $\eta_\alpha-1\leq |\mathcal{S}|\leq d-1$, there exists an $i\in [r+1]$ for which $2^{\tilde{k}_i-2}\leq|\mathcal{S}|\leq2^{\tilde{k}_i-1}$.
On the other hand, using Lemma~\ref{BBTD} we know that $\forall i\in[r+1]$, $\mathbf{B}_i$ has at least $p\tilde{\gamma}_i2^{\tilde{t}}$ rows that satisfy~\eqref{eq:BBTD1} and~\eqref{eq:BBTD2} for $\eta_{_T}=\eta_{_1}-1$ and $2^{\tilde{k}_i-2}\leq|\mathcal{S}|\leq2^{\tilde{k}_i-1}$, where $\tilde{\gamma}_i=\max\{1,2^{\tilde{k}_i-\tilde{l}_i}2^{\tilde{k}_i}/\left(10(\eta_{_1}-1)\right)\}$. 

In the first step of the construction, $\forall i\in[r+1]$, we formed $\mathbf{B}_i'\in {[2]}^{2^{r-i}m_i\times n}$ by repeating $\mathbf{B}_i$ $2^{r-i}$ times according to the rule $\mathbf{B}_i'={[{\mathbf{B}_i}^T,{\mathbf{B}_i}^T,\dots,{\mathbf{B}_i}^T]}^T$. As a result, $\forall i\in[r+1]$, $\mathbf{B}_i'$ has at least $p\tilde{\gamma}_i2^{\tilde{t}+r-i}$ rows that satisfy~\eqref{eq:BBTD1} and~\eqref{eq:BBTD2} for $\eta_{_T}=\eta_{_1}-1$ and $2^{\tilde{k}_i-2}\leq|\mathcal{S}|\leq2^{\tilde{k}_i-1}$. Since $\forall i\in[r+1]$, one also has
\begin{align}\nonumber
2^{r-i}\tilde{\gamma}_i&=2^{r-i}\max\{1,2^{\tilde{k}_i-\tilde{l}_i}2^{\tilde{k}_i}/(10(\eta_{_1}-1))\}\\\nonumber
&\geq\max\left\{1,2^{\tilde{k}_i-\tilde{l}_i}\frac{d-1}{5(\eta_{_1}-1)}\right\}\\\nonumber
&\geq\max\left\{1,\frac{d-1}{5(\eta_{_1}-1)}\min_{i\in[r+1]}\{2^{\tilde{k}_i-\tilde{l}_i}\}\right\},
\end{align}
then $\mathbf{B}_i'$ contains at least $p2^{\tilde{t}}\tilde{\gamma}'$ rows satisfying~\eqref{eq:BBTD1} and~\eqref{eq:BBTD2} for $\eta_{_T}=\eta_{_1}-1$ and $2^{\tilde{k}_i-2}\leq|\mathcal{S}|\leq2^{\tilde{k}_i-1}$, where 
\begin{align}\nonumber
\tilde{\gamma}'=\max\left\{1,\frac{d-1}{5(\eta_{_1}-1)}\min_{i\in[r+1]}\{2^{\tilde{k}_i-\tilde{l}_i}\}\right\}.
\end{align}
This result, in addition to the fact that for any set of column indices $\mathcal{S}$ for which $\eta_\alpha-1\leq |\mathcal{S}|\leq d-1$, there exists a $i\in [r+1]$ for which $2^{\tilde{k}_i-2}\leq|\mathcal{S}|\leq2^{\tilde{k}_i-1}$, implies that $\mathbf{C}'$ has at least $e'=p2^{\tilde{t}}\tilde{\gamma}'$ rows that satisfy
\begin{align}\label{eq:BBTD3}
&\sum_{k\in\mathcal{S}}\mathbf{C}'(j,k)=\eta_{_1}-1,\\\label{eq:BBTD4}
& \sum_{k\in\mathcal{N}}\mathbf{C}'(j,k)=0,
\end{align}
for any set $\mathcal{S}$ and $\mathcal{N}$, where $\eta_\alpha-1\leq |\mathcal{S}|\leq d-1$, $|\mathcal{N}|\leq|\mathcal{S}|$ and $\mathcal{S}\cap\mathcal{N}=\varnothing$. 

In order for $\mathbf{C}$ to be a $[2;Q;\boldsymbol{\eta};(\eta_\alpha:d);e]$-SQ-separable code\footnote{Although this construction resembles the construction of threshold disjunct codes in~\cite{C10}, one should notice that the matrix $\mathbf{C}'$ generated in Step 2 of Construction 8 is not a regular matrix (i.e. it is neither a $(d-1,e';\eta_1-1)$-regular matrix, nor a $(d-1,e';\eta_{\alpha}-1)$-regular matrix). Consequently,~\cite[Lemma 6]{C10} cannot be used directly to show that $\mathbf{C}$ is a SQ-separable code.}, we need to show that for any two distinct sets of codewords, i.e. columns of $\mathbf{C}$, denoted by $\mathcal{X}_1$ and $\mathcal{X}_2$, for which $\eta_\alpha\leq|\mathcal{X}_2|\leq|\mathcal{X}_1|\leq d$, one has $\mathbf{y}_{\!_{\mathcal{X}_1}}\neq\mathbf{y}_{\!_{\mathcal{X}_2}}$. Note that this constraint is weaker than the conditions~\eqref{sep_bin1}-\eqref{sep_bin3}. Without loss of generality, we made the assumption that $|\mathcal{X}_2|\leq|\mathcal{X}_1|$. 

Let $\mathcal{S}_1$ and $\mathcal{S}_2$ be the set of column-indices corresponding to $\mathcal{X}_1$ and $\mathcal{X}_2$, respectively. Since $\mathcal{S}_1\neq\mathcal{S}_2$ and $|\mathcal{S}_1|\geq|\mathcal{S}_2|$, the set $\mathcal{S}_1\backslash\mathcal{S}_2$ is nonempty. Let $l\in\mathcal{S}_1\backslash\mathcal{S}_2$. Given that $|\mathcal{S}_2|\leq d$, it follows from the definition of binary $d$-disjunct matrices that for the set $\mathcal{S}_2\cup\{l\}$ there exists a set of row indices of $\mathbf{D}$, denoted by $\mathcal{R}_{\mathbf{D}}$, with size at least $2e_1+1$, such that 
\begin{align}\label{Dcond1}
\sum_{k\in\mathcal{S}_2}\mathbf{D}(j,k)=0,\  \  \  \  \  &\forall j\in\mathcal{R}_{\mathbf{D}},\\\label{Dcond2}
\mathbf{D}(j,l)=1,\   \  \  \  \  \  \  \  \  &\forall j\in\mathcal{R}_{\mathbf{D}}.
\end{align}
Let $\mathcal{S}=\mathcal{S}_1\backslash\{l\}$. 
Also, if $\mathcal{S}_1\cap\mathcal{S}_2=\varnothing$ and $|\mathcal{S}_1|=|\mathcal{S}_2|$, let $\mathcal{N}=\mathcal{S}_2\backslash \{k_0\}$ where $k_0$ is an arbitrary column-index of $\mathcal{S}_2$. Otherwise, let $\mathcal{N}=\mathcal{S}_2\backslash\mathcal{S}_1$. 
Clearly, $|\mathcal{N}|\leq|\mathcal{S}|$. 

Next, let $\mathcal{R}_{\mathbf{C}'}$ be the set of row-indices of $\mathbf{C}'$ for which~\eqref{eq:BBTD3} and~\eqref{eq:BBTD4} are satisfied for the sets $\mathcal{S}$ and $\mathcal{N}$. Consider some $i\in\mathcal{R}_{\mathbf{C}'}$ and some $j\in\mathcal{R}_{\mathbf{D}}$. The $(j+(i-1)m_d)^{\text{th}}$ row of $\mathbf{C}$ is formed by finding the bit-wise OR of the $i^{\text{th}}$ row of $\mathbf{C}'$ and the $j^{\text{th}}$ row of $\mathbf{D}$. Consequently, 
\begin{align}\label{Dcond4}
&\sum_{k\in\mathcal{S}_1}\mathbf{C}(j+(i-1)m_d,k)=\sum_{k\in\mathcal{S}}\mathbf{C}(j+(i-1)m_d,k)+\mathbf{C}(j+(i-1)m_d,l)=\eta_1-1+1=\eta_1,\\\label{Dcond3}
&\sum_{k\in\mathcal{S}_2}\mathbf{C}(j+(i-1)m_d,k)<\eta_1,
\end{align}
where $\mathbf{C}(j+(i-1)m_d,l)=1$ follows from~\eqref{Dcond2}, and~\eqref{Dcond3} is a consequence of the following argument. First, note that using~\eqref{eq:BBTD2} and~\eqref{Dcond1}, one has $\sum_{k\in\mathcal{N}}\mathbf{C}(j+(i-1)m_d,k)=0$. As a result, if $\mathcal{S}_1\cap\mathcal{S}_2=\varnothing$ and $|\mathcal{S}_1|=|\mathcal{S}_2|$, then 
\begin{align}\nonumber
\sum_{k\in\mathcal{S}_2}\mathbf{C}(j+(i-1)m_d,k)=\sum_{k\in\mathcal{N}}\mathbf{C}(j+(i-1)m_d,k)+\mathbf{C}(j+(i-1)m_d,k_0)\leq 1<\eta_1.
\end{align}
Otherwise, one has
\begin{align}\nonumber
\sum_{k\in\mathcal{S}_2}\mathbf{C}(j+(i-1)m_d,k)&=\sum_{k\in\mathcal{N}}\mathbf{C}(j+(i-1)m_d,k)+\sum_{k\in\mathcal{S}_2\cap\mathcal{S}_1}\mathbf{C}(j+(i-1)m_d,k)\\\nonumber
&=\sum_{k\in\mathcal{S}_2\cap\mathcal{S}_1}\mathbf{C}(j+(i-1)m_d,k)\leq \sum_{k\in\mathcal{S}_1\backslash \{l\}}\mathbf{C}(j+(i-1)m_d,k)=\eta_1-1<\eta_1.
\end{align}
Since $|\mathcal{R}_{\mathbf{C}'}|\geq e'$ and $|\mathcal{R}_{\mathbf{D}}|\geq 2e_1+1$, $\mathbf{C}$ has a set of row indices $\mathcal{R}$, $|\mathcal{R}|\geq e'(2e_1+1)$, for which~\eqref{Dcond4} and~\eqref{Dcond3} are satisfied. 
This implies that $\forall j\in\mathcal{R}$, $\mathbf{y}_{\!_{\mathcal{X}_1}}(j)>\mathbf{y}_{\!_{\mathcal{X}_2}}(j)$, and therefore $\mathbf{C}$ is a $[2;Q;\boldsymbol{\eta};(\eta_\alpha:d);e]$-SQ-separable code, where $e=\lfloor\frac{(2e_1+1)p2^{\tilde{t}}\tilde{\gamma}'-1}{2}\rfloor$. Note that $\mathbf{C}$ is an $m\times n$ matrix, where $n=2^{\tilde{n}}$, and 
\begin{align}\nonumber
m = m_d\cdot\left(\sum_{i=0}^r2^{r-i}m_i\right)&\approx m_d\left(\sum_{i=0}^r2^{r+\tilde{t}+\log_2(\eta_1-1)+1}O_{\eta_{_1}}\!\!\left(2^{(\eta_{_1}-1)\!(\tilde{l}_i-\tilde{k}_i)}\right)\right)\\\nonumber
&= 2^{\tilde{t}}m_d(d-1)\frac{\eta_1-1}{\eta_\alpha-1}\left(\sum_{i=0}^rO_{\eta_{_1}}\!\!\left(2^{(\eta_{_1}-1)\!(\tilde{l}_i-\tilde{k}_i)}\right)\right).
\end{align}
\end{IEEEproof}

\begin{remark}
A comparison between the rate of the code described in Construction 8, denoted by $R_{SQ8}$, and the rate of the threshold disjunct code described in~\cite{C10} for $\eta_T=\eta_1$, denoted by $R_{TD}$, reveals that
\begin{align}\nonumber
\frac{R_{SQ8}}{R_{TD}}=\frac{\eta_{\alpha}-1}{\eta_1-1}.
\end{align}
\end{remark}
In order to compute this ratio, one needs to carefully calculate $R_{TD}$, keeping track of the constant values that may be hidden in the asymptotic expressions. It turns out that if the same $d$-disjunct binary matrix $\mathbf{D}$ is used in both constructions, $n_{SQ8}=n_{TD}$, and $m_{TD}=\frac{\eta_{\alpha}-1}{\eta_1-1} m_{SQ8}$.

%%%%%%%%%%%%%%%%%%%%%%%%%%%%%%
\subsection{Construction of SQ-separable codes for arbitrary number of defectives}\label{subsec:arb_d}
The constructions described up to this point are able to identify up to $d$ defectives in a pool of $n$ subjects whenever $d$ is significantly smaller than $n$, say $d=o(n)$ or $d$ constant. 
It is also of interest to address the same questions when $d$ is not constrained in size, so that one allows $0\leq d\leq n$. This ``dense'' testing regime may be of use whenever no bound on the number of defectives is known a priori or when the number of defectives is inherently large.

In~\cite{L75}, Lindstr{\"o}m described a binary construction for the adder model capable of identifying up to $n$ defectives. In the next theorem we describe a generalization of this construction that employs a $q$-ary alphabet; using this generalization, we construct a SQ-separable code capable of identifying up to $n$ defectives in an equidistant SQGT model. 
Extensions of~\cite{L75} to a $q$-ary alphabet were also addressed in~\cite{J95}. Multiplying these codes with $\eta$ results in a SQ-separable code with the same rate as our construction. But unlike our direct and very simple approach, the methods of~\cite{J95} and~\cite{CW99} may only be used in a recursive and rather complicated manner.

Before describing our construction, we state a lemma from~\cite{L75} that will be useful in proving the next theorem.
\begin{lemma}\label{lemma:Lind}
Let $\mathcal{F}$ be a collection of sets such that if $\mathcal{B}\in\mathcal{F}$, then $\mathcal{F}$ contains all the subsets of $\mathcal{B}$ as well. In other words, $\forall\mathcal{B}\in\mathcal{F}$, if $\mathcal{A}\subset\mathcal{B}$, then $\mathcal{A}\in\mathcal{F}$. Let $g:\mathcal{F}\mapsto\{0,1\}$ be a function defined on $\mathcal{F}$ such that for some fixed set $\mathcal{S}\in\mathcal{F}$, one has $g(\mathcal{A}\cap\mathcal{S})=g(\mathcal{A})$ whenever $\mathcal{A}\in\mathcal{F}$. If $\mathcal{C}\in\mathcal{F}$ and $\mathcal{C}\nsubset\mathcal{S}$, then
\begin{align}\nonumber
\sum_{\substack{\mathcal{A}\  \subseteq\  \mathcal{C}\\ |\mathcal{A}|\  \text{is odd}}}g(\mathcal{A})=\sum_{\substack{\mathcal{A}\  \subseteq\  \mathcal{C}\\ |\mathcal{A}|\  \text{is even}}}g(\mathcal{A}).
\end{align}
\end{lemma}

\begin{IEEEproof}
See~\cite{L75}.
\end{IEEEproof}

\begin{theorem}[\textbf{Construction 9}]\label{const9}
Let $\kappa\in\mathbb{Z}^{+}$ and $m=2^\kappa-1$. Consider the set  $\llbracket\kappa\rrbracket$ and label each of its non-empty subsets by $\mathcal{S}_i$,  $i\in\llbracket m\rrbracket$, such that for any  two subsets $\mathcal{S}_{i_1},\mathcal{S}_{i_2}\subseteq\llbracket\kappa\rrbracket$, the inequality $|\mathcal{S}_{i_1}|<|\mathcal{S}_{i_2}|$ implies $i_1<i_2$. Let $q'=\lfloor\frac{q-1}{\eta}\rfloor+1$ and $q''=\left\lfloor\log_2\lfloor\frac{q-1}{\eta}\rfloor\right\rfloor$; for each $\mathcal{S}_{i}$, form a matrix $\mathbf{C}_i\in{[q' ]}^{m\times (q''+|\mathcal{S}_i|)}$ as follows. For $j\in\llbracket m\rrbracket$ and $k\in\llbracket q''+1\rrbracket$, set 
\begin{align}\label{rule1}
\mathbf{C}_i(j,k)=
\left\{
     \begin{array}{ll}
         2^{^{q''-k+1}},   & \textnormal{if}\  \  \  |\mathcal{S}_i\cap\mathcal{S}_j| \  \  \textnormal{is odd},\\
              0,   & \textnormal{if} \  \  \  |\mathcal{S}_i\cap\mathcal{S}_j| \  \  \textnormal{is even.}
     \end{array}
   \right.
\end{align}
Let $\mathcal{T}_{i,q''+1}=\mathcal{S}_i$. For $k\in\{q''+2,q''+3,\ldots,q''+|\mathcal{S}_i|\}$, fix \emph{any} $\mathcal{T}_{i,k}\subset\mathcal{T}_{i,k-1}$ of size $|\mathcal{T}_{i,k}|=|\mathcal{S}_i|-k+q''+1$. Set
\begin{align}\label{rule2}
\mathbf{C}_i(j,k)=
\left\{
     \begin{array}{ll}
         1,   & \textnormal{if}\  \  \  \mathbf{C}_i(j,k-1)>0 \  \textnormal{and}\   |\mathcal{S}_j\cap\mathcal{T}_{i,k}| \  \  \textnormal{is odd,}\\
              0,   & \textnormal{otherwise,}
     \end{array}
   \right.
\end{align}
where $j\in\llbracket m\rrbracket$. Form a matrix $\mathbf{C}'=\eta\mathbf{C}$ where $\mathbf{C}=[\mathbf{C}_1,\mathbf{C}_2,\ldots,\mathbf{C}_m]$. The matrix $\mathbf{C}'$ is a $[q;Q;{\eta};(1\!:\!n);0]$-SQ-separable code of length $m=2^{\kappa}-1$ and size $n=\kappa2^{\kappa-1}+q''(2^{\kappa}-1)$.
\end{theorem}

\begin{IEEEproof}
As before, we define $\mathbf{w}\in{[2]}^n$ to be a binary vector such that its $l^\textnormal{th}$ coordinate is equal to $1$ if the $l^\text{th}$ subject is defective, and $0$ otherwise. From the construction, the matrix $\mathbf{C}$ is formed from $m$ sub-matrices $\mathbf{C}_i$, each corresponding to a subset of $\llbracket \kappa\rrbracket$, $\mathcal{S}_i$. This implies that each $\mathcal{S}_i$ corresponds to a set of variables, i.e. coordinates of $\mathbf{w}$. In addition, we label rows of $\mathbf{C}$ using subsets $\mathcal{S}_i$, $i\in\llbracket m\rrbracket$, such that the $i^{\text{th}}$ row is labeled by $\mathcal{S}_i$. Since each row of $\mathbf{C}$ corresponds to an equation in $\mathbf{y}=\mathbf{C}\mathbf{w}$, each $\mathcal{S}_i$ corresponds to \emph{exactly} one equation. 

The decoding includes $m$ steps, and in each step one solves for the variables corresponding to $\mathcal{S}_i$, given all the variables corresponding to $\mathcal{S}_{i+1},\mathcal{S}_{i+2},\dots, \mathcal{S}_{m}$. 
To find the variables corresponding to $\mathcal{S}_i$, we form two equations: the first equation is obtained by adding all the equations corresponding to the odd subsets of $\mathcal{S}_i$ while the second equation is obtained by adding all the equations corresponding to the even subsets of $\mathcal{S}_i$. These two equations can be represented by ${\mathbf{s}_{\text{odd}_{i}}}^T\mathbf{w}=y_{\text{odd}_{i}}$ and ${\mathbf{s}_{\text{even}_{i}}}^T\mathbf{w}=y_{\text{even}_{i}}$, respectively. Finally, we form the equation 
\begin{equation}\label{eq:maineq}
({\mathbf{s}_{\text{odd}_{i}}}-{\mathbf{s}_{\text{even}_{i}}})^T\mathbf{w}=y_{\text{odd}_{i}}-y_{\text{even}_{i}}.
\end{equation} 

For simplicity, let $w_{i_k}$ be the $k^{\text{th}}$ variable corresponding to $\mathcal{S}_i$, where $k\in\llbracket q''+|\mathcal{S}_i|\rrbracket$. The key in the proof of the theorem is to show that~\eqref{eq:maineq} is of the form 
\begin{align}\nonumber
2^{q''+|\mathcal{S}_i|-1}w_{i_1}+2^{q''+|\mathcal{S}_i|-2}w_{i_{2}}+\dots+w_{i_{q''+|\mathcal{S}_i|}}=a,
\end{align}
where $a$ is a scalar that depends on $\mathbf{y}$ and the \emph{known} variables corresponding to $\mathcal{S}_{i+1},\mathcal{S}_{i+2},\dots, \mathcal{S}_{m}$. This implies that all the coefficients of the variables corresponding to $\mathcal{S}_1,\mathcal{S}_2,\dots,\mathcal{S}_{i-1}$ are zero; also, given that $w_{i_k}\in[2]$ for all $k\in\llbracket q''+|\mathcal{S}_i|\rrbracket$, the unknown variables can be determined by finding the unique binary representation of $a$.
Note that the coefficient of the variable $w_{l_k}$, $l\leq i$, in the aforementioned expression equals 
\begin{align}\nonumber
\sum_{\substack{j:\  \mathcal{S}_j\  \subseteq\  {\mathcal{S}_i}\\ |\mathcal{S}_j|\  \text{is odd}}}\mathbf{C}_l(j,k)-\sum_{\substack{j:\  \mathcal{S}_j\  \subseteq\  {\mathcal{S}_i}\\ |\mathcal{S}_j|\  \text{is even}}}\mathbf{C}_l({j,k}).
\end{align}
%%%%%%%%%%%%%%%%%%%%%%%%%%%%%%%%%%%%%%%%%

We now show that $\forall l<i$, the coefficients of the variables in $\mathcal{S}_l$ of~\eqref{eq:maineq} are all zero. 
Although Lemma~\ref{lemma:Lind} cannot be directly applied to our problem since the matrix $\mathbf{C}$ is not binary, we make use of this lemma in our proof as follows. 

Let $\mathcal{F}={\{\mathcal{S}\}}_{1}^m$; this set satisfies the condition of Lemma~\ref{lemma:Lind}. Let $l<i$; due to the specific ordering of the elements of $\mathcal{F}$, we have $\mathcal{S}_i\nsubseteq\mathcal{S}_l$, and can consequently set $\mathcal{C}=\mathcal{S}_i$ and $\mathcal{S}=\mathcal{S}_l$. 
Consider the $k^{\text{th}}$ column of $\mathbf{C}_l$, where $k\in\{q''+1,q''+2,\dots,q''+|\mathcal{S}_l|\}$. For this column, let $g_{l,k}({\mathcal{S}_j})=\mathbf{C}_l(j,k)$. Careful inspection shows that $g_{l,k}(\mathcal{S}_j\cap\mathcal{S}_l)=g_{l,k}(\mathcal{S}_j)$, $\forall j\in\llbracket m\rrbracket$, and $g_{l,k}(\cdot)\in\{0,1\}$. Using Lemma~\ref{lemma:Lind}, we conclude that 
\begin{align}\label{eq:lemsum}
\sum_{\substack{j:\  \mathcal{S}_j\  \subseteq\  {\mathcal{S}_i}\\ |\mathcal{S}_j|\  \text{is odd}}}g_{l,k}(\mathcal{S}_j)=\sum_{\substack{j:\  \mathcal{S}_j\  \subseteq\  {\mathcal{S}_i}\\ |\mathcal{S}_j|\  \text{is even}}}g_{l,k}(\mathcal{S}_j).
\end{align}
Next, consider the $k^{\text{th}}$ column of $\mathbf{C}_l$, where $k\in\llbracket q''\rrbracket$. For this column, let $g_{l,k}({\mathcal{S}_j})=\mathbf{C}_l(j,k)$. Since $g_{l,k}({\mathcal{S}_j})=2^{q''-k+1}g_{l,q''+1}({\mathcal{S}_j})$, using~\eqref{eq:lemsum} one obtains
\begin{align}\nonumber
\sum_{\substack{j:\  \mathcal{S}_j\  \subseteq\  {\mathcal{S}_i}\\ |\mathcal{S}_j|\  \text{is odd}}}g_{l,k}(\mathcal{S}_j)&=2^{q''-k+1}\sum_{\substack{j:\  \mathcal{S}_j\  \subseteq\  {\mathcal{S}_i}\\ |\mathcal{S}_j|\  \text{is odd}}}g_{l,q''+1}(\mathcal{S}_j)\\\nonumber
&=2^{q''-k+1}\sum_{\substack{j:\  \mathcal{S}_j\  \subseteq\  {\mathcal{S}_i}\\ |\mathcal{S}_j|\  \text{is even}}}g_{l,q''+1}(\mathcal{S}_j)\\\nonumber
&=\sum_{\substack{j:\  \mathcal{S}_j\  \subseteq\  {\mathcal{S}_i}\\ |\mathcal{S}_j|\  \text{is even}}}g_{l,k}(\mathcal{S}_j).
\end{align}
As a result, $\forall l<i$ and $k\in\llbracket q''+|\mathcal{S}_l|\rrbracket$ one has
 \begin{align}\label{eq:lem1}
\sum_{\substack{j:\  \mathcal{S}_j\  \subseteq\  {\mathcal{S}_i}\\ |\mathcal{S}_j|\  \text{is odd}}}\mathbf{C}_l(j,k)-\sum_{\substack{j:\  \mathcal{S}_j\  \subseteq\  {\mathcal{S}_i}\\ |\mathcal{S}_j|\  \text{is even}}}\mathbf{C}_l({j,k})=0.
\end{align}

To complete the proof, consider the $k^{\text{th}}$ column of $\mathbf{C}_i$, where $k\in\llbracket q''+1\rrbracket$. Since~\eqref{eq:maineq} is formed using the rows labeled by odd and even subsets of $\mathcal{S}_i$, the coefficient of $w_{i_k}$ is equal to 
\begin{align}\label{eq:lem2}
\sum_{\substack{j:\  \mathcal{S}_j\  \subseteq\  {\mathcal{S}_i}\\ |\mathcal{S}_j|\  \text{is odd}}}\mathbf{C}_i(j,k)-\sum_{\substack{j:\  \mathcal{S}_j\  \subseteq\  {\mathcal{S}_i}\\ |\mathcal{S}_j|\  \text{is even}}}\mathbf{C}_i({j,k})=2^{q''-k+1}\cdot 2^{|\mathcal{S}_i|-1}-0=2^{q''+|\mathcal{S}_i|-k},
\end{align}
where $2^{|\mathcal{S}_i|-1}$ is the number of odd subsets of $\mathcal{S}_i$.
Next, consider  the $k^{\text{th}}$ column of $\mathbf{C}_i$, where $k\in\{ q''+2,q''+3,\dots,q''+|\mathcal{S}_i|\}$. From the definition of $\mathcal{T}_{i,k}$ and its relationship to $\mathcal{T}_{i,k-1}$, it can be shown that the coefficient of the variable $w_{i_k}$ equals 
\begin{align}\nonumber
&\sum_{\substack{j:\  \mathcal{S}_j\  \subseteq\  {\mathcal{S}_i}\\ |\mathcal{S}_j|\  \text{is odd}}}\mathbf{C}_i(j,k)-0=\sum_{\substack{j:\  \mathcal{S}_j\  \subseteq\  {\mathcal{S}_i}\\ |\mathcal{S}_j|\  \text{is odd}}}\mathbf{C}_i(j,k)\\\nonumber
&=\sum_{\substack{j:\  \mathcal{S}_j\  \subseteq\  {\mathcal{S}_i}\\ |\mathcal{S}_j|\  \text{is odd}}}\mathbf{1}\left[\{ |\mathcal{S}_j\cap\mathcal{T}_{i,q''+2}| \  \textnormal{is odd}\}   \cap \dots \cap \{ |\mathcal{S}_j\cap\mathcal{T}_{i,k-1}| \  \textnormal{is odd}\}\cap \{ |\mathcal{S}_j\cap\mathcal{T}_{i,k}| \  \textnormal{is odd}\}\right]\\\label{eq:lem3}
&=\frac{1}{2}\sum_{\substack{j:\  \mathcal{S}_j\  \subseteq\  {\mathcal{S}_i}\\ |\mathcal{S}_j|\  \text{is odd}}}\mathbf{1}\left[\{ |\mathcal{S}_j\cap\mathcal{T}_{i,q''+2}| \  \textnormal{is odd}\} \cap \dots \cap \{ |\mathcal{S}_j\cap\mathcal{T}_{i,k-1}| \  \textnormal{is odd}\}\right]=\dots=2^{q''+|\mathcal{S}_i|-k}.
\end{align}

Using~\eqref{eq:lem1},~\eqref{eq:lem2}, and~\eqref{eq:lem3}, one can write~\eqref{eq:maineq} in the form
\begin{align}\nonumber
\sum_{k=1}^{q''+|\mathcal{S}_i|}2^{q''+|\mathcal{S}_i|-k}w_{i_k}=a,
\end{align}
where $a$ depends on $\mathbf{y}$ and the known variables corresponding to $\mathcal{S}_{i+1},\mathcal{S}_{i+2},\dots, \mathcal{S}_{m}$.
This completes the proof of the claimed result.
\end{IEEEproof}

As an example, let $\kappa=3$, $\eta=2$, and $q=5$; consequently, $m=7$, $q'=9$, and $q''=2$. We label the non-empty subsets of $\llbracket 3\rrbracket$ as follows: $\mathcal{S}_1=\{1\}$, $\mathcal{S}_2=\{2\}$, $\mathcal{S}_3=\{3\}$, $\mathcal{S}_4=\{1,2\}$, $\mathcal{S}_5=\{1,3\}$, $\mathcal{S}_6=\{2,3\}$, $\mathcal{S}_7=\{1,2,3\}$. In construction $\mathbf{C}_7$, corresponding to $\mathcal{S}_7$, fix $\mathcal{T}_{7,4}=\{1,2\}$ and $\mathcal{T}_{7,5}=\{1\}$\footnote{Note that there exist other choices for $\mathcal{T}_{7,4}$ and $\mathcal{T}_{7,5}$ that provide for valid code constructions.}. Based on~\eqref{rule1} and~\eqref{rule2}, one has
\begin{align}\nonumber
\mathbf{C}_7=\begin{pmatrix}
4 & 2 & 1 & 1 & 1\\
4 & 2 & 1 & 1 & 0\\
4 & 2 & 1 & 0 & 0\\
0 & 0 & 0 & 0 & 0\\
0 & 0 & 0 & 0 & 0\\
0 & 0 & 0 & 0 & 0\\
4 & 2 & 1 & 0 & 0
\end{pmatrix}.
\end{align}

Using~\eqref{rule1} and~\eqref{rule2}, we obtain
\begin{align}\nonumber
\mathbf{C}=\bordermatrix{\text{}&\mathcal{S}_1&\mathcal{S}_2&\mathcal{S}_3&\mathcal{S}_4&\mathcal{S}_5&\mathcal{S}_6&\mathcal{S}_7\cr
                \mathcal{S}_1& 4\  2\  1& 0\  0\  0& 0\  0\  0& 4\  2\  1\   1& 4\  2\  1 \   0& 0\  0\  0\   0& 4\  2\  1\  1\  1 \cr
                \mathcal{S}_2& 0\  0\  0& 4\  2\  1& 0\  0\  0& 4\  2\  1\   0& 0\  0\  0 \   0& 4\  2\  1\   1& 4\  2\  1\  1\  0 \cr
                \mathcal{S}_3& 0\  0\  0& 0\  0\  0& 4\  2\  1& 0\  0\  0\   0& 4\  2\  1\   1& 4\  2\  1\   0& 4\  2\  1\   0\   0\cr
                \mathcal{S}_4& 4\  2\  1& 4\  2\  1& 0\  0\  0& 0\  0\  0\   0& 4\  2\  1\   0& 4\  2\  1\   1& 0\  0\  0\   0\   0\cr
                \mathcal{S}_5& 4\  2\  1& 0\  0\  0& 4\  2\  1& 4\  2\  1\   1& 0\  0\  0\   0& 4\  2\  1\   0& 0\  0\  0\   0\   0\cr
                \mathcal{S}_6& 0\  0\  0& 4\  2\  1& 4\  2\  1& 4\  2\  1\   0& 4\  2\  1\   1& 0\  0\  0\   0& 0\  0\  0\   0\   0\cr
                \mathcal{S}_7& 4\  2\  1& 4\  2\  1& 4\  2\  1& 0\  0\  0\   0& 0\  0\  0\   0& 0\  0\  0\   0& 4\  2\  1\   0\   0}
\end{align}
In order to prove that $\mathbf{C}'=2\mathbf{C}$ is a SQ-separable code, we only need to show that $\mathbf{C}$ is a separable code for an adder model. 

Let $\mathbf{w}\in{[2]}^n$ be a binary vector such that its $l^\textnormal{th}$ coordinate is equal to $1$ if the $l^\text{th}$ subject is defective and $0$ otherwise. In the adder model, the vector of test results equals $\mathbf{y}=\mathbf{C}\mathbf{w}$, which is a system of linear equations with $n$ variables and $m$ equations. Note that each set $\mathcal{S}_i$ corresponds to $q''+|\mathcal{S}_i|$ variables. 

We solve the system of equations in a recursive manner, by first solving for variables corresponding to $\mathcal{S}_m$, subtracting their effect on the syndrome and then solving for variables corresponding to $\mathcal{S}_{m-1}$, and so on. 

Returning to our example, we can solve for the variables corresponding to $\mathcal{S}_7$ as follows. Add all the equations corresponding to \emph{odd} subsets of $\mathcal{S}_{7}$. The result is an equation of the form
\begin{subequations}
\begin{align}\nonumber
\mathbf{s}_{\text{odd}_{7}}^T\mathbf{w}=\mathbf{y}(1)+\mathbf{y}(2)+\mathbf{y}(3)+\mathbf{y}(7),
\end{align}
where
\begin{align}\nonumber
\mathbf{s}_{\text{odd}_{7}}=(8\  \  4\  \  2\  \  8\  \  4\  \  2\  \  8\  \  4\  \  2\  \  8\  \  4\  \  2\  \  1\  \  8\  \  4\  \  2\  \  1\  \  8\  \  4\  \  2\  \  1\  \  16\  \  8\  \  4\  \  2\  \  1)^T.
\end{align}
\end{subequations}
Also, add all the equations corresponding to \emph{even} subsets of $\mathcal{S}_{7}$. The result is an equation of the form
\begin{subequations}
\begin{align}\nonumber
\mathbf{s}_{\text{even}_{7}}^T\mathbf{w}=\mathbf{y}(4)+\mathbf{y}(5)+\mathbf{y}(6),
\end{align}
where
\begin{align}\nonumber
\mathbf{s}_{\text{even}_{7}}=(8\  \  4\  \  2\  \  8\  \  4\  \  2\  \  8\  \  4\  \  2\  \  8\  \  4\  \  2\  \  1\  \  8\  \  4\  \  2\  \  1\  \  8\  \  4\  \  2\  \  1\  \  0\  \  0\  \  0\  \  0\  \  0)^T.
\end{align}
\end{subequations}
Since the first $21$ entries of  $\mathbf{s}_{\text{odd}_{7}}$ and $\mathbf{s}_{\text{even}_{7}}$ are identical, one has
\begin{align}\nonumber
(\mathbf{s}_{\text{odd}_{7}}-\mathbf{s}_{\text{even}_{7}})^T\mathbf{w}&=16\mathbf{w}(22)+8\mathbf{w}(23)+4\mathbf{w}(24)+2\mathbf{w}(25)+\mathbf{w}(26)\\\label{binrep}
&=\mathbf{y}(1)+\mathbf{y}(2)+\mathbf{y}(3)+\mathbf{y}(7)-\mathbf{y}(4)-\mathbf{y}(5)-\mathbf{y}(6).
\end{align}
The equation in~\eqref{binrep} provides a binary representation of the integer $\mathbf{y}(1)+\mathbf{y}(2)+\mathbf{y}(3)+\mathbf{y}(7)-\mathbf{y}(4)-\mathbf{y}(5)-\mathbf{y}(6)$. Therefore, the variables $\mathbf{w}(22)$, $\mathbf{w}(23)$, $\mathbf{w}(24)$, $\mathbf{w}(25)$, and $\mathbf{w}(26)$ are uniquely determined by the equation. Now, given these variables, one can add all the equations corresponding to odd and even subsets of $\mathcal{S}_6$ to similarly identify $\mathbf{w}(18)$, $\mathbf{w}(19)$, $\mathbf{w}(20)$, and $\mathbf{w}(21)$. This process can be applied iteratively until all the variables are uniquely determined. 

\begin{remark}
Construction~10 provides codes capable of identifying any number of defectives among $n=\kappa2^{\kappa-1}+q''(2^{\kappa}-1)$ subjects, using $m=2^\kappa-1$ experiments. It can be easily shown that the same approach applied for an arbitrary number of subjects. For a fixed value of $q''$, one can find the \emph{smallest} number $\kappa$ such that $n\leq \kappa2^{\kappa-1}+q''(2^{\kappa}-1)$. Removing the $(\kappa2^{\kappa-1}+q''(2^{\kappa}-1)-n)$ right most columns of $\mathbf{C}'$ in Construction~10 results in a SQ-separable code of size $n$ and length $m=2^\kappa-1$.
\end{remark}

%%%%%%%%%%%%%%%%%%%%%%%%%%%%%%%%%%
\subsection{Comparison of different SQGT code constructions}\label{sec:comparison}
The constructions described in this section were designed for a variety of code parameters and different modeling assumption for SQGT schemes. For example, the codes described in Constructions 1-4 and Construction 6 are capable of identifying an arbitrary number of defectives as long as $1\leq |\mathcal{D}|\leq d$, but require a non-binary alphabet; on the other hand, the codes described in Constructions 7 and 8 use binary test matrices, but are restricted by $|\mathcal{D}|$ being larger than a lower bound. Construction 9 introduced a family of codes capable of identifying an arbitrary number of defectives, i.e. $1\leq|\mathcal{D}|\leq n$. 

We summarized different properties of the constructions, including number of measurements, alphabet size, bounds on the number of defectives in Tables~\ref{table:compare1} and~\ref{table:compare2}. Since several constructions were based on classical binary $d$-disjunct and $d$-separable codes, we explicitly included these underlying building blocks (BBs) in ``Features''. In these cases, the number of tests $m$ as a function of $d$, $e$ and $n$, depends on the specific BBs used. Given that there are many different constructions for classical binary $d$-disjunct and $d$-separable codes available in the literature, a comprehensive survey of all possible SQGT codes would be well beyond the scope of this paper. We therefore focused on a small set of classical binary disjunct and separable codes well-documented in the literature, e.g.~\cite{DH00} and~\cite{DH06}. In addition, for cases where a reduction in the value of $m$ was achieved by using a particular method of concatenating the BBs, we emphasized such improvements by explicitly providing the parameters in the expression for $m$. For example, using a binary $d$-disjunct code of length $m_b$ and size $n_b$ as a BB in Construction 3, we constructed SQ-separable codes of length $m_b$ and size $Kn_b$. Since a typical bound for $m_b$ is $m_b=O\!\left(ed^2\log_2 (n_b/d)\right)$, we used $m=O\!\left(ed^2\log_2 (n/dK)\right)$ to emphasize that the number of allowed test subjects in the SQGT codes was increased by a factor of $K$. 

Construction 8 used BBTD matrices as BBs, the parameters of which depend on the underlying lossless condenser. Different forms of condensers were discussed in~\cite{C10}, and we refer an interested reader to this paper for more information. For an asymptotic bound on the number of measurements $m$ obtained via Construction 8, we used the parameters and condensers outlined in Construction M8 of~\cite[Table 1]{C10}. 

Note that in all the aforementioned code constructions, we assumed that $q$ is fixed and does not grow with $n$. For example, in Construction 9, we had $m=2^{\kappa}-1$ and $n=\kappa2^{\kappa-1}+q''(2^{\kappa}-1)$, where $q''=\left\lfloor\log_2\lfloor\frac{q-1}{\eta}\rfloor\right\rfloor$. Consequently, $n=1/2(m+1)\log_2(m+1)+q''m$, and for $q''=o(\log_2m)$, one has
\begin{align}\nonumber
\lim_{\kappa\rightarrow\infty}\frac{m}{2n/\log_2n}=\lim_{m\rightarrow\infty}\frac{\log_2\left(1/2(m+1)\log_2(m+1)+q''m\right)}{\log_2(m+1)+2q''+\frac{\log_2(m+1)}{m}}=\lim_{m\rightarrow\infty}\left(1+o(1)\right)=1. 
\end{align}
On the other hand, if $q=\eta {2}^{\kappa\alpha}$, for some fixed $\alpha>0$, similar calculations reveal that 
\begin{align}\nonumber
m\sim\left(\frac{2}{1+2\alpha}\right)\frac{n}{\log_2 n}.
\end{align}
In addition, if $q$ grows faster than exponential with $\kappa$ (or equivalently, $q''$ grows faster than logarithmic with $m$), then $m\sim\frac{1}{q''} n$. 

\begin{table}[t!]\centering
 \caption{A comparative summary of SQGT codes described in Constructions 1-5}
{\renewcommand{\arraystretch}{1.5}{\footnotesize\begin{tabular}{|l|c|c|c|c|c|}
			\hline 
			\textbf{Code} & Construction 1 & Construction 2 & Construction 3 & Construction 4 & Construction 5 \\
			\hline\hline
		
			\textbf{Parameters} & $[q;Q;\boldsymbol{\eta};(1\!:\!d);e]$& $[q;Q;\eta;(1\!:\!d);e]$ & $[q;Q;\eta;(1\!:\!d);e]$ & $[q;Q;\boldsymbol{\eta};(1\!:\!d);e]$&  $[q;Q;\eta;d;0]$\\[0pt]\hline
			\textbf{Type} & SQ-disjunct & SQ-disjunct & SQ-separable & SQ-separable & SQ-separable \\\hline
			\textbf{Thresholds} & Arbitrary & Equidistant & Equidistant & Arbitrary& Equidistant \\\hline			
			\textbf{Construction} & Explicit  & Probabilistic & Explicit & Explicit & Explicit \\\hline
			\textbf{Num. Tests} & $O(ed^2\log_2 \frac{n}{d})$& $O\!\left(\frac{\pi_1}{\pi_I}\!\left(d^2\log_2 \frac{n}{d}\!+\!2ed\right)\!\right)$ & $O\!\left(ed^2\log_2 \frac{n}{dK}\right)$& $O(ed^2\log_2 \frac{n}{d})$& $O(d\log_{q'}n)$ \\\hline
			 \textbf{Features} & Efficient decoder of & Efficient decoder of & Efficient decoder of & BB: separable (CGT)& Number theoretic \\
				                       & complexity $O(mn)$,&complexity $O(mn)$, &complexity $O(mn)$, & & (Bose-Chowla)  \\
						            & BB: disjunct (CGT) & &BB: disjunct (CGT) &  &  \\\hline

			\end{tabular}}}
			\label{table:compare1}
%			\vspace*{-15pt}	
\end{table}

\begin{table}[t!]\centering
 \caption{A comparative summary of SQGT codes described in Constructions 6-9}
{\renewcommand{\arraystretch}{1.5}{\footnotesize\begin{tabular}{|l|c|c|c|c|}
			\hline 
			\textbf{Code} & Construction 6 & Construction 7 & Construction 8 & Construction 9  \\
			\hline\hline
		
			\textbf{Parameters}  & $[q;Q;\eta;(1\!:\!d);e]$ & $[2;Q;\boldsymbol{\eta};(\eta_{\alpha}\!:\!d);e]$&$[2;Q;\boldsymbol{\eta};(\eta_{\alpha}\!:\!d);e]$ & $[q;Q;\eta;(1\!:\!n);0]$  \\[0pt]\hline
			\textbf{Type}  & SQ-separable  & SQ-separable & SQ-separable & SQ-separable  \\\hline
			\textbf{Thresholds}  & Equidistant  & Arbitrary &  Arbitrary & Equidistant  \\\hline			
			\textbf{Construction}    & Explicit  & Probabilistic & Explicit & Explicit  \\\hline
			\textbf{Num. Tests}  &$O\!\left(ed^2\log_2 \frac{n}{dK}\right)$  & $O_{\!e,\!\boldsymbol{\eta}}(d^2\log_2d\log_2\frac{n}{d})$ & $O_e\!\left(\frac{\eta_1-1}{\eta_\alpha-1}d^3\log_2 d \log^3(\log n) \log n\right)$& $\sim\frac{2n}{\log_2 n}$\\\hline
			\textbf{Features} & BB: separable (CGT),& Binary test matrix & Binary test matrix, & No restriction on $d$,\\
			 & BB: separable (QGT) & & Based on strong lossless condensers& Efficient decoder\\\hline			
			\end{tabular}}}
			\label{table:compare2}

%			\vspace*{-15pt}	
\end{table}

%%%%%%%%%%%%%%%%%%%%%%%%%%%%%%%%%%%%%%%%%%%%%%%%%%%%%%%%%
\section{Belief Propagation Decoders for SQGT} \label{sec:bp}

In the previous sections, we introduced different codes for SQGT. SQ-disjunct codes, as well as the codes described in Construction 3  have a decoding procedure with complexity $O(mn)$. SQ-separable codes described in Construction 9 have an iterative decoding procedure, outlined in the proof of Theorem~\ref{const9}. On the other hand, decoders for CGT were extensively investigated in the literature (e.g.~\cite{M10}-\cite{LCJS12}). Although these algorithms perform well for CGT schemes, due to the more complicated nature of SQGT, their direct application to SQGT does not appear to be plausible. Hence, we still do not know of any efficient \emph{universal} decoding method for SQ-separable codes.

One observation is in place: since most proposed SQGT codes are sparse, methods based on belief propagation (BP)~\cite{KFL01} emerge as viable decoding options. In particular, we focus on BP decoders suitable for SQGT codes based on probabilistic constructions (such as Constructions 2 and 8). The theoretical guarantees for these codes hold in the asymptotic domain, and when the number of subjects is small, these guarantees may not apply. 
Nevertheless, in what follows, we show that BP decoders perform reasonably well even for a small number of subjects and large coding rates and their performance may be further improved by tailoring the SQGT constructions to the decoder. 

BP is an iterative message passing algorithm for inference on graphical models, and it is centered around calculating the marginal distributions of the variables corresponding to the vertices of the underlying graph. BP decoding for binary disjunct codes was originally proposed by one of the authors in~\cite{HM08}. Later on, BP decoding was also considered in~\cite{SJ10} for CGT decoding. Motivated by these two methods, we propose a BP decoder for SQGT, which performs an approximate bitwise maximum a posteriori (MAP) decoding of SQGT codes in the presence of errors. Note that BP decoding can be used for different error models and assumptions; however, in the rest of this section, we focus on the following model.
 
Consider a SQGT model with thresholds $\boldsymbol{\eta}$ as defined in Section~\ref{sec:model}. Assume that each subject is defective with probability $d/n$ independent of other subjects. Note that one consequence of this assumption is that the number of defectives $|\mathcal{D}|$ is a random variable. Consider a set of $n$ subjects and let $W\in{[2]}^n$ be a random vector representing the incidence vector of defectives. Also, let $\mathbf{w}_t\in{[2]}^n$ denote the true incidence vector of defectives, i.e. the realization of $W$ that we want to reconstruct. Also, let $\mathbf{C}\in{[q]}^{m\times n}$ and $\mathbf{z}\in{[Q]}^m$ be the test matrix and the observed vector of (possibly) erroneous test results, respectively. 

%Throughout this section and the next, we use upper-case letters to denote random matrices and random vectors; we use bold-face upper-case and bold-face lower-case letters to denote specific realizations of random matrices and random vectors, respectively. 

The messages passed in a BP decoder depend on the message error model. We focus on one simple substitution error model for the test results. Let $Y\in {[Q]}^m$ and $Z\in {[Q]}^m$ be the random vectors corresponding to the error-free test results and the erroneous test results, respectively. We model the effect of false positives and false negatives using two probabilities, $\gamma_p$ and $\gamma_n$, respectively. In other words, for the $t^{\text{th}}$ test, if $Y(t)\in\{1,2,\dots,Q-2\}$ then $Z(t)=Y(t)$ with probability $1-\gamma_p-\gamma_n$, $Z(t)=Y(t)+1$ with probability $\gamma_p$, and $Z(t)=Y(t)-1$ with probability $\gamma_n$. If $Y(t)=0$ then $Z(t)=Y(t)$ with probability $1-\gamma_p$ and $Z(t)=Y(t)+1$ with probability $\gamma_p$. 
Finally, if $Y(t)=Q-1$, then $Z(t)=Y(t)$ with probability $1-\gamma_n$ and $Z(t)=Y(t)-1$ with probability $\gamma_n$. BP decoders for other substitution error models can be designed using similar arguments.

For the $i^{\text{th}}$ subject, we consider a bitwise MAP estimator, i.e. 
\begin{align}\label{eq:MAP}
\hat{\mathbf{w}}_{\text{MAP}}(i)=\arg\max_{\mathbf{w}(i)\in\{0,1\}} P_{W(i)|Z}\left(\mathbf{w}(i)|\mathbf{z}\right),
\end{align}
where $P_{W(i)|Z}(\cdot|\cdot)$ denotes the conditional probability distribution of $W(i)$ given $Z$. Henceforth, we use $P(\cdot)$ as a generic symbol for probability distribution and for simplicity, do not explicitly display the random variables in the subscript of $P(\cdot)$. 

Using the definition of conditional probability, $P\left(\mathbf{w}(i)|\mathbf{z}\right)=\frac{P\left(\mathbf{z},\mathbf{w}(i)\right)}{P\left(\mathbf{z}\right)}.$ Since the maximization in~\eqref{eq:MAP} is performed over different values of $\mathbf{w}(i)$, the value of $P\left(\mathbf{z}\right)$ does not affect $\hat{\mathbf{w}}_{\text{MAP}}(i)$. For a function $f(\mathbf{w}):{[2]}^n\mapsto\mathbb{R}$, let the sum of $f(\mathbf{w})$ over all configurations of the variables \emph{other than} $\mathbf{w}(i)$ be denoted by $\sum_{\sim\mathbf{w}(i)}f(\mathbf{w})$. In this case, one has
\begin{align}\nonumber
\hat{\mathbf{w}}_{\text{MAP}}(i)&=\arg\max_{\mathbf{w}(i)\in\{0,1\}} P\left(\mathbf{w}(i)|\mathbf{z}\right)\\\nonumber
&=\arg\max_{\mathbf{w}(i)\in\{0,1\}} {P\left(\mathbf{z},\mathbf{w}(i)\right)}\\\label{eq:MAP2}
&=\arg\max_{\mathbf{w}(i)\in\{0,1\}}\sum_{\sim \mathbf{w}(i)} {P\left(\mathbf{z},\mathbf{w}\right)},
\end{align}
where the last equality follows by marginalizing out all the $\mathbf{w}(j)$'s, $j\neq i$, from ${P\left(\mathbf{z},\mathbf{w}\right)}$. 

Since the result of the tests are independent of each other conditioned on $W=\mathbf{w}$, it holds that 
${P\left(\mathbf{z}|\mathbf{w}\right)}=\prod_{t=1}^{m}P\left(\mathbf{z}(t)|\mathbf{w}\right)$. Substituting this equality in~\eqref{eq:MAP2} yields
\begin{align}\nonumber
\hat{\mathbf{w}}_{\text{MAP}}(i)&=\arg\max_{\mathbf{w}(i)\in\{0,1\}}\sum_{\sim \mathbf{w}(i)} {P\left(\mathbf{z},\mathbf{w}\right)}\\\nonumber
&=\arg\max_{\mathbf{w}(i)\in\{0,1\}}\sum_{\sim \mathbf{w}(i)} {P\left(\mathbf{z}|\mathbf{w}\right)} {P\left(\mathbf{w}\right)}\\\nonumber
&=\arg\max_{\mathbf{w}(i)\in\{0,1\}}\sum_{\sim \mathbf{w}(i)}\left[ {\prod_{t=1}^{m}P\left(\mathbf{z}(t)|\mathbf{w}\right)} {P\left(\mathbf{w}\right)}\right]\\\nonumber
&=\arg\max_{\mathbf{w}(i)\in\{0,1\}}\sum_{\sim \mathbf{w}(i)}\left[\prod_{t=1}^{m}P\left(\mathbf{z}(t)|\mathbf{w}\right)\prod_{j=1}^{n}P(\mathbf{w}(j))\right],
\end{align}
where the last equality follows since we assumed that the event that a subject is defective is independent of the even of other subjects being defective. Finally, given that each subject is defective with probability $d/n$, one obtains
\begin{align}\label{MAP}
\hat{\mathbf{w}}_{\text{MAP}}(i)=\arg\max_{\mathbf{w}(i)\in\{0,1\}}\sum_{\sim \mathbf{w}(i)}\left[\prod_{t=1}^{m}P\left(\mathbf{z}(t)|\mathbf{w}\right)\prod_{j=1}^{n}\left(\frac{d}{n}\  \mathbb{I}\left(\mathbf{w}(j)=1\right)+\left(1-\frac{d}{n}\right)\  \mathbb{I}\left(\mathbf{w}(j)=0\right)\right)\right],
\end{align}
where $\mathbb{I}\left(\cdot\right)$ denotes the indicator function, equal to $1$ if the statement in the brackets holds, and equal to $0$ otherwise. 

Using~\eqref{MAP}, we can form a factor graph that corresponds to the bitwise MAP estimator with $n$ variable nodes and $m$ factor nodes; a factor node corresponding to test $t$ is only connected to variable nodes corresponding to subjects present in the $t^{\text{th}}$ test. Similarly, a variable node corresponding to the $i^{\text{th}}$ subject is only connected to the factor nodes corresponding to the tests in which the $i^{\text{th}}$ subject is used. As a result, the complexity of the BP decoder depends on the sparsity of the code matrix, $\mathbf{C}$. Designing specialized sparse SQGT codes amenable to BP decoding is a problem we plan to address in a companion paper. 

Let $\mathcal{N}(t)$ denote the neighbors of the factor node corresponding to test $t$ in the factor graph. Also, let $\mathcal{N}(i)$ denote the neighbors of the variable node corresponding to the $i^{\text{th}}$ subject. Let $\chi_{i\rightarrow t}^{(l)}(\mathbf{w}(i))$ denote the message from the $i^{\text{th}}$ variable node to the $t^{\text{th}}$ factor node in the ${l}^\text{th}$ iteration, $1\leq l\leq L$. Similarly, let $\hat{\chi}_{t\rightarrow i}^{(l )}(\mathbf{w}(i))$ denote the message at the ${l}^\text{th}$ iteration from the $t^{\text{th}}$ factor node to the $i^{\text{th}}$ variable node. The BP message update rules for finding the marginal distributions of each subject according to the MAP estimator of~\eqref{MAP} take the form:
\begin{align}
\chi_{i\rightarrow t}^{(l+1)}(\mathbf{w}(i))\propto\left(\frac{d}{n}\  \mathbb{I}(\mathbf{w}(i)=1)+\left(1-\frac{d}{n}\right)\  \mathbb{I}(\mathbf{w}(i)=0)\right)\prod_{\tau\in\mathcal{N}(i)\backslash\{t\}}\hat{\chi}_{\tau\rightarrow i}^{(l)}(\mathbf{w}(i)),
\end{align}
and 
\begin{align}\label{message2}
\hat{\chi}_{t\rightarrow i}^{(l+1)}(\mathbf{w}(i))\propto \sum_{\sim \mathbf{w}(i)}\left[P\left(\mathbf{z}(t)|\mathbf{w}\right)\prod_{j\in\mathcal{N}(t)\backslash\{i\}}{\chi}_{j\rightarrow t}^{(l)}(\mathbf{w}(j))\right],
\end{align}
where $\propto$ denotes ``equal up to a multiplicative constant''. For an in-depth explanation regarding message updates for marginals of a distribution, we refer the interested reader to~\cite{KFL01} and the references therein.

In order to get an explicit form for the message updates, we need to calculate the term $P\left(\mathbf{z}(t)|\mathbf{w}\right)$ in~\eqref{message2} for different values of $\mathbf{z}(t)$. For this purpose, let $\omega_i:=\sum_{l=1,l\neq i}^n\mathbf{w}(l)\mathbf{C}(t,l)$. 
Then, one has
\begin{align}\nonumber
P_{Z(t)|W}\left(0|\mathbf{w}\right)=
\left\{
     \begin{array}{ll}
         \gamma_n\   \mathbb{I}\left(\eta_1\leq \omega_i<\eta_2\right)+(1-\gamma_p)\  \mathbb{I}(\omega_i<\eta_1),   & \textnormal{if}\  \  \  \mathbf{w}(i)=0, \\
         \\
              \gamma_n\   \mathbb{I}\left(\eta_1-\mathbf{C}(t,i)\leq \omega_i<\eta_2-\mathbf{C}(t,i)\right)   & \textnormal{if} \  \  \  \mathbf{w}(i)=1,\\
            \  \  \  \  \  \  +(1-\gamma_p)\  \mathbb{I}(\omega_i<\eta_1-\mathbf{C}(t,i)),   &
                   \end{array}
   \right.
\end{align}
\begin{align}\nonumber
P_{Z(t)|W}\left(Q-1|\mathbf{w}\right)=
\left\{
     \begin{array}{ll}
        (1-\gamma_n)\   \mathbb{I}\left(\eta_{Q-1}\leq \omega_i<\eta_{Q}\right)+\gamma_p\  \mathbb{I}(\eta_{Q-2}\leq \omega_i<\eta_{Q-1}),  & \textnormal{if}\  \  \  \mathbf{w}(i)=0, \\
        & \\
             (1-\gamma_n)\   \mathbb{I}\left(\eta_{Q-1}-\mathbf{C}(t,i)\leq \omega_i<\eta_{Q}-\mathbf{C}(t,i)\right)   & \textnormal{if} \  \  \  \mathbf{w}(i)=1,\\
             \  \  \  \  \  \  +\gamma_p\  \mathbb{I}(\eta_{Q-2}-\mathbf{C}(t,i)\leq \omega_i<\eta_{Q-1}-\mathbf{C}(t,i)),  &
                   \end{array}
   \right.
\end{align}
and for $\mathbf{z}(t)=r$ and $r\in\{1,2,\dots,Q-2\}$, one has
\begin{align}\nonumber
P_{Z(t)|W}\left(r|\mathbf{w}\right)=
\left\{
     \begin{array}{ll}
        (1-\gamma_n-\gamma_p)\   \mathbb{I}\left(\eta_{r}\leq \omega_i<\eta_{r+1}\right),  & \textnormal{if}\  \  \  \mathbf{w}(i)=0, \\
        \  \  \  \  \  \  +\gamma_p\  \mathbb{I}(\eta_{r-1}\leq \omega_i<\eta_{r})+\gamma_n\  \mathbb{I}\left(\eta_{r+1}\leq \omega_i<\eta_{r+2}\right), & \\
        \\
            (1-\gamma_n-\gamma_p)\   \mathbb{I}\left(\eta_{r}-\mathbf{C}(t,i)\leq \omega_i<\eta_{r+1}-\mathbf{C}(t,i)\right)   & \textnormal{if} \  \  \  \mathbf{w}(i)=1.\\
            \  \  \  \  \  \  \  \  \  \  +\gamma_p\  \mathbb{I}(\eta_{r-1}-\mathbf{C}(t,i)\leq \omega_i<\eta_{r}-\mathbf{C}(t,i)) & \\
            \  \  \  \  \  \  \  \  \  \  +\gamma_n\  \mathbb{I}\left(\eta_{r+1}-\mathbf{C}(t,i)\leq \omega_i<\eta_{r+2}-\mathbf{C}(t,i)\right). &
                   \end{array}
   \right.
\end{align}
Using standard BP message independence assumptions, the marginal distribution of the $i^\text{th}$ subject after the $L^\text{th}$ iteration may be written as:
\begin{align}\nonumber
P^{^{(L)}}_{W(i)|Z}\left(\mathbf{w}(i)|\mathbf{z}\right)\propto\left(\frac{d}{n}\  \mathbb{I}(\mathbf{w}(i)=1)+\left(1-\frac{d}{n}\right)\  \mathbb{I}(\mathbf{w}(i)=0)\right)\prod_{\tau\in\mathcal{N}(i)}\hat{\chi}_{\tau\rightarrow i}^{(L)}(\mathbf{w}(i)).
\end{align}
Upon computing the marginals, the set of defectives may be determined based on the following two methods. In the first method,
\begin{align}\label{method1}
\hat{\mathcal{D}}=\left\{\  i\  : \  P^{^{(L)}}_{W(i)|Z}\left(1|\mathbf{z}\right)>P^{^{(L)}}_{W(i)|Z}\left(0|\mathbf{z}\right)\right\},
\end{align}
while in the second method
\begin{align}\label{method2}
\hat{\mathcal{D}}=\left\{\  i\  : \  \text{$S_i$ has one of the $d$ largest $P^{^{(L)}}_{W(i)|Z}\left(1|\mathbf{z}\right) $}\right\}.
\end{align}
Note that the complexity of this BP decoder can be further reduced by adapting approaches such as the ones described in the context of $q$-ary BP decoding in~\cite{HDYW06}-\cite{DF07}, which will be discussed elsewhere. 

\begin{figure}
\includegraphics[width=0.6\textwidth]{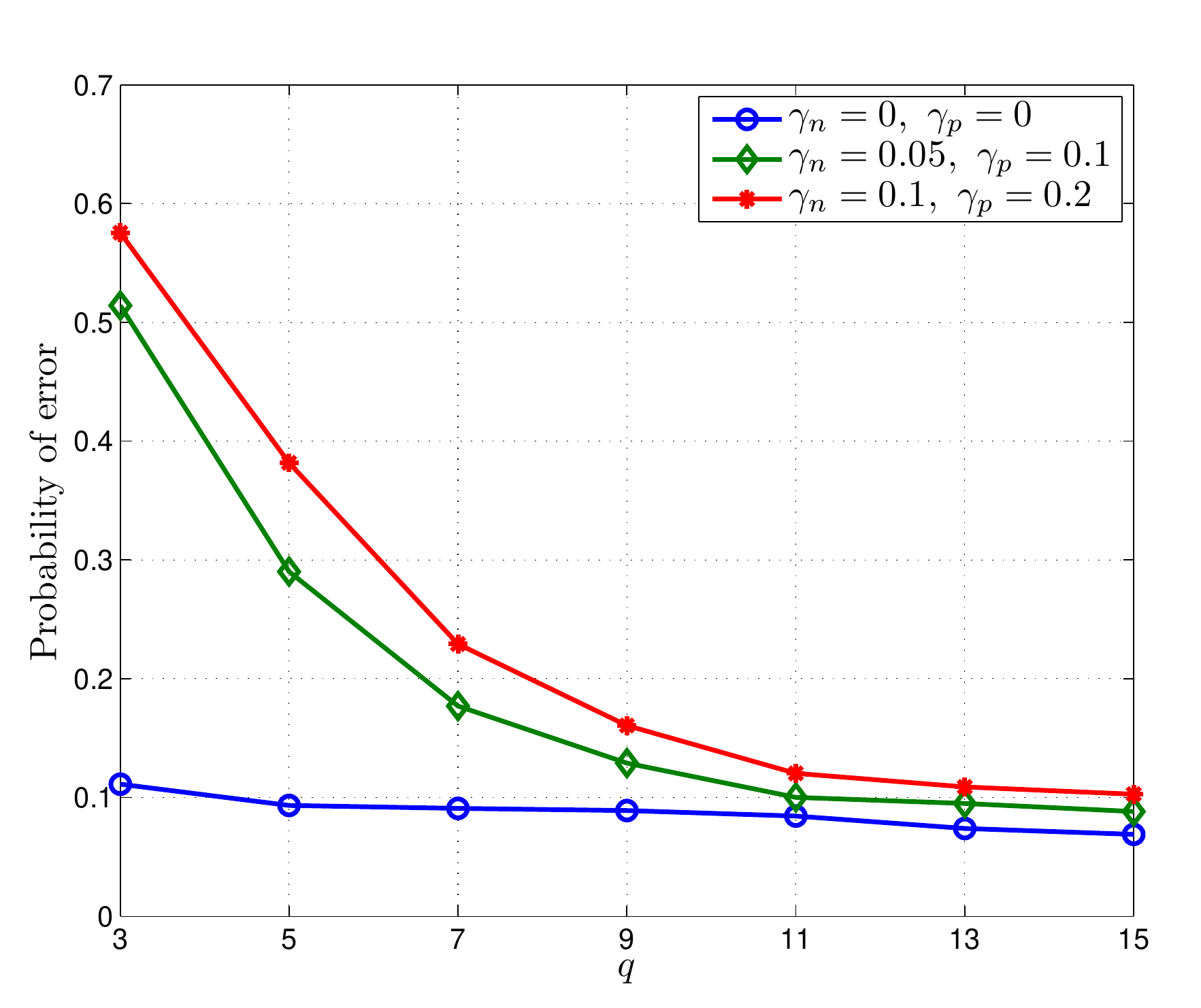}
\centering
\caption{Probability of error as a function of the test matrix alphabet size $q$, for different choices of noise parameters. In the model, we fixed $\eta=2$, $n=100$, $d=15$, and $m=50$. }
\label{fig:BPresult1}
\end{figure}

\begin{figure}
\includegraphics[width=0.6\textwidth]{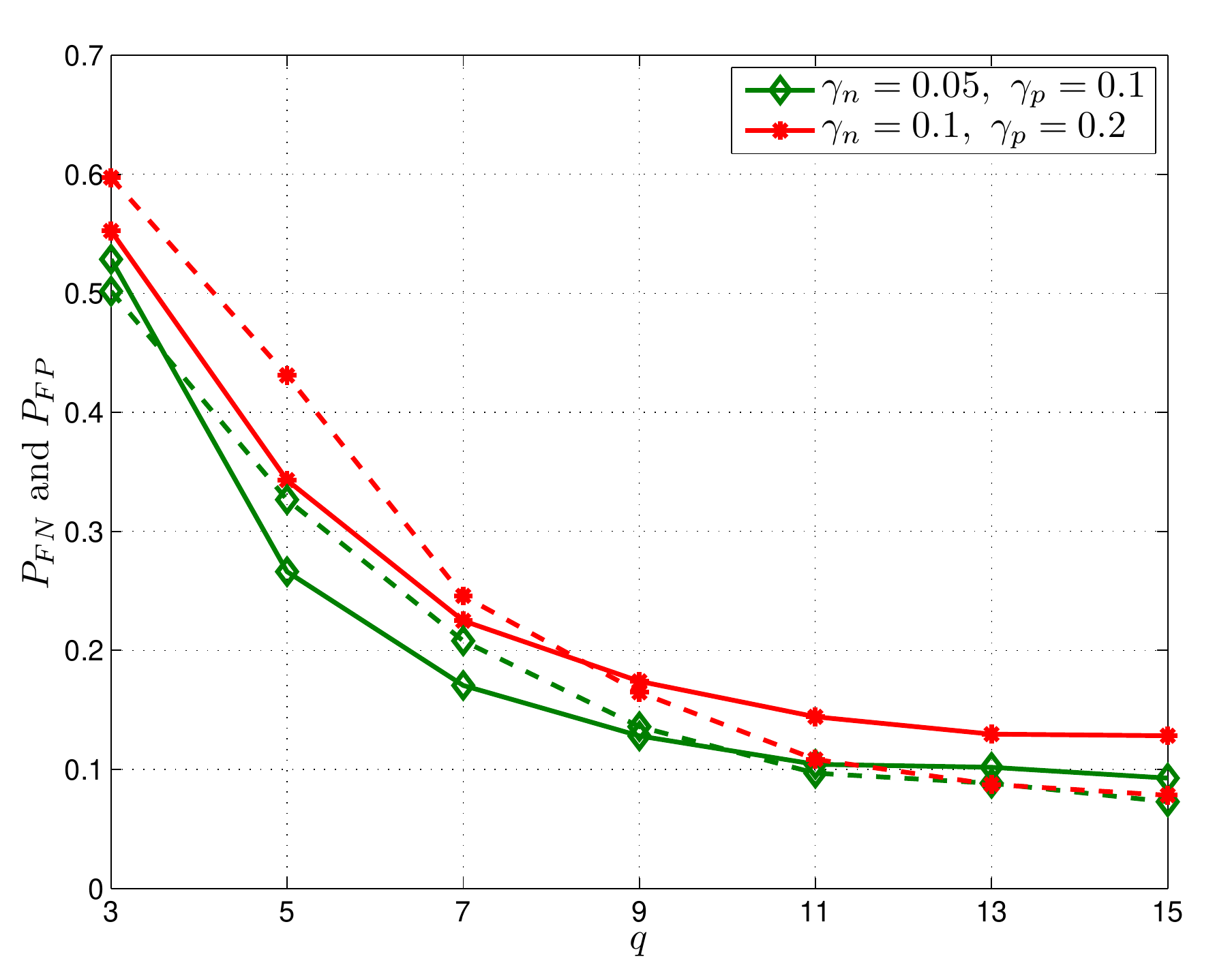}
\centering
\caption{Probability of false negatives and false positives as a function of the test matrix alphabet size $q$, for different choices of noise parameters. The solid lines represent the probability of false negatives, while the dashed lines represent the probability of false positives. We fixed $\eta=2$, $n=100$, $d=15$, and $m=50$.}
\label{fig:BPresult2}
\end{figure}

For demonstrative purposes, we applied the BP algorithm to an equidistant SQGT model with $\eta=2$. We used Construction 2 to generate codes with $n=100$ and $d=15$, which represent reasonable parameter choices for the application at hand. In Fig.~\ref{fig:BPresult1} we plotted the probability of error, $P_{e}$, as a function of $q$ for different values of $\gamma_p$ and $\gamma_n$, when $m=50$. We generated $400$ different sets of defectives (trials) for each choice of $q$, and fixed the number of iterations in the BP algorithm to $L=20$. The set of defectives was obtained using~\eqref{method2}. Fig.~\ref{fig:BPresult2} shows the performance of the BP algorithm in a similar setting when~\eqref{method1} was used to obtain the set of defectives. To keep the waterfall curves sufficiently uncluttered, we only reported on noisy SQGT performance. Note that the probability of false negatives, $P_{FN}$, is defined as the probability that a defective is not detected, while the probability of false positives, $P_{FP}$, is defined as the probability that a non-defective subject is detected as defective. Note that in method~\eqref{method2}, $P_e=P_{FN}=P_{FP}$.

As may be seen from the simulation results, there is a clear advantage of using codes with $q \geq 3$ from the perspective of BP decoding in the presence of errors. Unfortunately, this effect is accompanied by an increase in the complexity of non-binary BP decoding, which may be mitigated by applications of the aforementioned methods of~\cite{HDYW06}-\cite{DF07}. One may also notice that the decoding error probability of the BP decoder for the codes with the considered parameters remains bounded above a value close to $0.1$. 
We believe that this phenomenon is not a result of the unsuitability of BP decoding in SQGT, but rather a consequence of the fact that testing matrices constructed in the paper were not optimized with respect to the requirements of loopy BP. Furthermore, the high probability of error may also be attributed to the fact that the random codes were generated for parameters that are not in the range of values that guarantee high probability for the SQ disjunctness property\footnote{Testing the SQ disjuctness property for large matrices is computationally demanding and we did not attempt to determine the exact parameters of the SQGT code through simulation.}. Particularly, in Construction 2, the asymptotic guarantees were results of an upper bound on the probability that $\mathbf{C}$ is not a $[q;Q;\eta;(1\!:\!d);0]$-SQ-disjunct code. This bound took the form 
\begin{align}\nonumber
\Pro\left(\text{$\mathbf{C}$ is not $[q;Q;\eta;(1\!:\!d);0]$-SQ-disjunct}\right)\leq P_F={n\choose d+1}(d+1)(1-\pi_I)^m,
\end{align}
where $\pi_I$ was the probability of ``success'' of a row, as defined in the proof of Construction 2. However, as an example, when $n=100$, $m=50$, $\eta=2$, $q=11$, and $d=15$, this upper bound is larger than $1$, i.e. $P_F>1$, and we can therefore not guarantee that the code considered for these parameters is $[q;Q;\eta;(1\!:\!d);0]$-SQ-disjunct with high probability. A probability of error of approximately $0.15$ for $q\geq 11$ shows that even though the considered codes may not satisfy the distinctness property, one is still able to correctly identify the set of defectives with empirical probability approximately $0.85$, which is sufficiently high for the described genotyping applications. 

\begin{figure}
\includegraphics[width=0.6\textwidth]{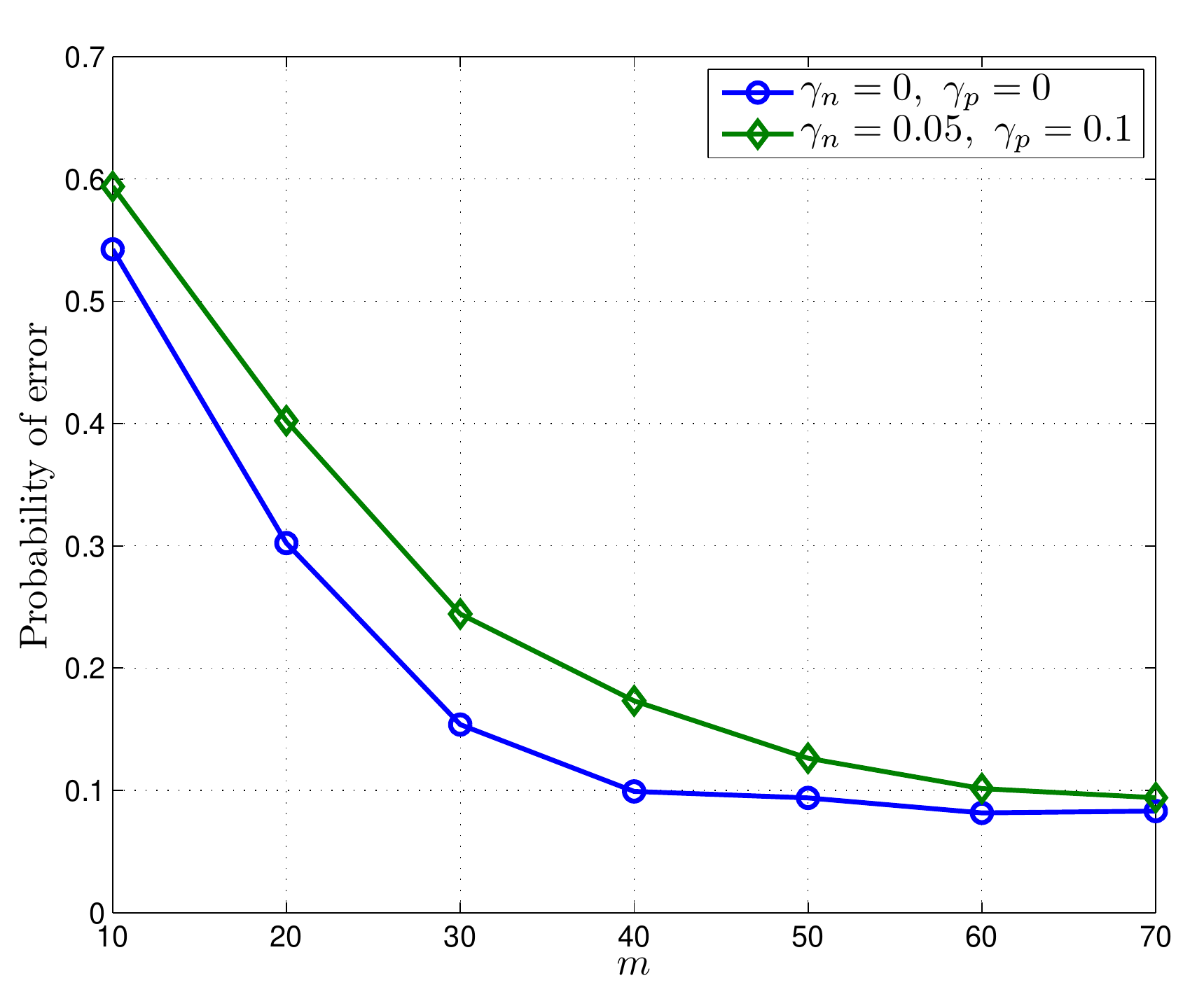}
\centering
\caption{Probability of error as a function of $m$ for different noisy and noise-free scenarios. In this model we fixed $\eta=2$, $n=100$, $d=15$, and $q=11$.}
\label{fig:BPresult3}
\end{figure}

\begin{figure}
\includegraphics[width=0.6\textwidth]{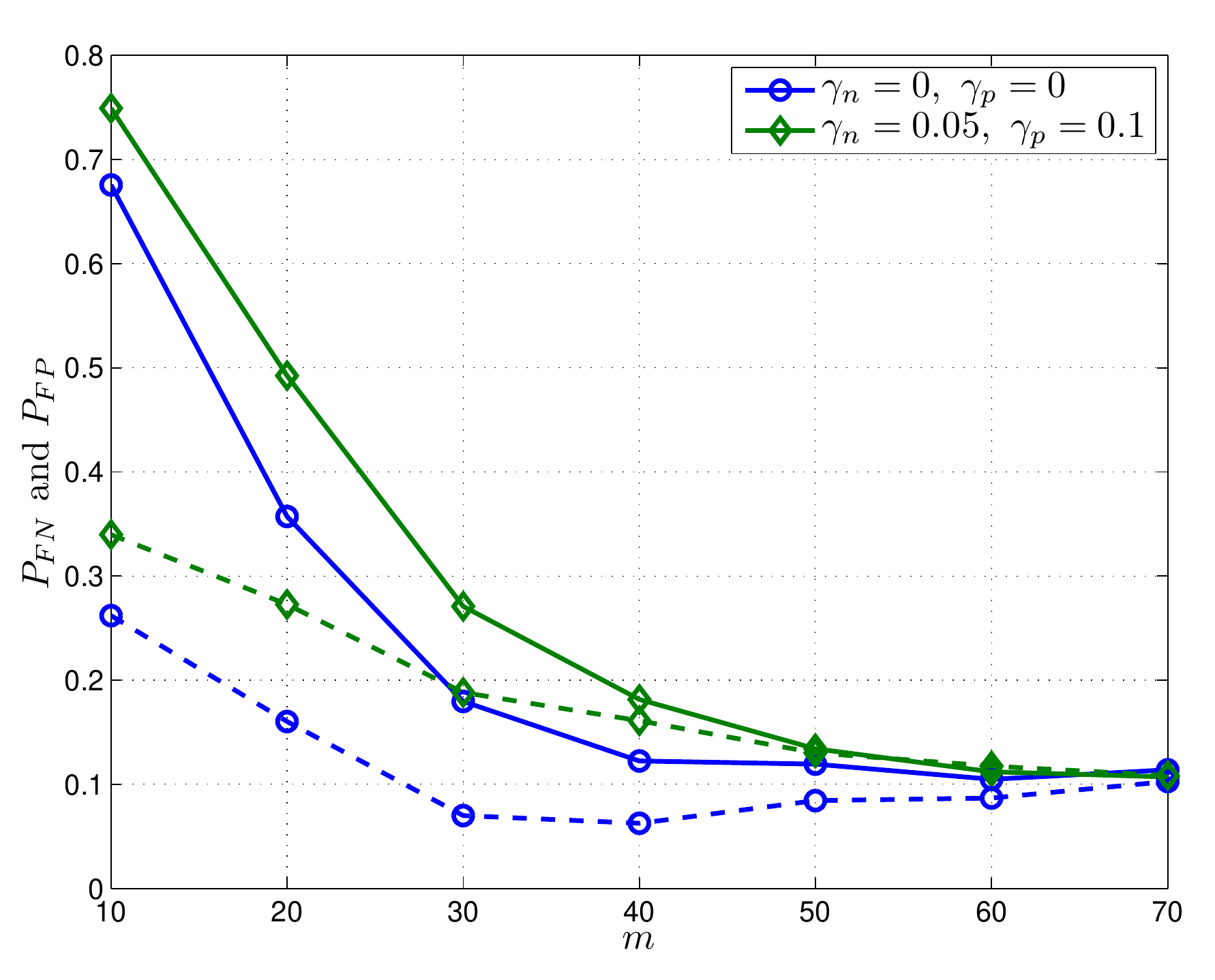}
\centering
\caption{Probability of false negatives and false positives as a function of $q$ for different noise parameters. The solid lines represent the probability of false negatives while the dashed lines represent the probability of false positives. In this model, we fixed $\eta=2$, $n=100$, $d=15$, and $q=11$.}
\label{fig:BPresult4}
\end{figure}

In order to demonstrate the effect of $m$ on the performance of the algorithm, we applied the BP algorithm on an equidistant SQGT model with $\eta=2$. Using Construction 2, we generated codes with $n=100$, $d=15$, and $q=11$. Fig.~\ref{fig:BPresult3} shows the probability of error as a function of $m$  for noisy and noise-free scenarios when~\eqref{method2} was used to obtain the set of defectives. For each $m$, the BP algorithm was applied on $400$ random codes and terminated with no more than $L=20$ iterations. Similarly, Fig.~\ref{fig:BPresult4} shows the probabilities of false negatives and false positives when~\eqref{method1} was used to find the set of defectives. 

%%%%%%%%%%%%%%%%%%%%%%%%%%%%%%%%%%%%%%%%%%%%%%%%%%%%%%%%%
\section{Capacity of SQGT}\label{sec:informationtheory}

In Section~\ref{sec:construction}, we described explicit and probabilistic constructions for SQGT test matrices capable of identifying defectives with \emph{zero probability of error}. On the other hand, a natural question to ask is what happens in an information-theoretic setting, where one is interested in identifying the defectives with an average probability of error that converges to zero. The answer to this question is closely related to Shannon's random coding theory. In particular, it is well-known that different models of group testing may be viewed as special instances of a multiple access channel (MAC). Using this connection, asymptotic information theoretic bounds were obtained on the number of tests needed to approach zero probability of error, see~\cite{M78}, \cite{MM80}, \cite{D04}, \cite{AS12}, \cite{SJ10}. Using these ideas, one can define the ``capacity'' of a group testing scheme similar to the capacity of a communication channel. 

Our goal in this section is not to derive new bounds on the number of tests for generalized MAC models, as substantial work was already performed for a number of different MAC models. Rather, we use the existing results and adapt them to the framework of SQGT while introducing novel ideas about \emph{optimal threshold selection} for the decimator. In other words, we introduce a problem from the area of source coding into the group testing framework -- the problem of designing the best quantization scheme for adder channels. Although one may argue that in genotyping applications the thresholds are usually fixed by the system design and architecture, it still remains an interesting theoretical problem to find the optimal thresholds when their number is fixed to some small value.

Although almost all information-theoretic approaches rely on using probabilistic constructions of \emph{binary} test matrices for CGT, the generalization of these methods to non-binary test matrices in a SQGT model is straightforward.  ``Probabilistic construction'' in these derivations refers to the test matrices being chosen in an i.i.d. manner, with probability of a subject being included in a test equal to $p$. The main difference in analysis arises in the form of the mutual information used to express the necessary and sufficient conditions on the number of tests that guarantee the average probability of error converges to zero.  

%Throughout this section, we use upper-case letters to denote random matrices and random vectors; we use bold-face upper-case and bold-face lower-case letters to denote specific realizations of random matrices and random vectors, respectively. 

Consider an SQGT model with parameters defined in Section~\ref{sec:model}. Assume that the test matrix is chosen probabilistically such that the sample amount of each subject in each test follows an i.i.d distribution $P_T$ over a $q$-ary alphabet. Let $C\in{[q]}^{m\times n}$ denote the random test matrix and let $\mathbf{C}$ denote a specific realization of $C$. Let $d$ denote the number of defectives, and let $\mathcal{P}_{d}(\llbracket n\rrbracket)$ be the set of all $d$-subsets of $\llbracket n\rrbracket$ with cardinality $|\mathcal{P}_{d}(\llbracket n\rrbracket)|={n\choose d}$. Assume that the set of defectives $\mathcal{D}$ is chosen uniformly at random from $\mathcal{P}_{d}(\llbracket n\rrbracket)$, independent of ${C}$, such that $\forall \tilde{\mathcal{D}}\in\mathcal{P}_{d}(\llbracket n\rrbracket)$, $P(\mathcal{D}=\tilde{\mathcal{D}}|C=\mathbf{C})=P(\mathcal{D}=\tilde{\mathcal{D}}|\mathbf{C})=P(\tilde{\mathcal{D}})=1/{n\choose d}$.

Let $Z\in{[Q]}^m$ denote the random vector of test results, and let $\mathbf{z}$ denote a specific realization of $Z$. Let $P\left(\mathbf{z}|\mathbf{C},\tilde{\mathcal{D}}\right)$ be the probability of observing $Z=\mathbf{z}$ given $C=\mathbf{C}$ and given the set of defectives $\mathcal{D}=\tilde{\mathcal{D}}$; this conditional probability may be viewed as the transition probability of the SQGT channel. Note that since the result of tests only depend on the codewords corresponding to the defectives, one has $P\left(\mathbf{z}|\mathbf{C},\tilde{\mathcal{D}}\right)=P\left(\mathbf{z}|\mathbf{C}_{\tilde{\mathcal{D}}},\tilde{\mathcal{D}}\right)$, where $\mathbf{C}_{\tilde{\mathcal{D}}}\in [q]^{m\times d}$ is the matrix formed using the columns of $\mathbf{C}$ indexed by $\tilde{\mathcal{D}}$. We assume that this channel is memoryless and that given the test matrix and the set of defectives, the test results are independent, i.e.,
\begin{align}\nonumber
P\left(\mathbf{z}|\mathbf{C},\tilde{\mathcal{D}}\right)=P\left(\mathbf{z}|\mathbf{C}_{\tilde{\mathcal{D}}},\tilde{\mathcal{D}}\right)=\prod_{k=1}^m P\left(\mathbf{z}(k)|\mathbf{t}_{\tilde{\mathcal{D}},k},\tilde{\mathcal{D}}\right).
\end{align}
Here, $\mathbf{z}(k)$ denotes the possibly erroneous result of the $k^{\text{th}}$ test, and $\mathbf{t}_{\tilde{\mathcal{D}},k}$ is a row vector of length $d$ corresponding to the sample amount of the defectives in the $k^{\text{th}}$ test. In other words, $\mathbf{t}_{\tilde{\mathcal{D}},k}$ is the $k^{\text{th}}$ row of $\mathbf{C}_{\tilde{\mathcal{D}}}$. Note that we implicitly made the above conditional independence assumptions in our derivations of the BP decoding method.

\begin{figure}
        \centering
        \begin{subfigure}[b]{0.3\textwidth}
                \centering
                \includegraphics[width=\textwidth]{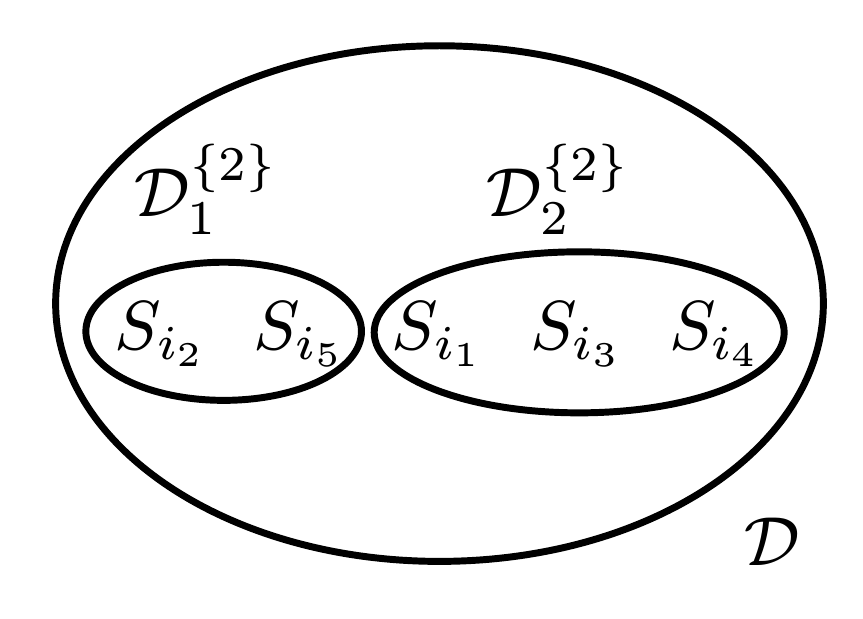}
                \caption{}
                \label{subfig:partition}
        \end{subfigure}%
        \qquad\qquad\qquad %add desired spacing between images, e. g. ~, \quad, \qquad etc. 
        \begin{subfigure}[b]{0.27\textwidth}
                \centering
                \includegraphics[width=\textwidth]{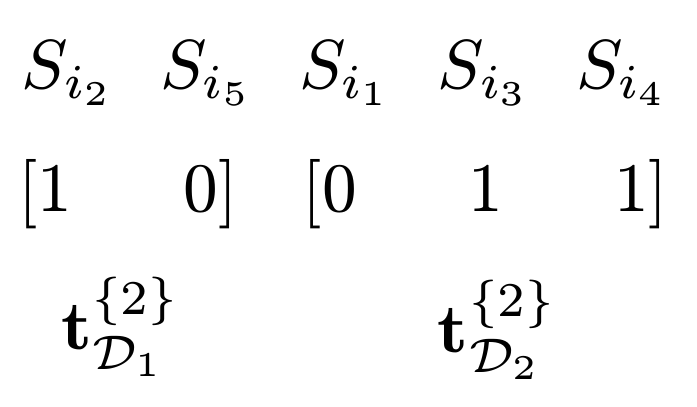}
                \caption{}
                \label{subfig:partition_vec}
        \end{subfigure}
\caption{One choice of $(\mathcal{D}^{\{2\}}_1,\mathcal{D}^{\{2\}}_2)$ and the corresponding sets $\mathbf{t}_{\mathcal{D}_1}^{\{2\}}$ and $\mathbf{t}_{\mathcal{D}_2}^{\{2\}}$ for a binary test design for $d=5$. }
\label{fig:partition}
\end{figure}

Using this model, one can define the capacity of the SQGT channel as follows. Let $\mathcal{D}^{\{i\}}_1$ and $\mathcal{D}^{\{i\}}_2$, $i\in\llbracket d\rrbracket$, form a partitions of the set of defectives, $\mathcal{D}$, such that $|\mathcal{D}^{\{i\}}_1|=i$ and $|\mathcal{D}^{\{i\}}_2|=d-i$; we denote by $\mathcal{A}_{\mathcal{D}}^{\{i\}}$ the set of all possible pairs $(\mathcal{D}^{\{i\}}_1,\mathcal{D}^{\{i\}}_2)$. 
For a single test with a possibly erroneous result $z$,  we define $\mathbf{t}_{\mathcal{D}_j}^{\{i\}}$ (where $j=1,2$) to be a row-vector of length $|\mathcal{D}^{\{i\}}_j|$, with its $k^{\textnormal{th}}$ entry equal to the sample amount of the $k^{\textnormal{th}}$ defective of $\mathcal{D}^{\{i\}}_j$ in the test. Fig.~\ref{fig:partition} shows a choice of $(\mathcal{D}^{\{2\}}_1,\mathcal{D}^{\{2\}}_2)$ and their corresponding vectors $\mathbf{t}_{\mathcal{D}_1}^{\{2\}}$ and $\mathbf{t}_{\mathcal{D}_2}^{\{2\}}$ for the case when $d=5$ and $q=2$.  

Also, let $I(\mathbf{t}_{\mathcal{D}_1}^{\{i\}};\mathbf{t}_{\mathcal{D}_2}^{\{i\}},z)$ denote the mutual information between $\mathbf{t}_{\mathcal{D}_1}^{\{i\}}$ and $(\mathbf{t}_{\mathcal{D}_2}^{\{i\}},z)$. Note that $\mathbf{t}_{\mathcal{D}_1}^{\{i\}}$ and $\mathbf{t}_{\mathcal{D}_2}^{\{i\}}$ are \emph{random} vectors. Since the amount of each subject in each test is chosen independently and with the same probability distribution, the value of $I(\mathbf{t}_{\mathcal{D}_1}^{\{i\}};\mathbf{t}_{\mathcal{D}_2}^{\{i\}},z)$ does not depend on the specific choice of $(\mathcal{D}^{\{i\}}_1,\mathcal{D}^{\{i\}}_2)$ and only depends on $i$, $P_T$, and $d$.
Let $R=\frac{\log_2 n}{m}$ denote the rate of a SQGT test matrix. Note that the frequently used alternative definition of the rate $\frac{\log_q n}{m}$ only introduces a change in the multiple constant, given that in all our derivations we assumed that the alphabet size $q$ is fixed. Using this notation, 
the capacity of a channel corresponding to the SQGT scheme is defined as follows.
 
%\begin{figure}
%\centering
%\subfigure[][]{
%\hspace{-20pt}
%\includegraphics[width=0.3\textwidth]{partition}
%\label{subfig:partition}}
%\subfigure[][]{
%\hspace{20pt}
%\includegraphics[width=0.27\textwidth]{partition_vec}
%\label{subfig:partition_vec}}
%\caption{One choice of $(\mathcal{D}^{\{2\}}_1,\mathcal{D}^{\{2\}}_2)$ and their corresponding $\mathbf{t}_{\mathcal{D}_1}^{\{2\}}$ and $\mathbf{t}_{\mathcal{D}_2}^{\{2\}}$ in a binary test design for $d=5$. }
%\label{fig:partition}
%\end{figure}

\begin{defin}[Capacity of SQGT channel]\label{def:capacity}
The capacity of a SQGT channel equals
\begin{equation}\label{eq:capacity}
C_{_\text{SQGT}}=\supr_{P_T,\boldsymbol{\eta}}{\alpha(d,P_T,\boldsymbol{\eta})},
\end{equation}
where $\alpha(d,P_T,\boldsymbol{\eta})=\min_{i=1,2,\dots,d}\frac{I(\mathbf{t}_{\mathcal{D}_1}^{\{i\}};\mathbf{t}_{\mathcal{D}_2}^{\{i\}},z)}{i}$, $\boldsymbol{\eta}=[\eta_0=0,\eta_1,\eta_2,\dots,\eta_Q]^T$, and $Q$ is fixed (i.e. the number of thresholds is fixed and is not an optimization variable).
\end{defin}
If the thresholds $\boldsymbol{\eta}$ are determined a priori by the resolution of the test equipment, the only design parameter to optimize over is $P_T$. 
On the other hand, if one is able to control the thresholds, $\boldsymbol{\eta}$ becomes a design parameter that clearly exhibits a strong influence on the capacity of the testing scheme. Henceforth, we mostly focus on the case when $\boldsymbol{\eta}$ are design parameters whose number is upper bounded
by some fixed control parameter. 

Definition~\ref{def:capacity} is a direct consequence of some modifications of the bounds on the number of tests that guarantee convergence to zero of the average probability of errors in~\cite{AS12}, namely the sufficient condition of the form  
\begin{equation}\label{sufficient}
m>\max_{i:(\mathcal{D}^{\{i\}}_1,\mathcal{D}^{\{i\}}_2)\in\mathcal{A}_{\mathcal{D}}^{\{i\}}}\frac{\log_2{{n-d}\choose i}{d\choose i}}{I(\mathbf{t}_{\mathcal{D}_1}^{\{i\}};\mathbf{t}_{\mathcal{D}_2}^{\{i\}},z)}\ \ \ \ i=1,2,\dots,d,
\end{equation}
and the necessary condition of the form
\begin{equation}\label{necessary}
m\geq\max_{i:(\mathcal{D}^{\{i\}}_1,\mathcal{D}^{\{i\}}_2)\in\mathcal{A}_{\mathcal{D}}^{\{i\}}}\frac{\log_2{{n-d+i}\choose i}}{I(\mathbf{t}_{\mathcal{D}_1}^{\{i\}};\mathbf{t}_{\mathcal{D}_2}^{\{i\}},z)}\ \ \ \ i=1,2,\dots,d.
\end{equation}
For completeness, we have provided the proof of these inequalities for the case of non-binary SQGT in Appendix~\ref{appendix:proof_ineq} and~\ref{appendix:proof_ineq2}. Further simplifications are possible by noting that for a fixed distribution $P_T$ and for fixed $\boldsymbol{\eta}$,
\begin{align}\label{eq:ineq}
\frac{I(\mathbf{t}_{\mathcal{D}_1}^{\{d\}};\mathbf{t}_{\mathcal{D}_2}^{\{d\}},z)}{d}\leq \frac{I(\mathbf{t}_{\mathcal{D}_1}^{\{d-1\}};\mathbf{t}_{\mathcal{D}_2}^{\{d-1\}},z)}{d-1}\leq \dots\leq {I(\mathbf{t}_{\mathcal{D}_1}^{\{1\}};\mathbf{t}_{\mathcal{D}_2}^{\{1\}},z)},
\end{align}
which is proved in~\cite{MM80} and~\cite{D04} for a general MAC model; since SQGT can be considered a special case of such MAC models, these inequalities hold for SQGT as well. 
The next theorem further clarifies the use of the term ``capacity'' in Definition~\ref{def:capacity}. 

\begin{theorem}\label{thm:capacity}
For the SQGT channel, the capacity equals $C_{_\text{SQGT}}=\supr_{P_T,\boldsymbol{\eta}}{I(\mathbf{t}_{\mathcal{D}_1}^{\{d\}};\mathbf{t}_{\mathcal{D}_2}^{\{d\}},z)}/{d}$, and all rates bellow capacity are achievable. 
In other words, for every rate $R<C_{_\text{SQGT}}$, there exists a test design for which the average probability of error converges to zero. Conversely, any test design with average probability of error approaching zero 
must asymptotically satisfy $R<C_{_\text{SQGT}}$.
\end{theorem}
\begin{IEEEproof}
One way to prove this theorem is by adapting the steps in the proofs given in~\cite{MM80} and~\cite{D04}. Equivalently, one can use~\eqref{sufficient}-\eqref{eq:ineq} -- we used the latter approach and provided the full proof of the claim in Appendix~\ref{app:2}. 
\end{IEEEproof}

The mutual information $I(\mathbf{t}_{\mathcal{D}_1}^{\{d\}};\mathbf{t}_{\mathcal{D}_2}^{\{d\}},z)$ in this theorem may be evaluated as follows. 
Let $W_1$ denote the $l_1$-norm of $\mathbf{t}_{\mathcal{D}_1}^{\{d\}}$. Then in the absence of noise,
\begin{align}\nonumber
&I_{\textnormal{SQ}}(\mathbf{t}_{\mathcal{D}_1}^{\{d\}};\mathbf{t}_{\mathcal{D}_2}^{\{d\}},z)=H(z|\mathbf{t}_{\mathcal{D}_2}^{\{d\}})-H(z|\mathbf{t}_{\mathcal{D}_1}^{\{d\}},\mathbf{t}_{\mathcal{D}_2}^{\{d\}})=H(z).
\end{align}
On the other hand, $\forall l\in[Q]$,
\begin{align}\nonumber
P(z = l) = P(\eta_l\leq W_1 <\eta_{l+1} )=\sum_{w_1=\eta_l}^{\eta_{l\!+\!1}-1}\!P_{W_1}(w_1),
\end{align}
where $P_{W_1}(w_1)$ is the probability mass function (PMF) of $W_1$ and can be found using 
\begin{equation}\nonumber
P_{W_1}(w_1)=P_T(t_1)*P_T(t_2)*\dots*P_T(t_d),
\end{equation}
where ``$*$'' denotes convolution of probability distributions. 
Note that when $q=2$, 
\begin{align}\nonumber
P(z = l) = \sum_{j=\eta_l}^{\eta_{l+1}-1}{d\choose j}p^j(1\!-\!p)^{d-j},
\end{align} 
with $p$ denoting the probability that a subject is present in a test.

Due to the complicated expression for the mutual information of an arbitrary distribution, a closed-form expression for the test capacity cannot be obtained. We therefore evaluated~\eqref{eq:capacity} numerically using a simple search procedure that allows us to quickly determine 
a lower bound on the capacity. Fig. 2 shows the obtained lower bound on the capacity when $q=3$, and $Q=2$ or $Q=3$. Table~\ref{table:capacity} shows one set of probability distributions and thresholds achieving this bound for $Q=3$. 

\begin{figure}
\centering
\includegraphics[width=0.5\textwidth]{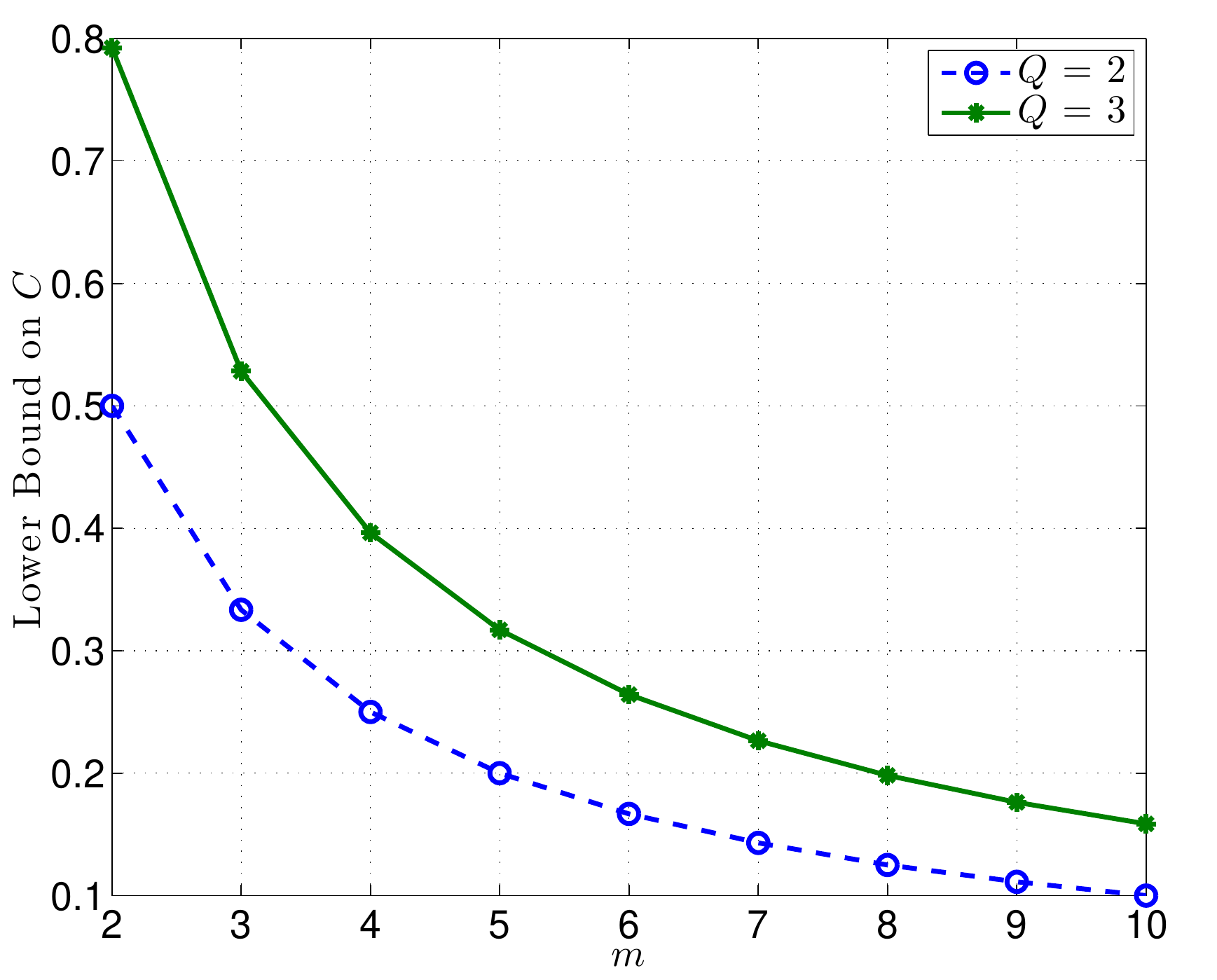}
\caption{Numerically obtained lower bounds for the capacity of SQGT schemes with $q=3$, depending on $d$. }
\vspace{-0.3cm}\label{fig:capacity}
\end{figure}

Finding the values of the thresholds that minimize the number of tests when the number of thresholds is fixed is equivalent to finding the best quantizer applied to the output of an adder MAC channel with predetermined number of quantization regions. The table in~\ref{table:capacity} reveals an interesting property of the quantizers found through numerical search: there exists at least one quantization region that consists of one or two elements only.
What this finding implies is that in order to reduce the number of tests as much as possible, some regions of the adder MAC output must be preserved with high precision. For example, by having a quantizer that assigns a unique value to an input region consisting of only one element, one is able to resolve a large amount of uncertainty about the identity of the test subjects. Furthermore, the most informative input that is left unaltered after quantization corresponds to the statistical average of the input symbols, reminiscent to the centroid of a quantization region. For example, when $d=3$, the statistical average of the adder MAC output, or, equivalently, the input of the quantizer is equal to $3\times(0\times 0.43+0.46\times 1+0.11\times 2)=2$, which is left unquantized. As another example, the input of the statistical average of the input of quantizer when $d=6$ is equal to $6\times(0\times 0.46+0.15\times 1+0.39\times 2)=5.58$ which is between the two points in the smallest cardinality quantization region $\{5,6\}$.

\begin{table}[t!]
	\centering
	\caption{A set of probability distributions and thresholds corresponding to $Q=3$ in Fig.~\ref{fig:capacity}.}
		\begin{tabular}{|c||c|c|}
			\hline 
			$d$ &  $P_T$ & quantizer\\ 
			\hline\hline
			
			$2$ & $[0.33\  0.34 \  0.33]$ & $\{0,1\}\{2\}\{3,4\}$\\
			\hline
			
			$3$ & $[0.43\  0.46\  0.11]$ & $\{0,1\}\{2\}\{3,4,5,6\}$\\			
			
			\hline
			
			$4$ & $[0.18\  0.64\  0.18]$ & $\{0,1,2,3\}\{4\}\{5,6,7,8\}$ \\
			
			\hline
			
			$5$ & $[0.15\  0.70\  0.15]$ & $\{0,1,2,3,4\}\{5\}\{6,7,8,9,10\}$\\
			
			\hline
			
			$6$ & $[0.46\  0.15\  0.39]$&  $\{0,1,2,3,4\}\{5,6\}\{7,8,\dots,12\}$\\
			
			\hline
			
			$7$ & $[0.34\  0.25\  0.41]$&  $\{0,1,\dots,6\}\{7,8\}\{9,10,\dots,14\}$\\	
			
			\hline
			
			$8$ & $[0.10\  0.80\  0.10]$&  $\{0,1,\dots,7\}\{8\}\{9,10,\dots,16\}$ \\	
			
			\hline
			
			$9$ & $[0.09\  0.82\  0.09]$ & $\{0,1,\dots,8\}\{9\}\{10,11,\dots,18\}$ \\							
			\hline
			
			$10$ & $[0.58\  0.28\  0.14]$ & $\{0,1,\dots,4\}\{5,6\}\{7,8,\dots,20\}$ \\	
			\hline					

			\end{tabular}\label{table:capacity}
			\vspace*{-10pt}	
\end{table}

%%%%%%%%%%%%%%%%%%%%%%%%%%%%%%%%%%%%%%%%%%%%%%%%%%%%%%%%%%%
\section{Conclusions}\label{sec:conclusion}

We introduced the notion of semi-quantitative group testing amenable for pooling schemes associated with high-throughput genotyping applications. We showed that the SQGT model can be considered as a unifying framework for group testing in the sense that most known group testing models are special cases of SQGT. For the novel (possibly) non-binary group testing framework, we generalized the notion of disjunct and separable codes and provided a number of combinatorial and probabilistic constructions for such codes. Furthermore, we developed a BP-decoding framework for semi-quantitative testing that may be used for testing schemes with measurement errors. 
Finally, we extended the notion of the capacity of group testing so that it applies to semi-quantitative testing, and we numerically evaluated this test invariant for a number of practical code parameters.

%%%%%%%%%%%%%%%%%%%%%%%%%%%%%%%%%%%%%%%%%%%%%%%%%%%

%%%%%%%%%%%%%%%%%%%%%%%%%%%%%%%%%%%%%%%%%%%%%%%%%%
\appendices

\section{Proof of~\eqref{sufficient} for the non-binary SQGT model}\label{appendix:proof_ineq}
The sufficient and necessary conditions in~\eqref{sufficient} and~\eqref{necessary} were proved for binary test matrices of a CGT model in~\cite{AS12}. A similar approach can be used to show that these inequalities also apply for non-binary SQGT models. For the sake of completeness, we provide a sketch of the proof of these inequalities for non-binary SQGT models using the approach of~\cite{AS12} and refer the interested reader for more details and discussions to~\cite{M78}-\cite{AS12} and~\cite{D04}.

%Throughout this section, we use upper-case letters to denote random matrices, random vectors and random variables; we use bold-face upper-case and bold-face lower-case letters to denote specific realizations of random matrices and random vectors, respectively. Consider an SQGT model with parameters defined in Section~\ref{sec:model}. 

For a matrix $\mathbf{C}\in{[q]}^{m\times n}$ and for an arbitrary set of indices $\mathcal{I}\subset \llbracket n\rrbracket$, we denote by $\mathbf{C}_{\mathcal{I}}\in{[q]}^{m\times |\mathcal{I}|}$ the submatrix consisting of the columns of $\mathbf{C}$ indexed by $\mathcal{I}$. More formally, if $\mathcal{I}=\{i_j\}_{j=1}^{|\mathcal{I}|}$ such that $i_1<i_2<\cdots<i_{|\mathcal{I}|}$, then the $j^{\text{th}}$ column of $\mathbf{C}_{\mathcal{I}}$ is equal to the ${i_j}^{\text{th}}$ column of $\mathbf{C}$, $1\leq j\leq |\mathcal{I}|$.
Similar to~\cite{AS12}, we consider a maximum likelihood (ML) decoder to find $\hat{\mathcal{D}}$ according to
\begin{align}
\hat{\mathcal{D}}=\arg\max_{\tilde{\mathcal{D}}\in\mathcal{P}_{d}(\llbracket n\rrbracket)} P\left(\mathbf{z}|\mathbf{C},\tilde{\mathcal{D}}\right)=\arg\max_{\tilde{\mathcal{D}}\in\mathcal{P}_{d}(\llbracket n\rrbracket)} P\left(\mathbf{z}|\mathbf{C}_{\tilde{\mathcal{D}}},\tilde{\mathcal{D}}\right).
\end{align}
By this definition, an error occurs if $\hat{\mathcal{D}}\neq \mathcal{D}_t$, where $\mathcal{D}_t$ is the true set of defectives. This maximization problem may not have a unique solution; therefore, we define the error event $E$ as the event that the decoder cannot find a unique set of defectives, or the event that the set recovered by the decoder is not equal to the set of true defectives. Let $E_i$, $1\leq i\leq d$, denote the event that there exists a set of subjects with cardinality $d$, differing from the true defective set in $i$ items, that is at least as likely as the true defective for the given decoder. Consequently, one has $E=\bigcup_{i=1}^d E_i$. Therefore,
\begin{align}
P(E)=P(\bigcup_{i=1}^d E_i)\leq\sum_{i=1}^d P(E_i),
\end{align}
where the inequality follows from the union bound.

On the other hand, due to the symmetry of the channel and the symmetry of code construction, $P(E_i)=P(E_i|\mathcal{D}={\mathcal{D}_t})=P(E_i|{\mathcal{D}_t})$. In other words, conditioned on $\mathcal{D}=\mathcal{D}_t$, the probability of $E_i$ does not depend on the labels chosen for the defectives, but rather depends on the codewords assigned to them; therefore, without loss of generality, one can assume that the set of defectives is a fixed set $\mathcal{D}_t$. 

For a set of defectives $\mathcal{D}_t$, let $\mathcal{G}_i(\mathcal{D}_t)$ (henceforth, $\mathcal{G}_i)$, $1\leq i\leq d$, be a set consisting of all the sets of subjects $\tilde{\mathcal{D}}\subset \llbracket n\rrbracket$, such that $|\tilde{\mathcal{D}}|=d$ and $|\tilde{\mathcal{D}}\backslash\mathcal{D}_t|=|\mathcal{D}_t\backslash\tilde{\mathcal{D}}|=i$. In other words, $\mathcal{G}_i$ is the set of all $d$-subsets of $\llbracket n\rrbracket$ that differ from $\mathcal{D}_t$ in exactly $i$ subjects. Note that $|\mathcal{G}_i|={d\choose i}{n-d\choose i}$. With this definition, conditioned on $\mathcal{D}=\mathcal{D}_t$, the error event $E_i$ can be defined as the event that there exists $\tilde{\mathcal{D}}\in\mathcal{G}_i$, such that $\tilde{\mathcal{D}}$ is at least as likely as $\mathcal{D}_t$ to the decoder. For any set $\tilde{\mathcal{D}}\in\mathcal{G}_i$, the occurrence of $E_i$ depends on the codewords assigned to the subjects in $\mathcal{D}_t$ and in $\tilde{\mathcal{D}}$. As a result, for a fixed $\mathbf{z}$ and $\mathcal{D}_t$ and for any $\tilde{\mathcal{D}}\in\mathcal{G}_i$, we can define a set of code matrices such that each code in this set assigns codewords to the subjects in a way that makes $\tilde{\mathcal{D}}$ at least as likely as $\mathcal{D}_t$ to the decoder. In order to take advantage of the results already established in~\cite{AS12}, we define this set conditioned on fixed realizations for $C_{\mathcal{D}_t\backslash\tilde{\mathcal{D}}}\in {[q]}^{m\times i}$ and $C_{\mathcal{D}_t\cap\tilde{\mathcal{D}}}\in {[q]}^{m\times (d-i)}$, namely $C_{\mathcal{D}_t\backslash\tilde{\mathcal{D}}}=\mathbf{C}_1$ and $C_{\mathcal{D}_t\cap\tilde{\mathcal{D}}}=\mathbf{C}_2$. 
For $1\leq i\leq d$ and for any $\tilde{\mathcal{D}}\in\mathcal{G}_i$, this set is denoted by $\tilde{\mathcal{E}}_{i}(\mathcal{D}_t,\tilde{\mathcal{D}},\mathbf{z},\mathbf{C}_1,\mathbf{C}_2)$, and defined as
\begin{align}\nonumber
\tilde{\mathcal{E}}_{i}(\mathcal{D}_t,\tilde{\mathcal{D}},\mathbf{z},\mathbf{C}_1,\mathbf{C}_2)=\{\mathbf{C}\  |\   \mathbf{C}_{\mathcal{D}_t\backslash\tilde{\mathcal{D}}}=\mathbf{C}_1,\   \mathbf{C}_{\mathcal{D}_t\cap\tilde{\mathcal{D}}}=\mathbf{C}_2,\  \text{and}\   P(\mathbf{z}|\mathbf{C},\tilde{\mathcal{D}})\geq P(\mathbf{z}|\mathbf{C},\mathcal{D}_t)\},
\end{align}
where $\mathbf{z}\in{[Q]}^{m}$, $\mathbf{C}\in{[q]}^{m\times n}$, $\mathbf{C}_1\in{[q]}^{m\times i}$, and $\mathbf{C}_2\in{[q]}^{m\times (d-i)}$. Now, let $\mathcal{E}_i(\mathcal{D}_t,\mathbf{z},\mathbf{C}_1,\mathbf{C}_2)$ be the union of all such sets over all $\tilde{\mathcal{D}}\in\mathcal{G}_i$, i.e.
\begin{align}\nonumber
\mathcal{E}_i(\mathcal{D}_t,\mathbf{z},\mathbf{C}_1,\mathbf{C}_2)=\bigcup_{\tilde{\mathcal{D}}\in\mathcal{G}_i}\tilde{\mathcal{E}}_{i}(\mathcal{D}_t,\tilde{\mathcal{D}},\mathbf{z},\mathbf{C}_1,\mathbf{C}_2).
\end{align}
Based on these definitions, $P(E_i)$ may be written as
\begin{align}\label{app:main}
P(E_i|\mathcal{D}_t)=\sum_{\mathbf{z}\in{[Q]}^{m}}\sum_{\mathbf{C}_1\in{[q]}^{m\times i}}\sum_{\mathbf{C}_2\in{[q]}^{m\times (d-i)}}P(\mathbf{z},\mathbf{C}_1,\mathbf{C}_2|\mathcal{D}_t) P(\mathcal{E}_i(\mathcal{D}_t,\mathbf{z},\mathbf{C}_1,\mathbf{C}_2)|\mathbf{z},\mathbf{C}_1,\mathbf{C}_2,\mathcal{D}_t).
\end{align}
At this point, one can directly apply~\cite[(A. 8)]{AS12} to obtain the following upper bound,
\begin{align}\label{app:upper1}
P(\mathcal{E}_i(\mathcal{D}_t,\mathbf{z},\mathbf{C}_1,\mathbf{C}_2)|\mathbf{z},\mathbf{C}_1,\mathbf{C}_2,\mathcal{D}_t)\leq |\mathcal{G}_i|\sum_{\mathbf{X}\in{[q]}^{m\times i}} P(\mathbf{X})\  \left(\frac{P(\mathbf{z},\mathbf{C}_2|\mathbf{X},\mathcal{D}_t)}{P(\mathbf{z},\mathbf{C}_2|\mathbf{C}_1,\mathcal{D}_t)}\right)^s,\  \  \  \  \forall s>0,
\end{align}
where ${\mathbf{X}\in{[q]}^{m\times i}} $ is a dummy variable with i.i.d. entries distributed according to $P_T$.
The proof of this inequality can be found in~\cite[(A. 8)]{AS12}. The proof exploits the symmetry of the channel and the symmetry of code construction, but is independent on the alphabet size assumption. 

A more general bound on $P(\mathcal{E}_i(\mathcal{D}_t,\mathbf{z},\mathbf{C}_1,\mathbf{C}_2)|\mathbf{z},\mathbf{C}_1,\mathbf{C}_2,\mathcal{D}_t)$ is of the form
\begin{align}\label{app:upper2}
P(\mathcal{E}_i(\mathcal{D}_t,\mathbf{z},\mathbf{C}_1,\mathbf{C}_2)|\mathbf{z},\mathbf{C}_1,\mathbf{C}_2,\mathcal{D}_t)\leq |\mathcal{G}_i|^{\rho}\left(\sum_{\mathbf{X}\in{[q]}^{m\times i}} P(\mathbf{X})\  \left(\frac{P(\mathbf{z},\mathbf{C}_2|\mathbf{X},\mathcal{D}_t)}{P(\mathbf{z},\mathbf{C}_2|\mathbf{C}_1,\mathcal{D}_t)}\right)^s\right)^{\rho},
\end{align}
for any $s>0$ and $0\leq \rho\leq 1$. The justification of this bound is as follows. Let $\Gamma$ be the upper bound on the right hand side of~\eqref{app:upper1}. If $\Gamma<1$, then ${\Gamma}^{^\rho}\geq \Gamma$ for $0\leq \rho\leq 1$. In this case, the bound in~\eqref{app:upper2} is looser than the bound in~\eqref{app:upper1}. On the other hand, if $\Gamma>1$, then $\Gamma>\Gamma^{^\rho}\geq 1$; however, since $P(\mathcal{E}_i(\mathcal{D}_t,\mathbf{z},\mathbf{C}_1,\mathbf{C}_2)|\mathbf{z},\mathbf{C}_1,\mathbf{C}_2,\mathcal{D}_t) \leq 1$, the bound in~\eqref{app:upper2} still holds.

Now, we combine~\eqref{app:upper2} and~\eqref{app:main} with the choice of $s=1/(1+\rho)$ to obtain 
\begin{align}\nonumber
P(E_i|\mathcal{D}_t)&\leq |\mathcal{G}_i|^{\rho}\sum_{\mathbf{z}}\sum_{\mathbf{C}_2}\sum_{\mathbf{C}_1}P(\mathbf{C}_1)P(\mathbf{z},\mathbf{C}_2|\mathbf{C}_1,\mathcal{D}_t)^{\frac{1}{1+\rho}}  \left(\sum_{\mathbf{X}} P(\mathbf{X})\  {P(\mathbf{z},\mathbf{C}_2|\mathbf{X},\mathcal{D}_t)}^{\frac{1}{1+\rho}}\right)^{\rho}\\\nonumber
&=|\mathcal{G}_i|^{\rho}\sum_{\mathbf{z}}\sum_{\mathbf{C}_2}\sum_{\mathbf{C}_1}\left(P(\mathbf{C}_1)P(\mathbf{z},\mathbf{C}_2|\mathbf{C}_1,\mathcal{D}_t)^{\frac{1}{1+\rho}} \right)^{1+\rho},
\end{align}
where the last equality follows since $\mathbf{X}$ is a dummy variable and can be substituted by $\mathbf{C}_1$. Since the tests are constructed independently of each other, and since the sample amount distributions are  i.i.d., one has

\begin{align}\nonumber
P(E_i|\mathcal{D}_t)&\leq \left[{d\choose i}{n-d\choose i}\right]^{\rho}\sum_{\mathbf{z}}\sum_{\mathbf{C}_2}\sum_{\mathbf{C}_1}\left(P(\mathbf{C}_1)P(\mathbf{z},\mathbf{C}_2|\mathbf{C}_1,\mathcal{D}_t)^{\frac{1}{1+\rho}} \right)^{1+\rho}\\\label{app:bound}
&=\left[{d\choose i}{n-d\choose i}\right]^{\rho}\left[\sum_{{z}}\sum_{\mathbf{t}_2}\sum_{\mathbf{t}_1}\left(P(\mathbf{t}_1)P({z},\mathbf{t}_2|\mathbf{t}_1,\mathcal{D}_t)^{\frac{1}{1+\rho}} \right)^{1+\rho}\right]^{m},
\end{align}
where $0\leq \rho\leq 1$, $z\in [Q]$, and $\mathbf{t}_1$ and $\mathbf{t}_2$ are row vectors of length $i$ and $d-i$, respectively, over the alphabet $[q]$, such that $P(\mathbf{t}_1)=\prod_{j=1}^iP_T(\mathbf{t}_1(j))$ and $P(\mathbf{t}_2)=\prod_{j=1}^{d-i}P_T(\mathbf{t}_2(j))$. Let 
\begin{align}\nonumber
\Psi(\rho)=-\log_2 \left[\sum_{{z}}\sum_{\mathbf{t}_2}\sum_{\mathbf{t}_1}\left(P(\mathbf{t}_1)P({z},\mathbf{t}_2|\mathbf{t}_1,\mathcal{D}_t)^{\frac{1}{1+\rho}} \right)^{1+\rho}\right]-\rho\  \frac{\log_2 \left[{d\choose i}{n-d\choose i}\right]}{m}.
\end{align}
Then~\eqref{app:bound} can be written as 
\begin{align}\nonumber
P(E_i)=P(E_i|\mathcal{D}_t)\leq2^{-m\Psi(\rho)}.
\end{align}
Now we can use an argument described in~\cite{G68} (and also used in~\cite{AS12}), as follows. 

Observe that $\Psi:\mathbb{R}\mapsto\mathbb{R}$ is a continuous and differentiable function in the neighborhood of $\rho_0=0$. Also, $\Psi(0)=-\log_2 \left[\sum_{{z}}\sum_{\mathbf{t}_2}\sum_{\mathbf{t}_1}P(\mathbf{t}_1)P({z},\mathbf{t}_2|\mathbf{t}_1,\mathcal{D}_t) \right]=-\log_2 \left[\sum_{{z}}\sum_{\mathbf{t}_2}P({z},\mathbf{t}_2|\mathcal{D}_t) \right]=0$. The derivative of $\Psi(\rho)$ at $\rho_0=0$ is equal to
\begin{align}\nonumber
\Psi'(0)&=\sum_{{z}}\sum_{\mathbf{t}_2}\sum_{\mathbf{t}_1}P(\mathbf{t}_1)P({z},\mathbf{t}_2|\mathbf{t}_1,\mathcal{D}_t)\left[\log_2 P({z},\mathbf{t}_2|\mathbf{t}_1,\mathcal{D}_t) -\log_2 \sum_{\mathbf{t}_1}P(\mathbf{t}_1)P({z},\mathbf{t}_2|\mathbf{t}_1,\mathcal{D}_t)\right]\\\nonumber
&\hspace{310pt}-\frac{1}{m} \log_2 \left[{d\choose i}{n-d\choose i}\right]\\\nonumber
&=\sum_{{z}}\sum_{\mathbf{t}_2}\sum_{\mathbf{t}_1}P(\mathbf{t}_1)P({z},\mathbf{t}_2|\mathbf{t}_1,\mathcal{D}_t)\left[\log_2 \frac{P({z},\mathbf{t}_2|\mathbf{t}_1,\mathcal{D}_t)}{ \sum_{\mathbf{t}_1}P(\mathbf{t}_1)P({z},\mathbf{t}_2|\mathbf{t}_1,\mathcal{D}_t)} \right]-\frac{1}{m} \log_2 \left[{d\choose i}{n-d\choose i}\right]\\\nonumber
&=\sum_{{z}}\sum_{\mathbf{t}_2}\sum_{\mathbf{t}_1}P({z},\mathbf{t}_2,\mathbf{t}_1|\mathcal{D}_t)\left[\log_2 \frac{P({z},\mathbf{t}_2,\mathbf{t}_1|\mathcal{D}_t)}{ P(\mathbf{t}_1|\mathcal{D}_t)P({z},\mathbf{t}_2|\mathcal{D}_t)} \right]-\frac{1}{m} \log_2 \left[{d\choose i}{n-d\choose i}\right]\\\nonumber
&=I(\mathbf{t}_1;\mathbf{t}_2,z|\mathcal{D}_t)-\frac{1}{m} \log_2 \left[{d\choose i}{n-d\choose i}\right]=I(\mathbf{t}_{\mathcal{D}_1}^{\{i\}};\mathbf{t}_{\mathcal{D}_2}^{\{i\}},z)-\frac{1}{m} \log_2 \left[{d\choose i}{n-d\choose i}\right],
\end{align}
where the last equality follows since $z$ only depends on $\mathbf{t}_{\mathcal{D}_1}^{\{i\}}$ and $\mathbf{t}_{\mathcal{D}_2}^{\{i\}}$ and is independent of $\mathcal{D}_t$. Now if $m>\frac{\log_2{{n-d}\choose i}{d\choose i}}{I(\mathbf{t}_{\mathcal{D}_1}^{\{i\}};\mathbf{t}_{\mathcal{D}_2}^{\{i\}},z)}$, then $\Psi'(0)>0$. Since $\Psi(\rho)$ is a continuous function in the neighborhood of $\rho_0=0$, then there exists a $\delta:0<\delta<1$, such that $\Psi(\delta)>0$. Given that $P(E_i)\leq 2^{-m\Psi(\delta)}$, one has $P(E_i)\rightarrow 0$ as $m\rightarrow\infty$. This implies that for a fixed value of $d$, $P(E)\rightarrow 0$ as $m\rightarrow\infty$, provided~\eqref{sufficient} holds.

%%%%%%%%%%%%%%%%%%%%%%%%%%%%%%%%%%
\section{Proof of~\eqref{necessary} for a non-binary SQGT model}\label{appendix:proof_ineq2}
The proof of~\eqref{necessary} for a binary CGT model was presented in~\cite[Section IV]{AS12}. The proof of the non-binary SQGT analogue follows along similar lines and is provided for the sake of completeness. 

Consider a SQGT model in which $d$ denotes the number of defectives. Also, let $\mathcal{D}^{\{i\}}_1$ and $\mathcal{D}^{\{i\}}_2$, $i\in\llbracket d\rrbracket$, form a partition of the set of defectives, $\mathcal{D}$, such that $|\mathcal{D}^{\{i\}}_1|=i$ and $|\mathcal{D}^{\{i\}}_2|=d-i$; we denote by $\mathcal{A}_{\mathcal{D}}^{\{i\}}$ the set of all possible pairs $(\mathcal{D}^{\{i\}}_1,\mathcal{D}^{\{i\}}_2)$. The inequality~\eqref{necessary} follows directly from $d$ distinct lower bounds on the necessary number of tests, namely
\begin{equation}\label{app:nec}
m\geq\frac{\log_2{{n-d+i}\choose i}}{I(\mathbf{t}_{\mathcal{D}_1}^{\{i\}};\mathbf{t}_{\mathcal{D}_2}^{\{i\}},z)}\ \ \ \ i=1,2,\dots,d,
\end{equation}
where ${(\mathcal{D}^{\{i\}}_1,\mathcal{D}^{\{i\}}_2)\in\mathcal{A}_{\mathcal{D}}^{\{i\}}}$; in this formulation,  $z\in [Q]$ is the result of a test and $\mathbf{t}_{\mathcal{D}_j}^{\{i\}}$, $j=1,2$, is a row-vector of length $|\mathcal{D}^{\{i\}}_j|$ corresponding to the sample amounts of the test assigned to the subjects in $\mathcal{D}^{\{i\}}_j$.

The intuition behind the bounds in~\eqref{app:nec} is that for any $i=1,2,\dots,d$, if $d-i$ of the defectives are known \emph{a priori}, $m$ should be large enough to ensure that the average probability of error in estimating the set of the other $i$ defectives converges to zero asymptotically. 
More formally, $\forall i\in\llbracket d\rrbracket$, let $\hat{\mathcal{D}}=\mathcal{D}^{\{i\}}_2\cup \hat{\mathcal{D}}^{\{i\}}_1$ be the recovered set of defectives, where $\mathcal{D}^{\{i\}}_2$ is the set of known defectives, $\hat{\mathcal{D}}^{\{i\}}_1=f(Z,C,\mathcal{D}^{\{i\}}_2)$ is the estimate of $\mathcal{D}\backslash\mathcal{D}^{\{i\}}_2$, $Z\in {[Q]}^m$ is the random vector of test results, $C\in{[q]}^{m\times n}$ is the random test matrix, $f:{[Q]}^m\times{[q]}^{m\times n}\times\mathcal{P}_{d-i}(\llbracket n\rrbracket)\mapsto \mathcal{P}_{i}(\llbracket n\rrbracket\backslash \mathcal{D}^{\{i\}}_2)$ is a function that determines the decoding procedure, $\mathcal{P}_{d-i}(\llbracket n\rrbracket)$ is the set of all subsets of $\llbracket n\rrbracket$ with cardinality $d-i$, and $\mathcal{P}_{i}(\llbracket n\rrbracket\backslash \mathcal{D}^{\{i\}}_2)$ is the set of all subsets of $\llbracket n\rrbracket\backslash \mathcal{D}^{\{i\}}_2$ with cardinality $i$. 

Let $E$ denote the error event $\hat{\mathcal{D}}\neq\mathcal{D}$. Then, 
\begin{align}\nonumber
P(E)=P(\hat{\mathcal{D}}\neq\mathcal{D})=P(\hat{\mathcal{D}}^{\{i\}}_1\neq \mathcal{D}^{\{i\}}_1).
\end{align}
Consequently, using the Fano's inequality~\cite{CT91}, one has
\begin{align}\label{app:temp1}
H(\mathcal{D}^{\{i\}}_1|Z,C,\mathcal{D}^{\{i\}}_2)\leq 1+P(E)\  \log_2 |\mathcal{P}_{i}(\llbracket n\rrbracket\backslash \mathcal{D}^{\{i\}}_2)|= 1+ P(E)\  \log_2{n-d+i\choose i},
\end{align}
where $H(\cdot)$ denotes the entropy function. Since the set of defectives $\mathcal{D}$ is chosen uniformly at random from $\mathcal{P}_{d}(\llbracket n\rrbracket)$, and independent of ${C}$, 
\begin{align}\label{app:temp2}
{H(\mathcal{D}^{\{i\}}_1|{C},\mathcal{D}^{\{i\}}_2)}=\log_2|\mathcal{P}_{i}(\llbracket n\rrbracket\backslash \mathcal{D}^{\{i\}}_2)|=\log_2{n-d+i\choose i}.
\end{align}
Using the definition of mutual information,
\begin{align}\nonumber
H({\mathcal{D}^{\{i\}}_1}|C,\mathcal{D}^{\{i\}}_2)&=H({\mathcal{D}^{\{i\}}_1}|Z,C,{\mathcal{D}^{\{i\}}_2})+I({\mathcal{D}^{\{i\}}_1};Z|C,\mathcal{D}^{\{i\}}_2)\\\nonumber
&=H({\mathcal{D}^{\{i\}}_1}|Z,C,{\mathcal{D}^{\{i\}}_2})+H(Z|C,\mathcal{D}^{\{i\}}_2)-H(Z|C,\mathcal{D}^{\{i\}}_2,{\mathcal{D}^{\{i\}}_1})\\\nonumber
&\leq H({\mathcal{D}^{\{i\}}_1}|Z,C,{\mathcal{D}^{\{i\}}_2})+H(Z|C_{\mathcal{D}^{\{i\}}_2})-H(Z|C_{\mathcal{D}^{\{i\}}_2},C_{\mathcal{D}^{\{i\}}_1})\\\label{app:temp3}
&= H({\mathcal{D}^{\{i\}}_1}|Z,C,{\mathcal{D}^{\{i\}}_2})+I(Z;C_{\mathcal{D}^{\{i\}}_1}|C_{\mathcal{D}^{\{i\}}_2}),
\end{align}
where the inequality follows since the test results $Z$ only depend on the codewords assigned to the set $\mathcal{D}$ and hence $H(Z|C,\mathcal{D}^{\{i\}}_2,{\mathcal{D}^{\{i\}}_1})=H(Z|C_{\mathcal{D}^{\{i\}}_2},C_{\mathcal{D}^{\{i\}}_1})$. In addition, $C_{\mathcal{D}^{\{i\}}_2}$ is a function of $C$ and ${\mathcal{D}^{\{i\}}_2}$; therefore, $H(Z|C,\mathcal{D}^{\{i\}}_2)=H(Z|C,\mathcal{D}^{\{i\}}_2,C_{\mathcal{D}^{\{i\}}_2})\leq H(Z|C_{\mathcal{D}^{\{i\}}_2})$, since conditioning reduces entropy.

Substituting~\eqref{app:temp1} and~\eqref{app:temp2} in~\eqref{app:temp3} yields
\begin{align}\nonumber
\log_2{n-d+i\choose i}&\leq 1+ P(E)\  \log_2{n-d+i\choose i}+I(Z;C_{\mathcal{D}^{\{i\}}_1}|C_{\mathcal{D}^{\{i\}}_2})\\
\nonumber 
&\Rightarrow P(E)\geq 1-\frac{I(Z;C_{\mathcal{D}^{\{i\}}_1}|C_{\mathcal{D}^{\{i\}}_2})+1}{\log_2{n-d+i\choose i}}.
\end{align}
Therefore, a necessary asymptotic condition for $P(E)\rightarrow 0$ is 
\begin{align}\nonumber
1-\frac{I(Z;C_{\mathcal{D}^{\{i\}}_1}|C_{\mathcal{D}^{\{i\}}_2})+1}{\log_2{n-d+i\choose i}}\leq 0
\  \Rightarrow \   \log_2{n-d+i\choose i}\leq I(Z;C_{\mathcal{D}^{\{i\}}_1}|C_{\mathcal{D}^{\{i\}}_2}).
\end{align}
Since the tests are independent of each other, (i.e., the tests are designed independently and the result of each test is not affected by the results of other tests), it can be easily verified that this necessary condition simplifies to
\begin{align}\nonumber
 \log_2{n-d+i\choose i}\leq \sum_{j=1}^m I(Z_j;T_{\mathcal{D}^{\{i\}}_1\!,j}\:|\:T_{\mathcal{D}^{\{i\}}_2\!,j}), 
 \end{align}
where $\forall j\in \llbracket m\rrbracket$, $Z_j\in [Q]$ is the result of the $j^{\text{th}}$ test and $T_{\mathcal{D}^{\{i\}}_1\!,j}$ and $T_{\mathcal{D}^{\{i\}}_2\!,j}$ are the $j^{\text{th}}$ rows of $C_{\mathcal{D}^{\{i\}}_1}$ and $C_{\mathcal{D}^{\{i\}}_2}$, respectively. 
In addition, due to the i.i.d. distributions of the sample amounts of each subject and the symmetry of the channel, $\forall k, j\in \llbracket m\rrbracket$, $I(Z_j;T_{\mathcal{D}^{\{i\}}_1\!,j}\:|\:T_{\mathcal{D}^{\{i\}}_2\!,j})=I(Z_k;T_{\mathcal{D}^{\{i\}}_1\!,k}\:|\:T_{\mathcal{D}^{\{i\}}_2\!,k})=I(z;\mathbf{t}_{\mathcal{D}_1}^{\{i\}}|\mathbf{t}_{\mathcal{D}_2}^{\{i\}})$. Given that $\mathbf{t}_{\mathcal{D}_1}^{\{i\}}$ and $\mathbf{t}_{\mathcal{D}_2}^{\{i\}}$ are independent, one has $I(\mathbf{t}_{\mathcal{D}_1}^{\{i\}};\mathbf{t}_{\mathcal{D}_2}^{\{i\}})=0$. Therefore, $I(\mathbf{t}_{\mathcal{D}_1}^{\{i\}};\mathbf{t}_{\mathcal{D}_2}^{\{i\}},z)=I(\mathbf{t}_{\mathcal{D}_1}^{\{i\}};\mathbf{t}_{\mathcal{D}_2}^{\{i\}})+I(z;\mathbf{t}_{\mathcal{D}_1}^{\{i\}}|\mathbf{t}_{\mathcal{D}_2}^{\{i\}})=I(z;\mathbf{t}_{\mathcal{D}_1}^{\{i\}}|\mathbf{t}_{\mathcal{D}_2}^{\{i\}})$, which completes the proof of~\eqref{app:nec}.

%%%%%%%%%%%%%%%%%%%%%%%%%%%%%%%%%%%%
\section{Proof of Theorem~\ref{thm:capacity}}\label{app:2}

\textbf{Proof of Theorem~\ref{thm:capacity}:}
First, we prove that any rate $R<C_{_\text{SQGT}}$ is achievable. Since $C_{_\text{SQGT}}=\supr_{P_T,\boldsymbol{\eta}}{\alpha(d,P_T,\boldsymbol{\eta})}$, then $\forall \epsilon>0$ there exist  $P_T'$ and $\boldsymbol{\eta}'$ such that $C_{_\text{SQGT}}-\epsilon<{\alpha(d,P_T',\boldsymbol{\eta}')}\leq C_{_\text{SQGT}}$. Let $\epsilon=C_{_\text{SQGT}}-R$ and $\alpha'={\alpha(d,P_T',\boldsymbol{\eta}')}$; then there exists a test design with parameters $P_T'$ and $\boldsymbol{\eta}'$ such that $R<\alpha'$. Generate a random code of size $n$ and length $m$ according to $P_T'$ for a test with thresholds $\boldsymbol{\eta}'$. Let $0<\epsilon'<|\alpha'-R|$. Then, 
\begin{align}\nonumber
R+\epsilon'<\alpha'\Rightarrow \frac{\log_2 n}{m}+\epsilon'<\alpha'\Rightarrow m>\frac{\log_2 n+m\epsilon'}{\alpha'}.
\end{align}
For any choice of $\epsilon'$ and sufficiently large enough values of $m$ and $n$, $m\epsilon'>\log_2(d\e^2)+\log_2\left(1-\frac{d}{n}\right)$, so that
\begin{align}\nonumber
m>\frac{\log_2 n+m\epsilon'}{\alpha'}&>\frac{\log_2(d\e^2)+\log_2\left(1-\frac{d}{n}\right)+\log_2 n}{\alpha'}\\\nonumber
&=\frac{\max_i\log_2\left(\frac{(n-d)\e}{i}\frac{d\e}{i}\right)}{\alpha'}\\\nonumber
&=\frac{\max_i\frac{1}{i}\log_2\left(\frac{(n-d)\e}{i}\frac{d\e}{i}\right)^i}{\alpha'}\\\nonumber
&>\frac{\max_i\frac{1}{i}\log_2{n-d\choose i}{d\choose i}}{\alpha'}\\\nonumber
&\geq\max_{i:(\mathcal{D}^{\{i\}}_1,\mathcal{D}^{\{i\}}_2)\in\mathcal{A}_{\mathcal{D}}^{\{i\}}}\frac{\log_2{{n-d}\choose i}{d\choose i}}{I(\mathbf{t}_{\mathcal{D}_1}^{\{i\}};\mathbf{t}_{\mathcal{D}_2}^{\{i\}},z)}.
\end{align}
Using~\eqref{sufficient}, these inequalities imply that the average probability of error converges to zero as $m,n\rightarrow\infty$.
 
Conversely, if the average probability of error converges to zero, than for any $i\in\{1,2,\dots,d\}$ one has
\begin{align}\nonumber
m\geq\frac{\log_2{{n-d+i}\choose i}}{I(\mathbf{t}_{\mathcal{D}_1}^{\{i\}};\mathbf{t}_{\mathcal{D}_2}^{\{i\}},z)}\Rightarrow \frac{I(\mathbf{t}_{\mathcal{D}_1}^{\{i\}};\mathbf{t}_{\mathcal{D}_2}^{\{i\}},z)}{i}&\geq \frac{\log_2{n-d+i\choose i}}{im}> \frac{\log_2\left(\frac{n-d+i}{i}\right)}{m}.
\end{align}
Consequently,
\begin{align}\nonumber
\alpha>\min_i \frac{\log_2\left(\frac{n-d+i}{i}\right)}{m}=\frac{\log_2\left(\frac{n}{d}\right)}{m}=R-\frac{\log_2 d}{m},
\end{align}
which in the asymptotic regime simplifies to $R<\alpha$. As a result, the inequality $R<\supr_{P_T,\boldsymbol{\eta}}{\alpha}$ holds in the asymptotic regime and therefore $R<C_{_\text{SQGT}}$.

\end{document}